\documentclass[12pt]{article}

\usepackage{microtype} % Improves appearance in pdflatex
\usepackage{amsfonts}
\usepackage{amsmath}
\usepackage{amssymb,amsthm}
\usepackage{mathabx}
\usepackage{mathrsfs}
\usepackage[round]{natbib}
\usepackage[pdftex]{graphicx}
\usepackage[T1]{fontenc}
\usepackage{lmodern}
\usepackage{titlesec}
\usepackage{rotating,multirow}
\usepackage{geometry}
\usepackage{color}
\usepackage{longtable,lscape}
\usepackage[doublespacing]{setspace}
\usepackage{bbm}
\usepackage{caption}
\usepackage{subcaption}
\usepackage{tikz}
\usepackage{pgfplots}
\usepackage{accents}
\usepackage{float}
\usepackage[todonotes={textsize=tiny}, defaultcolor=red]{changes}
\usepackage{amsthm}
\usepackage{ulem}

\pgfplotsset{compat=newest}
\pgfplotsset{plot coordinates/math parser=false}
\newlength\figureheight
\newlength\figurewidth
\usepackage{epstopdf}
\usepackage{booktabs}
\usepackage[pdftitle={A Dynamic Model of Private Asset Allocation},
  pdfauthor={Chen,Scheidegger,Xu},
  pdftoolbar=true,
  colorlinks=true,
  linkcolor=blue,
  citecolor=blue,
  bookmarksopen=true,
  bookmarksnumbered=true]{hyperref}

  \usepackage{cleveref}

\usepgfplotslibrary{fillbetween}

\usepackage{makecell}
\usepackage{dcolumn}
\newcolumntype{d}[1]{D..{#1}} % for alignment of numbers on decimal marker

\crefformat{equation}{#2(#1)#3}
\theoremstyle{definition}

\newif\ifcoverLMI
\coverLMIfalse

\begin{document}

\title{A Dynamic Model of Private Asset Allocation\thanks{We thank Liberty Mutual Investments for generous support on this project.}}

\author{Hui Chen\footnote{MIT Sloan and NBER, huichen@mit.edu} \and Giovanni Gambarotta\footnote{Liberty Mutual Investments, giovanni.gambarotta@lmi.com} \and Simon Scheidegger\footnote{Department of Economics, HEC Lausanne, simon.scheidegger@unil.ch} \and Yu Xu\footnote{Lerner College of Business and Economics, University of Delaware, yuxu@udel.edu}}
\date{\today}

\maketitle

\setstretch{1.25}

%\begin{abstract}
%We build a state-of-the-art dynamic model of private asset allocation. Like existing models, we capture the illiquidity and capital commitment lags associated with private equity. Our model further considers important features of PE investing that challenge investors, including (1) business cycle conditions, (2) serial correlation in observed PE returns, and (3) regulation-induced portfolio constraints. We use cutting-edge machine learning methods to quantify the optimal investment policies over the life cycle of a fund. Our model can also be used by regulators to assessed the costs and benefits of different choices of risk-based capital charges.
%\end{abstract}

\begin{abstract}
We build a state-of-the-art dynamic model of private asset allocation that considers five key features of private asset markets: (1) the illiquid nature of private assets, (2) timing lags between capital commitments, capital calls, and eventual distributions, (3) time-varying business cycle conditions, (4) serial correlation in observed private asset returns, and (5) regulatory constraints on certain institutional investors' portfolio choices. We use cutting-edge machine learning methods to quantify the optimal investment policies over the life cycle of a fund. Moreover, our model offers regulators a tool for precisely quantifying the trade-offs when setting risk-based capital charges.
% 97 words
\end{abstract}

\bigskip


\medskip


\thispagestyle{empty}

\newpage

\setcounter{page}{1}
\setstretch{1.25}

\parskip 3pt

\section{Introduction}

Recent estimates put the size of global private markets' assets under management at \$13.1 trillion as of June 30, 2023 \citep{mckinsey2024}. A multitude of factors make private asset allocation decisions particularly challenging. First, the illiquid nature of private assets means that it takes considerable time and costs to establish, rebalance, and exit from a position, making careful long-term planning an integral part of private asset management. Second, this illiquidity can lead to mark-to-market induced serial correlation in observed private asset returns, which further complicates dynamic allocation decisions. Third, private market investments require investors to first commit capital and subsequently manage the liquidity needs arising from future capital calls. Fourth, some institutional investors such as insurance companies face additional regulatory constraints on their allocation choices. Fifth, investors must take all of these factors into account as market and business cycle conditions constantly change.

We develop a state-of-the-art dynamic model of private asset allocation. It captures (1) the illiquidity inherent in private asset investing, (2) the lag between initial capital commitments, subsequent capital calls, and eventual distributions, (3) time-varying market and business cycle conditions, (4) serial correlation in observed alternative asset returns, and (5) regulation-induced portfolio constraints. The interplay among these factors introduces substantial nonlinearities and expands the state space, rendering the model numerically intractable with traditional solution methods. To address this challenge, we use advanced machine learning techniques\textemdash Deep Kernel Gaussian Processes \citep{pmlr-v51-wilson16}\textemdash to solve the underlying dynamic programming problem. The solution quantifies the optimal private asset allocation policies over the life cycle of a fund, and accounts for the multitude of factors described above. To the best of our knowledge, (i) our model is the most comprehensive private asset allocation model to date in terms of the range of factors considered, and (ii) we are the first to use Deep Kernel Gaussian Processes in the context of solving dynamic models in economics and finance.

%Our main contribution is to characterize the optimal solution for dynamic private asset allocation where many key features are simultaneously considered. Like the existing literature, our model captures (1) illiquidity of PE investing, and (2) implementation lags in PE investing including the process from capital commitments, to capital calls, and finally distribution. We significantly extend the existing literature by simultaneously considering a number of realistic features. This includes (3) serial correlation in alternative asset returns \citep{getmansky/lo/makarov:2004}, (4) risk-based capital constraints that institutional investors typically face, and (5) fluctuations in all of the above factors over the business cycles. The number of simultaneous features increases the number of state variables and substantial nonlinearity to the problem which makes our model numerically intractable using traditional methods. To overcome this challenge, we use state-of-the-art Machine Learning methods to make our problem numerically feasible. Specifically, we use Deep Kernel Gaussian Processes \citep{pmlr-v51-wilson16} to solve our dynamic programming problem. To the best of our knowledge, we are the first to use Deep Kernel Gaussian Processes in the context of solving dynamic models in economics and finance.

We solve a dynamic portfolio optimization problem for a risk-averse limited partner (LP) investor with a fixed (but long) horizon. The LP chooses among stocks, bonds, and a private equity (PE) fund, subject to various constraints. To invest in the PE fund, the LP must first make capital commitments before capital is called and distributed from the PE fund. The illiquid nature of PE assets makes liquidity management central: the LP must finance capital calls out of its liquid wealth (i.e., stocks and bonds). Default occurs if the LP does not have sufficient liquid wealth to meet a capital call, in which case the LP's PE stake is sold at a discount on the secondary market and the LP loses future access to the PE fund. In addition to such liquidity-driven defaults, we also allow for strategic defaults on capital calls.

Our model further incorporates three realistic features that make PE investing challenging. First, a unique feature of alternative investments is the serial correlation in observed returns \citep{geltner1991smoothing,getmansky/lo/makarov:2004}. As \citet{getmansky/lo/makarov:2004} argue, sparsely-traded illiquid assets can lead to reported PE fund returns that appear smoother than their true economic counterparts, owing to mark-to-market practices. To capture this effect, we embed the \citet{getmansky/lo/makarov:2004} return-smoothing model into our PE return structure, allowing observed PE returns to depend on past lags of true economic returns. Consequently, the LP must make allocation decisions in the presence of mark-to-market induced smoothing, as is the case in reality.

Second, key determinants of PE investing depend on business cycle conditions. For example, PE returns, capital calls, and distributions can all vary over the business cycle (see, e.g., \citealt{neubergerberman2022}). We capture these time-varying conditions by modeling a macroeconomic state that evolves according to a Markov chain. All our model parameters, including those governing PE returns, call rates, and distributions, can depend on this macroeconomic state. Consequently, our framework provides guidance for optimal PE allocations over different phases of the business cycle.

Third, many PE investors are institutional investors that must comply with regulatory constraints on their asset holdings. For example, U.S. insurers' portfolio allocations are governed by Risk-Based Capital (RBC) standards overseen by the National Association of Insurance Commissioners (NAIC), while European insurers are subject to analogous RBC requirements under Solvency II (see \citealt{eling/holzmuller:2008} for an overview of RBC standards). Violations of these requirements can be extremely costly because they trigger regulatory interventions: in severe cases, an insurer may be placed under regulatory control, which can lead to rehabilitation or even liquidation. We capture these regulatory constraints by imposing a risk budget on the LP's portfolio, applying separate risk charges to stocks, bonds, and PE. As long as the LP’s total risk charge remains below a specified threshold, no costs arise; however, costs escalate once that threshold is breached. This modeling approach aligns with risk-charge frameworks commonly used by rating agencies to assess insurers’ RBC adequacy (see, e.g., \citealt{spglobal:23}).

%The comprehensive list of factors considered in our model makes it challenging to solve the LP's dynamic portfolio optimization problem. There are two main reasons. First, the interaction of these factors generate significant nonlinearities which must be resolved in order to obtain accurate allocation policies. Second, it expands the state space which makes the model numerical intractable for traditional methods that are subject to the curse of dimensionality. To overcome these challenges, we use Deep Kernel Gaussian Processes \citep{pmlr-v51-wilson16} to track the value and policy functions associated with the LP's dynamic portfolio optimization problem.

The comprehensive set of factors considered in our model makes it challenging to solve the LP's dynamic portfolio optimization problem for two main reasons. First, the interactions among these factors introduce substantial nonlinearities that must be resolved to obtain accurate allocation policies. Second, they enlarge the state space, making traditional methods infeasible due to their vulnerability to the curse of dimensionality.

To overcome these challenges, we employ Deep Kernel Gaussian Processes (DKGPs) to track the LP's value and policy functions within our dynamic optimization framework. As proposed by \citet{pmlr-v51-wilson16}, DKGPs combine the strengths of Gaussian Processes (GPs) and Neural Networks (NNs). While both GPs and NNs can address high-dimensional challenges, GPs are generally more cost-effective to train since they involve fewer hyperparameters. This is especially relevant in our setting where each sample point requires solving a constrained optimization problem with multi-dimensional controls and adaptive quadrature to accurately compute expectations. However, standard GPs often lack the flexibility of NNs in capturing complex nonlinear behavior. DKGPs resolve this issue by embedding NNs into GPs, thereby harnessing the ability to model highly nonlinear dynamics while reducing the need for expensive sample points. While we use DKGPs to solve our model, this methodology may be of independent interest, as many models in economics and finance face similar challenges.

%Model ingredients:
%\begin{itemize}
%  \item Choices: stocks, bonds, PE
%  \item Lags in PE investing: commitment, capital calls, distributions.
%  \item PE illiquidity and liquidity management. PE liquidation costs.
%  \item NEW: Macroeconomic conditions
%  \item NEW: Risk budget: motivated using risk-based capital
%  \item NEW: Serial correlation in PE returns. \citet{getmansky/lo/makarov:2004}
%  \item NEW: life cycle dynamics
%\end{itemize}

%\begin{itemize}
%  \item Investment insights.
%  \item unconditional life cycle
%  \item Bus cycle: keep PE portfolio steady. Adjust through stocks. Costly to ignore business cycles.
%  \item Return smoothing: may be a moot point once properly accounted for.
%  \item Risk budget analysis
%\end{itemize}

\ifcoverLMI

Our calibration assumes a 10-year investment horizon for the LP, and we use private equity (PE) data from a large anonymous institutional investor to calibrate our model. The calibrated model yields several quantitative insights into optimal asset allocation over a fund's life cycle\textemdash from the initial stage, where no PE investments are held, to the transition phase, during which the LP ramps up its PE exposure, and finally to the maintenance stage, where the desired balance between PE and public investments is achieved.

\else

Our calibration assumes a 10-year investment horizon for the LP, and we use private equity (PE) data from Liberty Mutual Investments to calibrate our model. The calibrated model yields several quantitative insights into optimal asset allocation over a fund's life cycle\textemdash from the initial stage, where no PE investments are held, to the transition phase, during which the LP ramps up its PE exposure, and finally to the maintenance stage, where the desired balance between PE and public investments is achieved.

\fi

First, the optimal allocation policy unfolds as follows. In the early years, the LP aggressively commits new capital to build up the fund, taking into account the various factors described above to ensure a smooth transition. At the same time, the LP adjusts its liquid portfolio by initially allocating heavily to stocks to achieve the desired overall aggregate risk exposure as PE investments ramp up. As PE commitments are called and the PE share of the portfolio increases, the LP gradually reduces stock exposure. This transition phase lasts approximately 4–5 years, after which the LP enters a maintenance phase, consistently managing the portfolio to preserve the optimal asset mix. Notably, under the optimal policy, the cumulative default rate is only 0.1\% over the ten-year investment horizon.

Although similar outcomes are observed in practice, the complexity of the problem has traditionally led practitioners to rely on heuristics. For instance, standard approaches often use static mean-variance analysis combined with heuristic adjustments for capital commitment lags (see, e.g., \citealt{TA:02}). Our contribution is to rigorously quantify the optimal policies in a realistic setting, moving beyond rules of thumb to help PE investors achieve improved outcomes. Indeed, we demonstrate that the optimal portfolio allocations obtained under our fully dynamic framework can differ dramatically from those derived using heuristic methods.

Second, the optimal allocation accounts for business cycle conditions as follows. To first order, the optimal allocation maintains the intended transition profile for the PE portfolio, as if short-term business cycle fluctuations did not exist. This is because adjusting PE allocations can be prohibitively costly given their illiquidity, commitment lags, and long planning horizons. Instead, the LP modulates overall risk exposure over the business cycle by adjusting its public stock holdings, lowering exposure during recessions when necessary. Since transaction costs for public stocks are negligible in comparison, this approach offers a cost-effective means of managing risk over the business cycle.

Failing to account for business cycles is extremely costly. To quantify these costs, we compare the outcomes for a naive LP that ignores business cycle variation against outcomes under the optimal policy. The naive LP adopts overly aggressive PE allocations, leading to a dramatic increase in default risk: the cumulative default frequency is 13.6\% for the naive approach, compared to only 0.1\% under the optimal policy. Ex-ante, this suboptimal strategy translates to a 9.3\% loss in initial wealth in certainty-equivalent terms.

Third, we examine whether it is necessary to ``unsmooth'' returns before making allocation decisions. Prior research has argued that mark-to-market accounting induces serial correlation, making observed alternative asset returns appear smoother than their underlying true economic performance \citep{getmansky/lo/makarov:2004}. This smoothing can cause private assets' true risk to be underestimated and, consequently, lead to suboptimal portfolio allocations. As a result, some have advocated unsmoothing returns prior to drawing portfolio inferences.

Our model revisits this issue from the perspective of a long-term PE investor and demonstrates that the need to unsmooth returns may be a moot point for such an investor. The reasons are as follows. First, long-term investors care primarily about the characteristics of long-horizon returns, which are less affected by short-term mark-to-market fluctuations. Second, even if serial correlation is an inherent feature of true PE returns, its benefits may be negated once realistic implementation lags and adjustment costs are considered. Indeed, in our baseline calibration\textemdash where observed PE returns have a quarterly autocorrelation of 0.2\textemdash we observe little difference in outcomes between a LP that optimizes its portfolio directly using the raw returns and one that first unsmooths the return series to remove autocorrelation while preserving the long-run moments.

%Fourth, we can use our model to assess the impact of changes in risk charges on the LP's portfolio outcomes. Our model quantifies the risk-return tradeoff of alternative risk-charges. In the baseline calibration, we set risk charges to 50\% for both equities (public and private) which are representative numbers of average (see, e.g., \citealt[Table 14]{spglobal:23}). This number differs across asset classes. When we increase this number to 100\% (which is at the upper end), long run realized returns decrease from 8.4\% in the baseline to 7.1\% while long run realized volatility also decreases 2.78\% to 2.35\%. Therefore, our model offers both institutional investors and regulators a tool with which to assess the impact of alternative values of risk charges.

Fourth, our model can be used to evaluate the impact of varying risk charges on the LP’s portfolio outcomes by quantifying the associated risk–return trade-off. In our baseline calibration, we impose a 50\% risk charge on both public and private equities, reflecting representative industry figures (see, e.g., \citealt[Table 14]{spglobal:23}). Because these charges differ across asset classes, we also consider a scenario in which the risk charge is increased to 100\%, near the upper end of observed ranges. Under this higher risk charge scenario, the calibrated model indicates that long-run realized returns decline from 8.4\% to 7.1\%, while long-run realized volatility decreases from 2.78\% to 2.35\%.\footnote{Note that the seemingly low standard deviation is because we are reporting the standard deviation of annualized long-horizon returns over the LP's ten year investment horizon. For example, if annual returns $r_{t+k}$ are iid with a volatility of $\sigma_1$, then the standard deviation of annualized returns over a horizon of $H$ years is $\sigma\left(\frac{1}{H}(r_{t+1}+r_{t+2}+...+r_{t+H})\right)=\sigma_1/H.$\label{footnote: remark annualized vol}}  These findings underscore how our framework can help both institutional investors and regulators assess the cost and benefits of alternative risk-charge levels.

%Fourth, our model can assess the impact of changes in risk charges on the LP's portfolio outcomes by quantifying the risk-return tradeoff under alternative scenarios. In our baseline calibration, we assume risk charges of 50\% for both public and private equities—a representative figure based on industry averages (see, e.g., \citealt[Table 14]{spglobal:23}). Note that risk charges differ across asset classes. When we increase the risk charge to 100\%—at the upper end of the observed range—long-run realized returns decrease from 8.4\% in the baseline to 7.1\%, while long-run realized volatility declines from 2.78\% to 2.35\%. Thus, our model provides both institutional investors and regulators with a valuable tool to evaluate the effects of different risk-charge levels on portfolio performance.

\paragraph{Related literature.}

%We contribute to two strands of literature. First, we contribute to the literature on optimal private asset allocation; see \citet{KW:22} for a recent survey of this literature. Early industry approaches for estimating exposures and cash flows include \citet{TA:02}. More recent works focus on dynamic models with stochastic shocks and has emphasized two key features of private asset investing: the inherent illiquidity of these assets \citep{ang/etal:14, sorensen/wang/yang:2014, dimmock/wang/yang:2024} and the delays between capital commitments, calls, and distributions \citep{GS:24, GPW:24}. We build upon these prior works by further considering important features of PE investing that challenge investors, including time-varying business cycle conditions, serial correlation in observed PE returns, and regulation-induced portfolio constraints. To the best of our knowledge, our framework considers the most comprehensive set of factors amongst stochastic models of private asset allocation that characterize the optimal allocation policy.

We contribute to two strands of literature. First, we advance the research on optimal private asset allocation (see, e.g., \citet{KW:22} for a recent survey). Early industry approaches to estimating exposures and cash flows include \citet{TA:02}. More recent work has focused on dynamic models with stochastic shocks, and highlight two key features of private asset investing: the inherent illiquidity of these assets \citep{ang/etal:14, sorensen/wang/yang:2014, dimmock/wang/yang:2024} and the delays between capital commitments, calls, and distributions \citep{GS:24, GPW:24}. We build on these prior works by incorporating additional factors that pose significant challenges for PE investors, including time-varying business cycle conditions, serial correlation in observed PE returns, and regulation-induced portfolio constraints. To the best of our knowledge, our model is the most comprehensive stochastic model of private asset allocation to date.

Second, we contribute to a growing literature that uses machine learning methods to tackle portfolio choice problems featuring higher dimensions, realistic trading frictions, and complex constraints. Most of this work has focused on liquid asset markets such as stocks and government bonds (see, e.g., \citealt{gaegauf/scheidegger/trojani:2023}; \citealt{duarte/duarte/silva:2024}), leaving alternative asset classes like private equity understudied in comparison\textemdash despite their rising importance in institutional portfolios. Our contribution is to extend machine learning techniques to the alternative asset allocation setting, enabling us to characterize optimal private asset allocation under a range of realistic features. To the best of our knowledge, we are the first to deploy Deep Kernel Gaussian Processes for solving dynamic models in economics and finance. This methodological advance is of independent interest, as similar challenges frequently arise in various settings in economics and finance. For example, \citet{duarte/fonseca/goodman/parker:2022} apply neural networks to a portfolio choice problem in the household context that incorporates multiple realistic features. While their methodology characterizes optimal portfolio choice under full commitment, our approach instead characterizes the optimal time-consistent policy.
%original text
%This methodological advance is of independent interest, as similar challenges can arise in many settings in economics and finance. \citet{duarte/fonseca/goodman/parker:2022} use neural networks to tackle a portfolio choice problem with many realistic ingredients in the household context. Their methodology characterizes optimal portfolio choice assuming full commitment; in contrast, our methodology characterizes the optimal time-consistent policy.

%Exciting recent developments in solution techniques involving various machine learning methods have enabled us to tackle problems with higher dimensions, more realistic trading frictions, and more complex portfolio constraints. However, much of the literature focuses on markets for liquid assets such as stocks and government bonds, while the

\section{Model}

% ADD OVERVIEW

% \subsection{Preferences and choices}

\begin{figure}[t]
    \centering
    \resizebox{1.0\textwidth}{!}{
    \begin{tikzpicture}
  \draw[line width = 1.2pt, ->] (0,0) -- (13,0) ;
\draw (0.5,0.1) -- (0.5,-0.1) node[below,font=\large]{$t$} ;
\draw (12.5,0.1) -- (12.5,-0.1) node[below,font=\large]{$t+1$} ;

\draw[<-] (0.5,0.4) -- (0.5,0.9) node[above,align=center,font=\normalsize]{Liquid wealth: $W_t$\\[1ex]Illiquid wealth: $P_t$\\[1ex]Uncalled commitments: $K_t$\\[1ex]Macroeconomic conditions: $s_t$\\[1ex]Expected returns on PE: $\mu_{P,t}$} ;

\draw[<-] (6.5,0.4) -- (6.5,0.9) node[above,align=center,font=\normalsize]{Allocation to stocks\\and bonds:\\$S_t$ and $B_t$\\[1em]New commitments\\to PE: $N_t$} ;

\draw[<-] (12.5,0.4) -- (12.5,0.9) node[above,align=center,font=\normalsize]{PE, stock, and\\bond returns:\\ $R_{P,t+1}, R_{S,t+1}, R_f(s_t)$\\[1em]PE Distribution:\\$\lambda_D(s_{t+1})P_tR_{P,t+1}$\\[1em]Capital calls:\\$\lambda_K(s_{t+1})K_t$ and $\lambda_N(s_{t+1})N_t$\\[1em]Default on capital calls?} ;
\end{tikzpicture}
}
    \caption{\textbf{Timing of the Model}}
    \label{fig: model timing}
\end{figure}
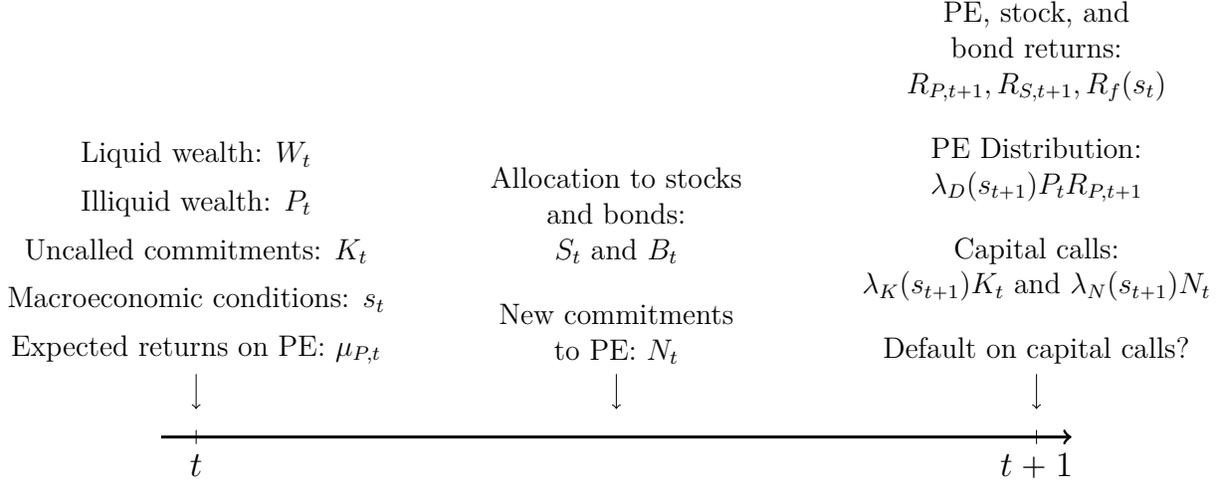

Time is discrete and runs from $t=0$ to the terminal date $t=T$. The limited partner (LP) investor maximizes utility over terminal wealth $W_T$,
\begin{equation}
u\left(W_T\right)=\frac{W_T^{1-\gamma}}{1-\gamma},
\end{equation}
where $\gamma$ is the coefficient of relative risk aversion. At each time period $t$, the LP can invest in a risk-free bond, a public stock index, and private equity (PE). The timing of the decisions is illustrated in \Cref{fig: model timing} and is described below.

\paragraph{Time-varying macroeconomic conditions.}
The LP's investment opportunities set varies over the business cycle. We capture time-varying business cycle conditions through a Markov process $s_t\in\{1,2\}$ that takes two values corresponding to recessions ($s_t=1$) and booms ($s_t=2$). We denote by $p_{ss^\prime}=\mbox{prob}(s_{t+1}=s^\prime|s_t=s)$ the probability of transitioning from state $s$ to state $s^\prime$.

\paragraph{Liquidity constraint.}
The LP enters period $t$ with liquid wealth $W_t\geq0$, uncalled PE commitments $K_t\geq0$, and illiquid wealth $P_t\geq0$. The latter corresponds to the net asset value (NAV) of the LP's previously called PE investments. The LP then makes three choices: (1) new PE commitments $N_t\geq0$, and its allocations to (2) stocks $S_t$ and (3) bonds $B_t$. These choices are subject to the following liquidity constraint:
\begin{equation}\label{eq: model liq constraint}
    S_t+B_t+\gamma_N (W_t+P_t)\left(\frac{N_t}{W_t+P_t}-\overline{n}\right)^2+\gamma_S(W_t+P_t)\left(\frac{S_t}{W_t+P_t} - \overline{s} \right)^2=W_t
\end{equation}
where $\gamma_N (W_t+P_t)\left(\frac{N_t}{W_t+P_t}-\overline{n}\right)^2$ and $\gamma_S(W_t+P_t)\left(\frac{S_t}{W_t+P_t}-\overline{s}\right)^2$ are adjustment costs on new PE commitments and stocks, respectively.\footnote{In \autoref{app: scaled value functions}, we show that the problem is homogeneous in total wealth $W_t+P_t$. The forms of these adjustment costs conveniently preserves this homogeneity property.} That is, stock and bond allocations and adjustment costs are paid out of liquid wealth.

Additionally, we show in \autoref{app: scaled value functions} that solvency concerns rule out shorting so that
\begin{equation}\label{eq: no shorting}
    B_t\geq0,\quad\quad S_t\geq0 .
\end{equation}

\paragraph{Risk budget.}

The LP's overall portfolio is subject to internal risk controls which we model through risk weights. Specifically, the risk weight of the LP's overall portfolio is defined as
\begin{equation}\label{eq: risk weight}
    \theta_t=\frac{\theta_B B_t + \theta_S \left[S_t + \gamma_S (W_t+P_t)\left(\frac{S_t}{W_t+P_t}\right)^2\right] + \theta_P P_t}{B_t + S_t + \gamma_S (W_t+P_t)\left(\frac{S_t}{W_t+P_t}\right)^2 + P_t}
\end{equation}
The interpretation of the risk weight \eqref{eq: risk weight} is as follows. The LP's combined portfolio consists of $B_t$ invested in bonds, $S_t + \gamma_S (W_t+P_t)\left(\frac{S_t}{W_t+P_t}\right)^2$ in stocks (including stock adjustment costs), and $P_t$ in PE. Bonds carry a risk weight of $\theta_B$, while stocks and PE carry risk weights of $\theta_S$ and $\theta_P$, respectively. The risk weight \eqref{eq: risk weight} is the value-weighted average over the risk weights of the three classes of investments in the LP's portfolio. We assume the risk weights are ordered as follows:
\begin{equation}\label{eq: risk weight ordering}
0=\theta_{B}<\theta_{S}\leq\theta_{P}.
\end{equation}
That is, risk-free bonds carry a zero risk weight as they are risk free while public stocks and PE carry positive risk weights. The risk weight for PE is no lower than that of public stocks.

The LP incurs a cost if it violates its risk budget. Specifically, the cost of violating its risk budget equals $\left(W_{t}+P_{t}\right)\times \Gamma\left(\theta_{t}\right)$ where the proportional cost equals
\begin{equation}\label{eq: proportional risk cost}
\Gamma\left(\theta_{t}\right)=\kappa\left(\theta_{t}-\overline{\theta}\right)^{2}1_{\left\{ \theta_{t}>\overline{\theta}\right\} }.
\end{equation}
That is, no costs are assessed if the LP's risk weight \eqref{eq: risk weight} falls below the threshold $\overline{\theta}$; quadratic costs are assessed above the threshold $\overline{\theta}$. Without loss of generality, we can take the threshold to be $\overline{\theta}=1$.\footnote{We can always rescale $\kappa$ and the risk weights $\theta_S$ and $\theta_P$ appropriately to effectively enforce $\overline{\theta}=1$.} The risk budget cost is paid during period $t+1$ in a manner we describe below.

% fill in the rest. include the following bullet points:
% 1. the risk weights reflect the different riskiness of the assets which matters for how RBCs are computed
% 2. think of $\kappa$ as the shadow cost of the contribution of the PE investments arm to the overal RBC of the company

Our modeling of the risk budget through equations \eqref{eq: risk weight} and \eqref{eq: proportional risk cost} is in line with regulatory constraints faced by institutional investors. For example, U.S. insurers' exposure to risky assets are subject to Risk-Based Capital (RBC) standards developed by the National Association of Insurance Commissioners (NAIC); similarly, the European Solvency II directive imposes RBC requirements on European insurers (see \citealt{eling/holzmuller:2008} for an overview of Risk-Based Capital standards). Such RBC requirements are also reflected in rating agencies' assessments of insurers' RBC adequecy (see, e.g., \citealt{spglobal:23}). The risk-weights $\theta_B$, $\theta_S$, and $\theta_P$ in our setting capture risk charges used to calculate RBC adequecy. The cost parameter $\kappa$ can be interpreted as the shadow cost of the contribution of an insurer's PE investments to its overall RBC adequency.

\paragraph{Laws of motion.}

New PE commitments affect future uncalled commitments through the law of motion
\begin{equation}\label{eq: law of motion K}
    K_{t+1}=K_t + N_t - \underbrace{\left[\lambda_N(s_{t+1})N_t + \lambda_K(s_{t+1})K_{t} \right]}_{\mbox{capital calls}}.
\end{equation}
New commitments $N_t$ are added to the outstanding pool of uncalled commitments $K_t$ before capital calls are subsequently made. A fraction $\lambda_N(s_{t+1})$ and $\lambda_K(s_{t+1})$ of new and previously uncalled commitments are called at the start of the next period $t+1$, respectively. Both fractions depend on next period's business cycle conditions through $s_{t+1}$. The amount of outstanding PE commitments at the start of period $t+1$, $K_{t+1}$, is then given by equation \eqref{eq: law of motion K}.

The next period's illiquid wealth is determined through the law of motion
\begin{equation}\label{eq: law of motion P}
P_{t+1}=P_{t}R_{P,t+1}-\lambda_{D}\left(s_{t+1}\right)P_{t}R_{P,t+1}+\lambda_{K}\left(s_{t+1}\right)K_{t}+\lambda_{N}\left(s_{t+1}\right)N_{t}.
\end{equation}
The NAV of the LP's PE investments at the start of period $t$, $P_t$, earns risky gross returns $R_{P,t+1}$ so that $P_t R_{P,t+1}$ is the updated NAV. Two adjustments are subsequently made. First, a fraction $\lambda_D(s_{t+1})$ of the updated NAV is distributed to the LP by its PE funds; these distributions depend on business cycle conditions. Second, the NAV increases by the LP's new PE contributions which equals capital calls $\lambda_N(s_{t+1})N_t + \lambda_K(s_{t+1})K_{t}$.

The LP's liquid wealth evolves as follows:
\begin{align}
W_{t+1}= &\,\lambda_{D}\left(s_{t+1}\right)R_{P,t+1}P_{t}+R_{S,t+1}S_{t}+R_{f}\left(s_{t}\right)B_{t}\nonumber\\
&\,-\lambda_{K}\left(s_{t+1}\right)N_{t}-\lambda_{N}\left(s_{t+1}\right)N_{t}-\left(W_{t}+P_{t}\right)\Gamma\left(\theta_{t}\right). \label{eq: law of motion W}
\end{align}
The next period's liquid wealth $W_{t+1}$ consists of distributions from its PE funds totaling $\lambda_{D}\left(s_{t+1}\right)R_{P,t+1}P_{t}$ and the proceeds from the LP's investments in public stocks $R_{S,t+1}S_{t}$ and risk-free bonds $R_{f}\left(s_{t}\right)B_{t}$. Here, $R_{S,t+1}$ and $R_f(s_t)$ denote the gross risky return on stocks and the gross risk-free return on bonds, respectively. Additionally, $W_{t+1}$ is decreased by capital calls totaling $\lambda_{K}\left(s_{t+1}\right)N_{t}+\lambda_{N}\left(s_{t+1}\right)N_{t}$ and risk budget costs $\left(W_{t}+P_{t}\right)\Gamma\left(\theta_{t}\right)$.

Finally, we assume that illiquid wealth can be converted into liquid wealth one-for-one at the terminal date $t=T$, and that any uncalled commitments at $t=T$ are discarded.

\paragraph{Default and liquidation.}

If the investor does not have sufficient liquid wealth to meet capital
calls at $t+1$, it is considered a ``default.'' This occurs if the realized return shocks at $t+1$ are such that next period's liquidity \eqref{eq: law of motion W} is negative: $W_{t+1}<0$. In case of default, all uncalled commitments are written off, the LP's PE holdings are liquidated at a discount $\alpha(s_{t+1})<1$, and the investor loses access to PE investments in the future. That is, the LP enters default with only liquid wealth totaling
\begin{align}
W_{D,t+1}= &\,\lambda_{D}\left(s_{t+1}\right)R_{P,t+1}P_{t}+R_{S,t+1}S_{t}+R_{f}\left(s_{t}\right)B_{t}\nonumber\\
&\,-\left(W_{t}+P_{t}\right)\Gamma\left(\theta_{D,t}\right)+\alpha(s_{t+1})\left[1-\lambda_D(s_{t+1})\right]R_{P,t+1}P_t. \label{eq: law of motion W default}
\end{align}
The differences between the post-default liquid wealth \eqref{eq: law of motion W default} and its counterpart in the absence of default \eqref{eq: law of motion W} are as follows. First, capital calls are not paid should the LP choose to default. Second, the liquid wealth \eqref{eq: law of motion W default} includes the proceeds from liquidating the LP's PE holdings, $\alpha(s_{t+1})\left[1-\lambda_D(s_{t+1})\right]R_{P,t+1}P_t$. Third, since PE holdings are liquidated and no longer a part of the LP's portfolio at $t+1$, the risk weight \eqref{eq: risk weight} is modified accordingly in default:
\begin{equation}\label{eq: risk weight default}
    \theta_{D,t}=\frac{\theta_B B_t + \theta_S \left[S_t + \gamma_S (W_t+P_t)\left(\frac{S_t}{W_t+P_t}\right)^2\right]}{B_t + S_t + \gamma_S (W_t+P_t)\left(\frac{S_t}{W_t+P_t}\right)^2 + P_t}.
\end{equation}
That is, compared to the risk weight \eqref{eq: risk weight}, the risk weight in default \eqref{eq: risk weight default} does away with the $\theta_P P_t$ term in the numerator. The risk cost \eqref{eq: proportional risk cost} is then calculated using the risk weight in default \eqref{eq: risk weight default}.\footnote{This is necessary for utilities to be well-defined\textemdash see \autoref{app: scaled value functions} for details.}

The LP can also choose to strategically default. Strategic default occurs when the LP has sufficient liquidity $W_{t+1}$ at $t+1$ to meet capital calls, but chooses not to do so. The LP's decision to strategically default is endogenously determined as part of the LP's optimization problem and is characterized in \Cref{sec: recursive formulation}.

\paragraph{Asset returns.}

The gross return at $t+1$ for the risk-free bond $R_{f}\left(s_{t}\right)$ depends on the macroeconomic state at time $t$, $s_t$, whose dynamics follows a first-order Markov chain.

The gross returns for private equity $R_{P,t+1}$ and the public stock index $R_{S,t+1}$ follow a log-normal distribution:
\begin{equation}
\log\left[\begin{array}{c}
R_{P,t+1}\\
R_{S,t+1}
\end{array}\right]\sim\mathcal{N}\left(\left[\begin{array}{c}
\mu_{P,t}\\
\mu_{S}\left(s_{t}\right)
\end{array}\right],\left[\begin{array}{cc}
\sigma_{P}\left(s_{t}\right)^{2} & \rho\left(s_{t}\right)\sigma_{P}\left(s_{t}\right)\sigma_{S}\left(s_{t}\right)\\
\rho\left(s_{t}\right)\sigma_{P}\left(s_{t}\right)\sigma_{S}\left(s_{t}\right) & \sigma_{S}\left(s_{t}\right)^{2}
\end{array}\right]\right).\label{eq: asset returns specification}
\end{equation}
Public stocks have expected return $\mu_S(s_t)$, volatility $\sigma_S(s_t)$, and correlation $\rho(s_t)$ with private equity returns; all of these quantities depend on macroeconomic conditions through $s_t$.

The expected return to investing in private equity follows the law of motion
\begin{equation}\label{eq: muP law of motion}
    \mu_{P,t} = \varrho_{P,1} \mu_{P,t-1} + \varrho_{P,2} \log R_{P,t} + \nu_P(s_t),
\end{equation}
where $\nu_P(s_t)$ captures business cycle variation in private equity returns. The specification \eqref{eq: muP law of motion} allows for serial correlation in private equity returns\textemdash a hallmark feature of alternative investments (see, e.g., \citealt{getmansky/lo/makarov:2004}). To see this, write $\log R_{P,t}=\mu_{P,t-1}+\sigma_{P}(s_{t-1})\varepsilon_{P,t}$ where $\varepsilon_{P,t}\equiv\left(\log R_{P,t}-\mu_{P,t-1}\right)/\sigma_{P}\left(s_{t-1}\right)\sim\mathcal{N}\left(0,1\right)$ is a standard normal shock. Equation \eqref{eq: muP law of motion} can then be expressed as
\begin{equation}
\mu_{P,t}=(\varrho_{P,1}+\varrho_{P,2})\mu_{P,t-1} + \varrho_{P,2}\sigma_P(s_{t-1})\varepsilon_{P,t}+\nu_P(s_t)\label{eq: muP AR1 process}
\end{equation}
from which we see $\varrho_{P,1}+\varrho_{P,2}$ is the autocorrelation coefficient for $\mu_{P,t}$. In \autoref{app: expected PE returns}, we show that equation \eqref{eq: muP AR1 process} can be motivated through the \citet{getmansky/lo/makarov:2004} model of smoothed returns for illiquid alternative investments.

\subsection{Recursive Formulation}\label{sec: recursive formulation}

The problem has six state variables: liquid wealth $W$, illiquid wealth $P$, uncalled PE commitments $K$, expected PE returns $\mu_P$, the macroeconomic state $s$, and time $t$. In the notation that follows, we drop time subscripts and use a prime to denote variables for the next period (e.g., $W^\prime$ denotes $W_{t+1}$).

The problem after defaulting can be expressed recursively as follows:
\begin{equation}\label{eq: default problem}
    V_D(t,W,s) = \max_{S\geq0,B\geq0} \mathbb{E}\left[V_D(t+1,W^\prime,s^\prime)\left|W,s\right.\right]
\end{equation}
subject to
\begin{align}
W^\prime & = R_S^\prime S+ R_f(s)B - W\Gamma(\theta), \nonumber\\
    \theta &=\frac{\theta_B B + \theta_S\left[S+\gamma_S W \left(\frac{S}{W}\right)^2\right]}{B+S+\gamma_S W \left(\frac{S}{W}\right)^2},\nonumber\\
    W & = S + B + \gamma_S W \left(\frac{S}{W}\right)^2,\nonumber
\end{align}
with the terminal condition being $$V_D(T,W,s)=\frac{W^{1-\gamma}}{1-\gamma}.$$
Note that the problem after defaulting \eqref{eq: default problem} does not depend on $K$, $P$, and $\mu_P$ because the LP loses access to PE investing after defaulting.

The problem before defaulting is given by
\begin{align}\label{eq: nondefault problem}
    & V(t,W,P,K,\mu_P,s)\\
    =&\max_{N\geq0, S\geq0, B\geq 0 } \mathbb{E}\left[\max\left\{V(t+1,W^\prime,P^\prime,K^\prime,\mu_P^\prime,s^\prime), V_D(t+1,W_D^\prime,s^\prime)\right\}\left|W,P,K,\mu_P,s\right.\right]\nonumber
\end{align}
subject to
\begingroup
\allowdisplaybreaks
\begin{align}
W^\prime & =\lambda_D(s^\prime)R_P^\prime P + R_S^\prime S + R_f(s) B - \lambda_K(s^\prime)K-\lambda_N(s^\prime)N - (W+P)\Gamma(\theta), \nonumber\\
P^\prime & =\left[1 - \lambda_D(s^\prime)\right]R_P^\prime P + \lambda_K(s^\prime)K + \lambda_N(s^\prime)N,\nonumber\\
K^\prime & =\left[1-\lambda_K(s^\prime)\right]K+\left[1-\lambda_N(s^\prime)\right]N,\nonumber\\
\mu_P^\prime&=\varrho_{P,1}\mu_P + \varrho_{P,2}\log R_P^\prime + \nu_P(s^\prime),\nonumber\\
W&=S+B+\gamma_N(W+P)\left(\frac{N}{W+P}-\overline{n}\right)^2+\gamma_S(W+P)\left(\frac{S}{W+P}\right)^2,\nonumber\\
\theta&=\frac{\theta_B B + \theta_S \left[S + \gamma_S (W+P)\left(\frac{S}{W+P}\right)^2\right] + \theta_P P}{B + S + \gamma_S (W+P)\left(\frac{S}{W+P}\right)^2 + P},\nonumber\\
W_D^\prime & = \left[\lambda_D(s^\prime)+\alpha(s^\prime)(1 - \lambda_D(s^\prime))\right]R_P^\prime P + R_S^\prime S + R_f(s)B-(W+P)\Gamma(\theta_D),\nonumber\\
\theta_D & = \frac{\theta_B B + \theta_S \left[S + \gamma_S (W+P)\left(\frac{S}{W+P}\right)^2\right]}{B + S + \gamma_S (W+P)\left(\frac{S}{W+P}\right)^2 + P},\nonumber
\end{align}
\endgroup
and the terminal condition
$$V(T,W,P,K,\mu_P,s)=\frac{\left(W+P\right)^{1-\gamma}}{1-\gamma}.$$

We additionally impose the boundary condition $V(t,W^\prime,P^\prime,K^\prime,\mu_P^\prime,s^\prime)=-\infty$ if $W^\prime<0$ since the LP has no option but to default if it does not have sufficient liquid wealth to meet capital calls. Finally, note that the $\max$ term in the objective function of problem \eqref{eq: nondefault problem} captures the LP's strategic default decision\textemdash even with sufficient liquidity to meet capital calls, the LP can choose to default if the value of defaulting next period is higher than the value of not defaulting.

\paragraph{Scaling.}

The value functions \eqref{eq: default problem} and \eqref{eq: nondefault problem} are homogeneous in total wealth $W+P$ and can be scaled as follows:
\begin{align}
    V_D(t,W,s) & = \frac{\left[Wv_D(t,s)\right]^{1-\gamma}}{1-\gamma}\label{eq: def scaled value fun default}\\
    V(t,W,P,K,\mu_P,s) & =\frac{\left[(W+P)v(t,w,k,\mu_P,s)\right]^{1-\gamma}}{1-\gamma}\label{eq: def scaled value fun not default}
\end{align}
where $w\equiv W/(W+P)$ is the liquid fraction of the LP's total wealth and $k\equiv K/(W+P)$ is the LP's uncalled commitments relative to its total wealth.

The scaled value function $v$ represents the certainty equivalent values for terminal wealth per unit of current total wealth, while the scaled value function $v_D$ provides the certainty equivalent values conditional on the LP being in default. Both scaled value functions are characterized by Bellman equations, as summarized in \Cref{app: scaled value functions}. We solve these scaled value functions using the numerical techniques described in \Cref{sec: solution technique}; \Cref{sec: numerical implementation} provides further details. 

\section{Solution Technique}\label{sec: solution technique}

Our goal is to compute the global solution for the scaled value functions outlined towards the end of \Cref{sec: recursive formulation}. This requires us to track the value function $v(t,w,k,\mu_P,s)$ over time; unfortunately, the large number of state variables make it challenging to do so using traditional methods. We instead use Machine Learning methods\textemdash Deep Kernel Gaussian Process dynamic programming \textemdash to make the problem computationally feasible. We outline the key steps below and refer readers to \citet{rasmussen/book:2005} for a textbook treatment of Gaussian Processes (GPs), and \citet{scheideggerbilinois_2019} and \citet{renner/scheidegger:2025} for an introduction to using GPs in a dynamic programming context.

To the best of our knowledge, we are the first to use Deep Kernel GPs in the context of solving dynamic models in economics and finance.

\subsection{Gaussian Process Regression}

Let $f:\mathbb{R}^D\rightarrow\mathbb{R}$ be a multivariate function of interest. In our context, $f(\cdot)$ is either a value function or a policy function. We can measure $f(\boldsymbol{x})$ for input $\boldsymbol{x}\in\mathbb{R}^D$ by querying an information source. We allow for noise in the information source: we measure $y=f(\boldsymbol{x})+\epsilon$ where $\epsilon\sim \mathcal{N}(0,\sigma_n^2)$ captures measurement noise.

In our setting, the information source is computer code that solves an optimization problem at point $x$ in the state space. We make queries at $N$ sample points $\boldsymbol{X}=\left\{\boldsymbol{x}^{(1)}, ..., \boldsymbol{x}^{(N)}\right\}$ and observe the corresponding measurements $\boldsymbol{y}=\left\{y^{(1)}, ..., y^{(N)}\right\}$. Computational constraints limit the number $N$ of measurements that we can make as individual function evaluations are computationally expensive. The key idea of Gaussian Process Regression (GPR) is to replace the computationally expensive function $f(\cdot)$ with a cheap-to-evaluate surrogate learned from the training inputs $\boldsymbol{X}$ and training targets $\boldsymbol{y}$.

GPR constructs the surrogate as follows. Before making any queries, we model our prior knowledge of $f(\cdot)$ by assigning it a GP prior:
$$f(\boldsymbol{x})\sim \mathcal{GP}\left(m(\boldsymbol{x};\boldsymbol{\theta}),k(\boldsymbol{x},\boldsymbol{x}^\prime;\boldsymbol{\theta})\right).$$
That is, $f(\boldsymbol{x})$ is a GP if any finite collection of function values $\{f(\boldsymbol{x}_1),...,f(\boldsymbol{x}_M)\}$ has a joint Gaussian distribution with mean function $m(\boldsymbol{x};\theta)=\mathbb{E}[f(\boldsymbol{x})]$ and covariance kernel $k(\boldsymbol{x},\boldsymbol{x}^\prime;\theta)=\mathbb{E}[(f(\boldsymbol{x})-m(\boldsymbol{x};\boldsymbol{\theta}))(f(\boldsymbol{x}^\prime)-m(\boldsymbol{x}^\prime;\boldsymbol{\theta}))]$. The vector $\boldsymbol{\theta}$ denotes the hyperparameters of the GP model which must be estimated.

A typical specification for the mean function is simply a constant,
\begin{equation}
    m(\boldsymbol{x};\theta)=m_0, \label{eq: constant mean function}
\end{equation}
while an often-used kernel is the Matern 5/2 kernel. The automatic relevance detection (ARD) Matern 5/2 kernel is defined by
\begin{equation}\label{eq: ardmatern52}
    % k(x,x^\prime)=\sigma_f^2\exp\left(-\frac{(x-x^\prime)^2}{2\ell^2}\right),
    %k(x,x^\prime)=\sigma_f^2\left(1+\frac{\sqrt{5}r}{\ell}+\frac{5r^2}{3\ell^2}\right)\exp\left(-\frac{\sqrt{5}r}{\ell}\right),
    k_{\mbox{matern52}}(\boldsymbol{x},\boldsymbol{x}^\prime)=\sigma_f^2\left(1+\sqrt{5}r+\frac{5}{3}r^2\right)\exp\left(-\sqrt{5}r\right),
\end{equation}
where
\begin{equation*}
    r=\sqrt{\sum_{d=1}^D\frac{(x_d-x_d^\prime)^2}{\ell_d^2}}
\end{equation*}
and $x_d$ is the value of $\boldsymbol{x}$ along dimension $d$. The parameter $\sigma_f$ modulates the output amplitude of the GP and is referred to as the signal standard deviation. The parameter $\ell_d$ is the characteristic lengthscale of along dimension $d$; it captures how quickly the GP changes as the input changes along dimension $d$. The resulting hyperparameters under specification \eqref{eq: constant mean function} and \eqref{eq: ardmatern52} are $\boldsymbol{\theta}=[m_0, \sigma_f, \ell_1, ..., \ell_D]$.

% In our numerical implementation, we take the mean function to be a constant; we discuss our choice of the covariance kernel in \Cref{sec: deep kernel}.

We use the training data $\boldsymbol{X}$ and $\boldsymbol{y}$ to update our beliefs regarding $f(\boldsymbol{x})$. Specifically, the posterior for $f(\boldsymbol{x})$ is still a GP with posterior mean and variance given by
\begin{align}
    \widetilde{m}(\boldsymbol{x};\boldsymbol{\theta})&=m(\boldsymbol{x};\boldsymbol{\theta}) + \boldsymbol{K}(\boldsymbol{x},\boldsymbol{X};\boldsymbol{\theta})^T\left[\boldsymbol{K}+\sigma^2_n \boldsymbol{I}_N\right]^{-1}(\boldsymbol{y}-\boldsymbol{m}),\label{eq: GP posterior mean}\\
    \mbox{and }Var\left(f(\boldsymbol{x})|\boldsymbol{X},\boldsymbol{y};\boldsymbol{\theta}\right)&=k(\boldsymbol{x},\boldsymbol{x};\boldsymbol{\theta})-\boldsymbol{K}(\boldsymbol{x},\boldsymbol{X};\boldsymbol{\theta})^T\left[\boldsymbol{K}+\sigma^2_n \boldsymbol{I}_N\right]^{-1}\boldsymbol{K}(\boldsymbol{x},\boldsymbol{X};\boldsymbol{\theta}),\label{eq: GP posterior variance}
\end{align}
respectively. Here, $\boldsymbol{K}(\boldsymbol{x},\boldsymbol{X};\boldsymbol{\theta})$ denotes a column vector with $i$th entry $k(\boldsymbol{x}, \boldsymbol{x}^{(i)};\boldsymbol{\theta})$, $\boldsymbol{K}$ denotes a $N\times N$ matrix with $ij$th entry $k(\boldsymbol{x}^{(i)}, \boldsymbol{x}^{(j)};\boldsymbol{\theta})$, $\boldsymbol{I}_N$ denotes the $N\times N$ identity matrix, and $\boldsymbol{m}$ is a column vector with $i$th entry $m(\boldsymbol{x}^{(i)};\boldsymbol{\theta})$.

The posterior mean \eqref{eq: GP posterior mean} serves as the cheap-to-evaluate surrogate for $f(\boldsymbol{x})$ that we seek. Finally, the hyperparameters $\boldsymbol{\theta}$ must be carefully chosen to ensure that the surrogate accurately represents $f(\boldsymbol{x})$. This is done by maximizing the marginal likelihood for the training data:
\begin{equation}
    \log p(\boldsymbol{y}|\boldsymbol{X})=-\frac{1}{2}\boldsymbol{y}^T\left[\boldsymbol{K}+\sigma^2_n \boldsymbol{I}_N\right]^{-1}\boldsymbol{y}-\frac{1}{2}\log\left|\boldsymbol{K}+\sigma^2_n \boldsymbol{I}_N\right|-\frac{N}{2}\log \left(2\pi\right).  \label{eq: GP marginal likelihood}
\end{equation}

\begin{figure}[t]
    \centering
    \includegraphics[width=0.8\linewidth]{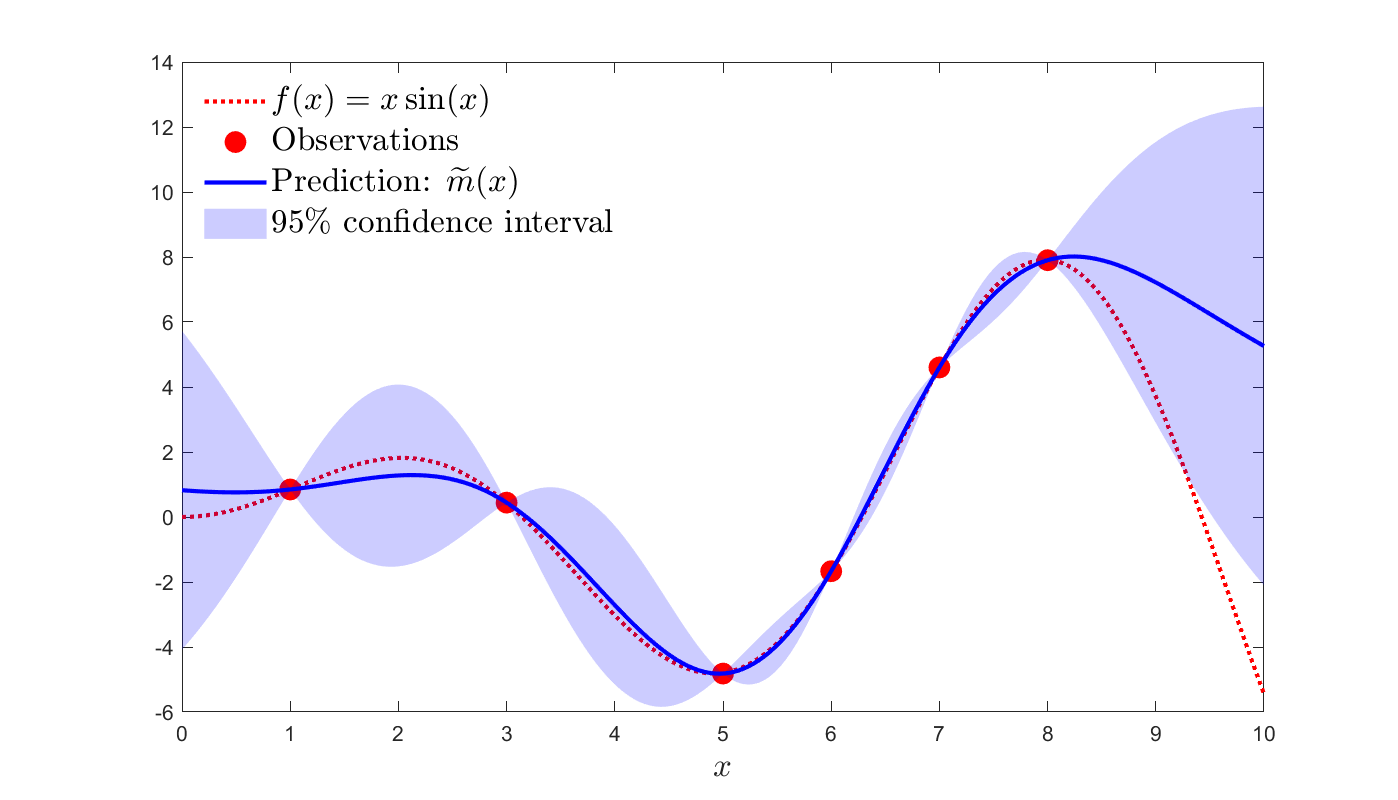}
    \caption{\textbf{Illustration of Gaussian Process Regression.} \small The dotted is the true function while the solid line is the posterior mean of the fitted GP. The shaded region plots the 95\% confidence interval.}
    \label{fig: illus GPR}
\end{figure}

\Cref{fig: illus GPR} illustrates a GPR in a one-dimensional setting in which the true function is $f(x)=x\sin(x)$. The example is for $N=5$ observations with observation noise $\sigma_n=0.01$. The true function is modeled as a GP with specification \eqref{eq: constant mean function} and \eqref{eq: ardmatern52} for the mean and covariance kernel, respectively. The posterior mean is calculated according to equation \eqref{eq: GP posterior mean} and is shown in the solid line. The shaded region plots the 95\% confidence interval around the posterior mean and is computed based on the posterior variance \eqref{eq: GP posterior variance}.

\subsection{Deep Kernel Gaussian Process}
\label{sec: deep kernel}

We use the Deep Kernel Gaussian Process (DKGP) proposed by \citet{pmlr-v51-wilson16}. A DKGP is a GP whose covariance kernel embeds a neural network component within a ``standard'' kernel as follows:
\begin{equation}\label{eq: def deep kernel}
    k_{\mbox{DK}}(\boldsymbol{x}, \boldsymbol{x}^\prime; \boldsymbol{\theta}, \boldsymbol{w}) = k\left(NN(\boldsymbol{x}; \boldsymbol{w}),NN(\boldsymbol{x}^\prime; \boldsymbol{w});\boldsymbol{\theta}\right)
\end{equation}
where $NN(\cdot; \boldsymbol{w})$ is a neural network with hyperparameters $\boldsymbol{w}$, and $k(\cdot, \cdot; \boldsymbol{\theta})$ is a ``regular'' kernel such as the Matern 5/2 kernel \eqref{eq: ardmatern52}. That is, a deep kernel \eqref{eq: def deep kernel} first uses a neural network $NN(\cdot;\boldsymbol{w})$ to process the input $\boldsymbol{x}$ before feeding the extracted feature(s) $NN(\boldsymbol{x};\boldsymbol{w})$ into a standard kernel $k(\cdot, \cdot; \boldsymbol{\theta})$.

%The hyperparameters of the neural network $\boldsymbol{w}$ are trained jointly with those of the regular kernel $\boldsymbol{\theta}$.

\begin{figure}[t]
    \centering
    \includegraphics[width=1.0\linewidth]{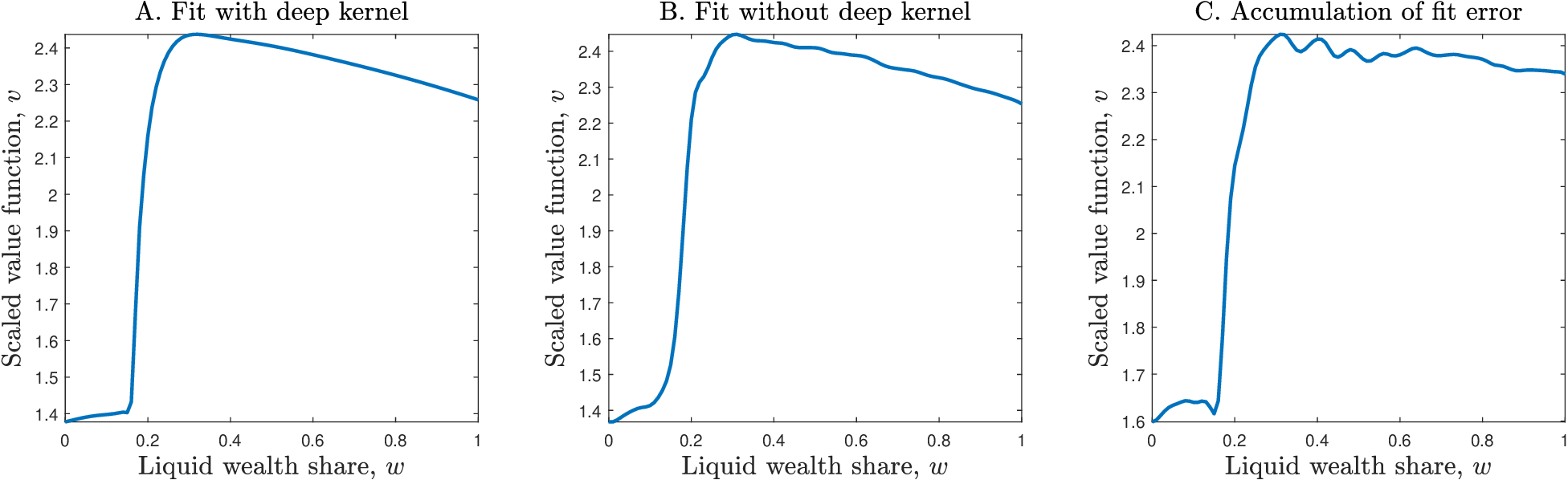}
    \caption{\textbf{Illustration of necessity of Deep Kernel Gaussian Process.} \small Panel A illustrates the value function when the model is solved using DKGP at every single step. Panel B illustrates the solution from a single backward induction step when the value function is fitted using a standard kernel. Panel C illustrates the accumulation of error over successive steps of backward induction when the value function is fitted using a standard kernel.}
    \label{fig: illus DKGP}
\end{figure}

To see why it is necessary to use a deep kernel in our setting, consider \Cref{fig: illus DKGP}. Panel A illustrates a slice of the scaled value function $v(t,w,k,\mu_P,s)$ along the liquid wealth share $w$ dimension (holding all other variables fixed). We see that the value function consists of two relatively flat regions, below $w=0.15$ and above $w=0.30$, with a sharp transition in the intermediate region between $w=0.15$ and $w=0.30$.

The sharp transition in the value function is a generic feature of our problem and is due to the possibility of default. The logic is as follows. Defaulting next period is likely when current liquidity $w$ is low; in this case, the current period's value function primarily depends on the value of default $v_D(t+1, s^\prime)$ next period which is low. Conversely, defaulting next period becomes unlikely when current liquidity $w$ is high and the current period's value function primarily depends on the much larger non-default value next period $v(t+1,w^\prime,k^\prime,\mu_P^\prime,s^\prime)$. The sharp transition for intermediate liquidity $w$ captures the transition between defaulting and not defaulting next period.

Sharp transitions are challenging for standard kernels to capture. The reason is that standard kernels such as the Matern kernel \eqref{eq: ardmatern52} feature a constant lengthscale along each dimension. In contrast, the value function in our setting features several lengthscales\textemdash long lengthscales are necessary to capture the slowly-varying value function when $w$ is low or high, while a short lengthscale is needed to capture the fast-varying value function for intermediate values of $w$. In addition, these lengthscales vary over the entire state space as both the location and speed of the sharp transition depend on all state variables.

Panels B and C of \Cref{fig: illus DKGP} illustrates the pitfalls of using a single lengthscale to fit a multi-lengthscale value function. Maximum likelihood estimation fits a short lengthscale to capture the sharp transition. The resulting fitted GP is fast-varying and contains wiggles (see panel B). These wiggles accumulate over successive steps of backward induction and ultimately results in a very noisy value function after sufficiently many steps (see panel C of \Cref{fig: illus DKGP}).

The deep kernel \eqref{eq: def deep kernel} captures state-dependant lengthscales through its neural network component. This can be seen in Panel A of \Cref{fig: illus DKGP} which plots the fitted value function when a DKGP is used.

\section{Quantitative Analysis}

%\textcolor{red}{[WORK IN SOMEWHERE]To capture return smoothing, we will calibrate local vol $\sigma_{P}\left(s_{t}\right)$
%to be smaller. The point is that return smoothing makes returns autocorrelated
%which still makes LONG TERM returns volatile. So pulling the wool
%over investors' eyes with return smoothing won't make much of a difference
%long term.}

We begin by calibrating the model in \Cref{sec: calibration} and describing the model's solution in \Cref{sec: value and policy functions}. We then report our model's implications for optimal private asset allocation. \Cref{sec: life cycle dynamics} and \Cref{sec: business cycle dynamics} describe the optimal allocations over the LP's life cycle and how business cycle conditions alter those allocations, respectively. \Cref{sec: serial correlation in returns} shows that, for long-term PE investors, whether or not PE returns must first be ``unsmoothed'' prior to making inferences and allocation decisions may be a moot point. \Cref{sec: risk budget} examines the impact of the risk weights on the LP's portfolio allocation. \autoref{sec: numerical implementation} provides details on the numerical implementation of the model.

\subsection{Calibration}
\label{sec: calibration}

We use the parameters listed in \Cref{tbl: parameters}. We calibrated the model at a quarterly frequency and take the investment horizon to be $T=40$ quarters or 10 years.

We take state $s=1$ to be a recessionary state and state $s=2$ to be an expansionary state. We set the transition probabilities for the macroeconomic state to $p_{12}=0.25$ and $p_{21}=0.05$ based on the duration of NBER recessions in the post-WWII sample. This implies recessions and expansions last for 4 and 20 quarters on average, respectively. In what follows, we use NBER recession dates when estimating parameter values that depend on the macroeconomic state.

\begin{table}[t]
  \centering
  \begin{tabular}{lccc}
    \toprule
    \multicolumn{1}{c}{Description} & Symbol & \multicolumn{2}{c}{Value}\\
    & & $s=1$ & $s=2$\\
    \midrule
    %\\[-1ex]
%    \multicolumn{2}{c}{(a) Constant parameters} &  & \\
%    \midrule
     Investment horizon, quarters & $T$ & \multicolumn{2}{c}{40}\\
%
%     [1em]
%
%    \multicolumn{2}{c}{(b) State-dependant parameters} & $s=1$ & $s=2$\\
%    \midrule
    Macro transition probability & $p_{ss^\prime}$ & 0.25 & 0.05 \\
    PE call rate, new commitments & $\lambda_N(s)$ & 0.18 & 0.047 \\
    PE call rate, existing commitments & $\lambda_K(s)$ & 0.050 & 0.078 \\
    PE distribution rate & $\lambda_D(s)$ & 0.028 & 0.071 \\
    PE liquidation discount & $\alpha(s)$ & 0.66 & 0.90 \\
    Risk-free rate & $\log R_f(s)$ & 0.0028 & 0.0051 \\
    Stock expected return & $\mu_S(s)$ & 0.0079 & 0.0238 \\
    Stock return volatility & $\sigma_S(s)$ & 0.1493 & 0.0829 \\
    Stock-PE return correlation & $\rho(s)$ & 0.9527 & 0.4575 \\
    PE return volatility & $\sigma_P(s)$ & 0.0768 & 0.0424 \\
    PE expected return autocorrelation & $\varrho_{P,1}+\varrho_{P,2}$ & \multicolumn{2}{c}{0.201} \\
    PE expected return state-dependance & $\nu_P(s)$ & 0.0024 & 0.0317 \\
    Risk aversion & $\gamma$ & \multicolumn{2}{c}{2}\\
     PE adjustment costs & $\gamma_N$ & \multicolumn{2}{c}{0.1}\\
     PE adjustment costs & $\overline{n}$ & \multicolumn{2}{c}{0}\\
     Stock adjustment costs & $\gamma_S$ & \multicolumn{2}{c}{0.01}\\
     Risk budget threshold & $\overline{\theta}$ & \multicolumn{2}{c}{1}\\
     Risk budget cost & $\kappa$ & \multicolumn{2}{c}{1}\\
     Risk weight, bonds & $\theta_B$ & \multicolumn{2}{c}{0}\\
     Risk weight, stocks & $\theta_S$ & \multicolumn{2}{c}{1.5}\\
     Risk weight, PE & $\theta_P$ & \multicolumn{2}{c}{1.5}\\
    \bottomrule
  \end{tabular}
  \caption{\textbf{Parameters.} \small We simulate the model at a quarterly frequency using the parameters in this table.}\label{tbl: parameters}
\end{table}

\ifcoverLMI
\else

We use quarterly data provided by Liberty Mutual Investments (LMI) on its PE holdings to calibrate various PE-related parameters. This data contains LMI's total PE exposure broken down by asset class\textemdash buyout, growth, and venture capital. For each asset class, the data details total exposure over time ($K_t+P_t$ in the model) and breaks this exposure down into outstanding commitments ($K_t$) and the NAV of current PE investments ($P_t$). The data also include information on LMI's contributions, distributions received from its PE investments, and the number of investees LMI invests in. In estimating parameters, we restrict attention to buyout funds from the moment when the number of buyout investees first reach 10. The resulting sample starts in 1996Q4 and ends in 2023Q4; all estimates are based on this sample period.

$\mbox{Contribution}_t$ in the LMI data corresponds to $\lambda_K(s_t)K_{t-1} + \lambda_N(s_t)N_{t-1}$ in the model. We estimate the call rates using the following regression:
\begin{equation}
  \frac{\mbox{Contribution}_t}{K_{t-1}} = a(s_t) + b(s_t)\frac{N_{t-1}}{K_{t-1}}+\epsilon_t,\label{eq: regression estimate call rates}
\end{equation}
where the constant $a(s_t)$ maps to $\lambda_K(s_t)$ and the slope coefficient $b(s_t)$ maps $\lambda_N(s_t)$. We estimate regression \eqref{eq: regression estimate call rates} separately for recessionary ($s=1$) and expansionary ($s=2$) regimes to obtain state-dependant estimates for the call rates. The resulting point estimates give $\lambda_K(1)=0.050$ and $\lambda_N(1)=0.18$ during recessions, and $\lambda_K(2)=0.078$ and $\lambda_N(2)=0.047$ during expansions.

We set the PE payout rate $\lambda_D(s)$ to be the average payout rate in the LMI dataset. This results in $\lambda_D(1)=0.028$ and $\lambda_D(2)=0.071$ during recessions and expansions, respectively.

\fi

We calibrate the discount from liquidating PE investments $\alpha(s)$ based on estimates of transaction costs in the secondary market for selling PE stakes from \citet[Table 1]{NADAULD2019158}. They find that the purchase price, expressed as a pecent of NAV, is 86\% on average and drops to 66\% during 2008-09 (the only recession in their sample). We set $\alpha(1)=0.66$ and $\alpha(2)=0.90$ during recessions and expansions, respectively, based on these estimates.

We calibrate the returns processes as follows. We take the series for the risk-free rate and aggregate stock returns from Kenneth French's website. We set the risk-free rate and expected stock returns to their sample averages during recessions and expansions; this results in $\log R_f(1)=0.0028$ and $\log R_f(2)=0.0051$ for the risk-free rate and $\mu_S(1)=0.0079$ and $\mu_S(2)=0.0238$ for expected stock returns.

\ifcoverLMI
  % This code will be included since \coverLMI is true.
  The anonymous intitutional investor's data includes quarterly mark-to-market information for the NAV of its PE investments; this allows us to compute the realized PE return $R_{P,t}$ as the ratio of the mark-to-market and the lagged NAV value. We use moments of the resulting time series for $R_{P,t}$ to estimate the PE return process. We set the stock-PE return correlation $\rho(s)$ to its sample correlation; this results in $\rho(1)=0.9527$ and $\rho(2)=0.4575$ during recessions and expansions, respectively. We restrict $\varrho_{P,1}=\varrho_{P,2}$ and choose the remaining parameters of the PE expected return process \eqref{eq: muP law of motion} and PE return volatility $\sigma_P(s)$ by matching the following data moments: (1) a PE expected return $\mathbb{E}\left[\log R_{P,t+1}|s_t\right]$ of 0.0052 and 0.0392 during recessions and expansions, respectively, (2) a PE return volatility $\sigma\left(\log R_{P,t+1}|s_t\right)$ of 0.0772 and 0.0427 during recessions and expansions, respectively, and (3) a PE return autocorrelation of $\rho\left(R_{P,t}, R_{P,t+1}\right)=0.1425$. This results in $\varrho_{P,1}=\varrho_{P,2}=0.1006$, $\nu_P(1)=0.0024$ and $\sigma_P(1)=0.0768$ during recessions, and $\nu_P(2)=0.0317$ and $\sigma_P(2)=0.0424$ during expansions.
\else
  % This code will be excluded.
  LMI data includes quarterly mark-to-market information for the NAV of its PE investments; this allows us to compute the realized PE return $R_{P,t}$ as the ratio of the mark-to-market and the lagged NAV value. We use moments of the resulting time series for $R_{P,t}$ to estimate the PE return process. We set the stock-PE return correlation $\rho(s)$ to its sample correlation; this results in $\rho(1)=0.9527$ and $\rho(2)=0.4575$ during recessions and expansions, respectively. We restrict $\varrho_{P,1}=\varrho_{P,2}$ and choose the remaining parameters of the PE expected return process \eqref{eq: muP law of motion} and PE return volatility $\sigma_P(s)$ by matching the following data moments: (1) a PE expected return $\mathbb{E}\left[\log R_{P,t+1}|s_t\right]$ of 0.0052 and 0.0392 during recessions and expansions, respectively, (2) a PE return volatility $\sigma\left(\log R_{P,t+1}|s_t\right)$ of 0.0772 and 0.0427 during recessions and expansions, respectively, and (3) a PE return autocorrelation of $\rho\left(R_{P,t}, R_{P,t+1}\right)=0.1425$. This results in $\varrho_{P,1}=\varrho_{P,2}=0.1006$, $\nu_P(1)=0.0024$ and $\sigma_P(1)=0.0768$ during recessions, and $\nu_P(2)=0.0317$ and $\sigma_P(2)=0.0424$ during expansions.
\fi

We set the risk weights for stocks and private equity to $\theta_S=1.5$ and $\theta_P=1.5$, respectively. Since the overall risk budget $\overline{\theta}$ is normalized to 1, these risk weights imply a 50\% risk charge for stocks and private equity. These values are in line with equity risk charges used by ratings agencies for assessing insurers' RBC adequency (see \citealt[Table 14]{spglobal:23}). The risk weight for risk-free bonds is $\theta_B=0$. We set the cost parameter for the proportional risk cost \eqref{eq: proportional risk cost} to $\kappa=1$.

We set risk aversion to $\gamma=2$ which is standard. The remaining parameters relate to adjustment costs. We set $\gamma_N=0.1$ and $\overline{n}$ for PE adjustment costs, and $\gamma_S=0.01$ for stock adjustment costs.

% \subsection{Results}

\subsection{Value and policy functions}
\label{sec: value and policy functions}

We begin by illustrating the solution for the scaled value function and the policy functions; recall that the scaled and unscaled value functions are related through equations \eqref{eq: def scaled value fun default} and \eqref{eq: def scaled value fun not default}.

\paragraph{Value functions.}

\begin{figure}[t]
  \centering
  \includegraphics[width=1.0\textwidth]{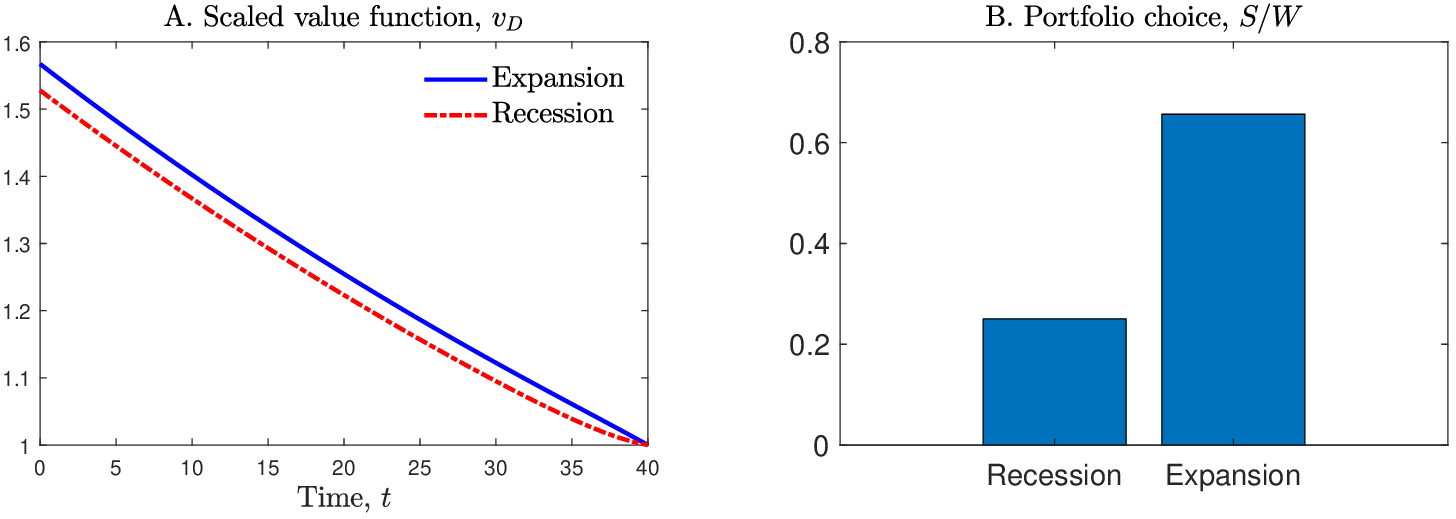}
  \caption{\textbf{Solution after defaulting.} \small Panel A plots the scaled value function after defaulting $v_D(t,s)$; it is related to the unscaled value function through equation \eqref{eq: def scaled value fun default}. Panel B plots the optimal allocation to stocks after defaulting.}\label{fig: illus default solution}
\end{figure}

\Cref{fig: illus default solution} plots the solution after defaulting. Panel A plots the scaled value function $v_D(t,s)$. The value function is declining over time since the investment horizon over which to accumulate wealth shrinks over time. The value function is also higher during expansions than recessions. Panel B plots the portfolio choice after defaulting. The investor's allocation depends only on the macroeconomic state and does not vary over time\textemdash the optimal allocation to stocks is 25\% and 66\% during recessions and expansions, respectively; all remaining wealth are allocated to bonds.

Next, we illustrate the solution before defaulting during which the scaled value function $v(t,w,k,\mu_P,s)$ depends on time $t$, the liquid fraction of total wealth $w=W/(W+P)$, uncalled commitments relative to total wealth $k=K/(W+P)$, the expected return for PE $\mu_P$, and the macroeconomic state $s$.

\begin{figure}[t]
  \centering
  \includegraphics[width=1.0\textwidth]{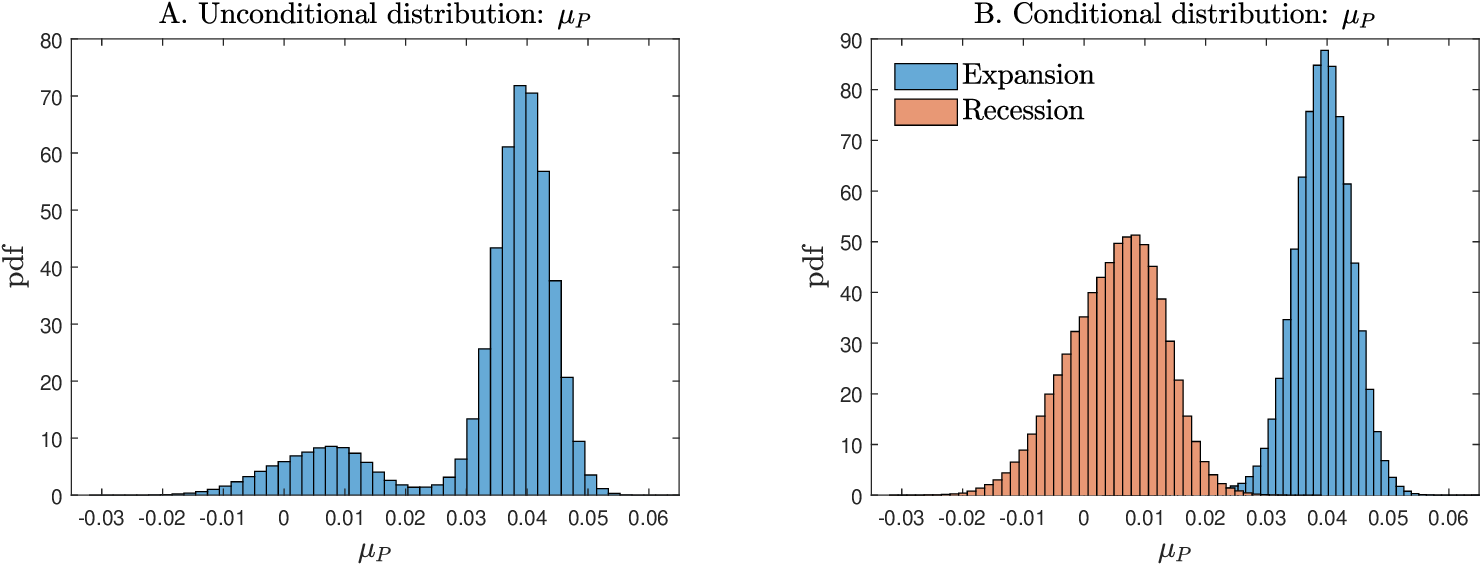}
  \caption{\textbf{Distribution for $\mu_P$.} \small Panel A plots the stationary distribution for $\mu_P$ whose law of motion is given by equation \eqref{eq: muP law of motion}. Panel B plots the distribution for $\mu_P$ conditional on the current macroeconomic state.}\label{fig: illus distribution muP}
\end{figure}

\Cref{fig: illus distribution muP} plots the distribution of the expected PE return $\mu_P$. Panel A plots the unconditional distribution while panel B plots the conditional distribution given the macroeconomic state. Expected returns for PE are higher during expansions under our baseline calibration with expected returns averaging $\mathbb{E}[\mu_{P,t}|s_t=1]=0.0052$ and $\mathbb{E}[\mu_{P,t}|s_t=2]=0.0392$ per quarter during recessions and expansions, respectively. From panel B, we also see that the distribution of expected PE returns is wider during expansions than recessions; this is on account of higher PE return volatilities during recessions.

\begin{figure}[htbp]
  \centering
  % Subfigure 1 (top)
  \begin{subfigure}[t]{\textwidth}
    \centering
    % Replace 'example-image-a' with your actual image file name
    \includegraphics[width=\textwidth]{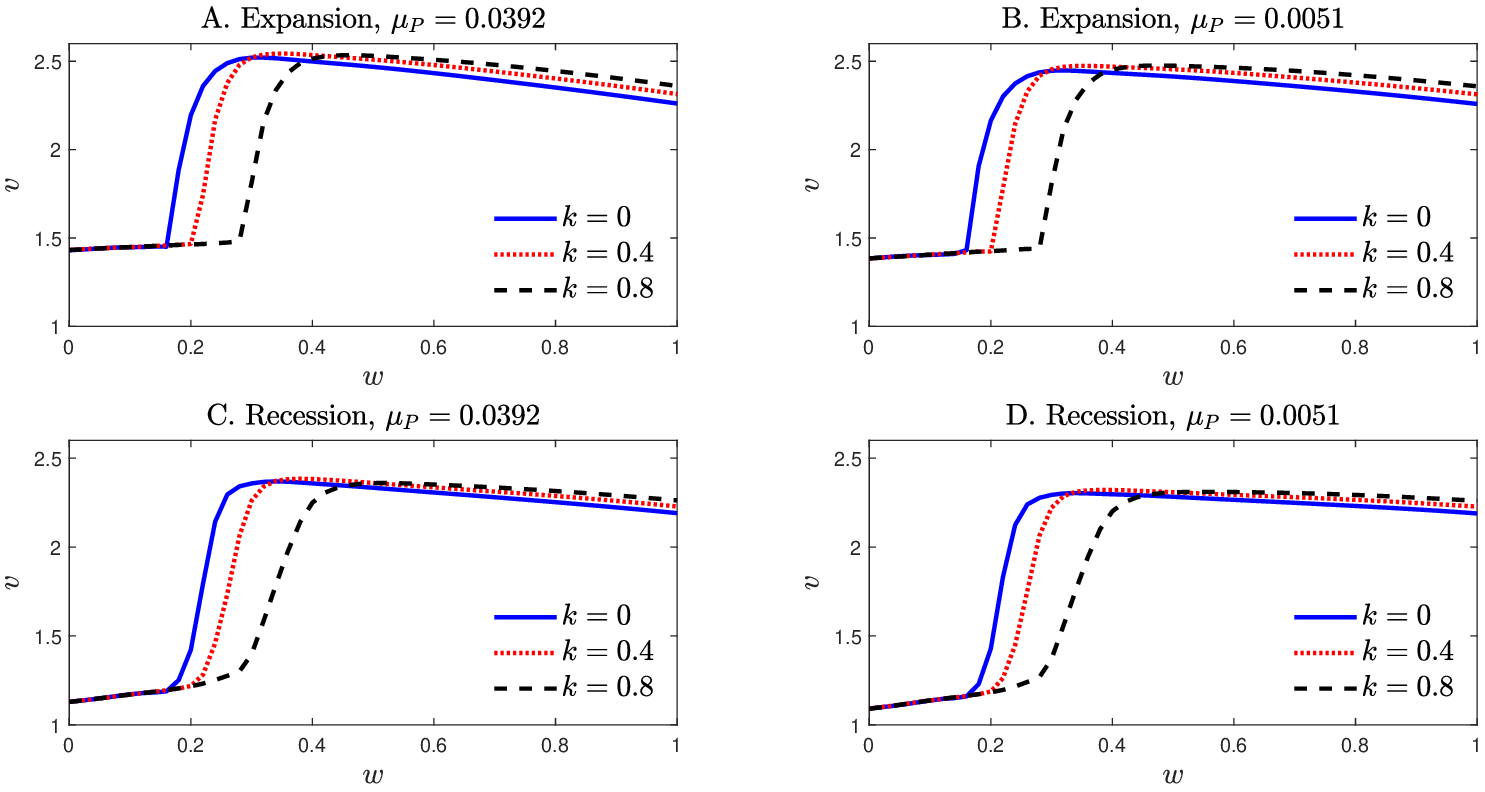}
    \caption{Slices of the value function along the $w$ dimension.}
    \label{fig: illus value fun 2D}
  \end{subfigure}

  \vspace{1em} % Optional extra vertical space between subfigures

  % Subfigure 2 (bottom)
  \begin{subfigure}[b]{\textwidth}
    \centering
    % Replace 'example-image-b' with your actual image file name
    % \includegraphics[width=\textwidth]{figures/illus_value_fun_3D.eps}
    \includegraphics[width=\textwidth]{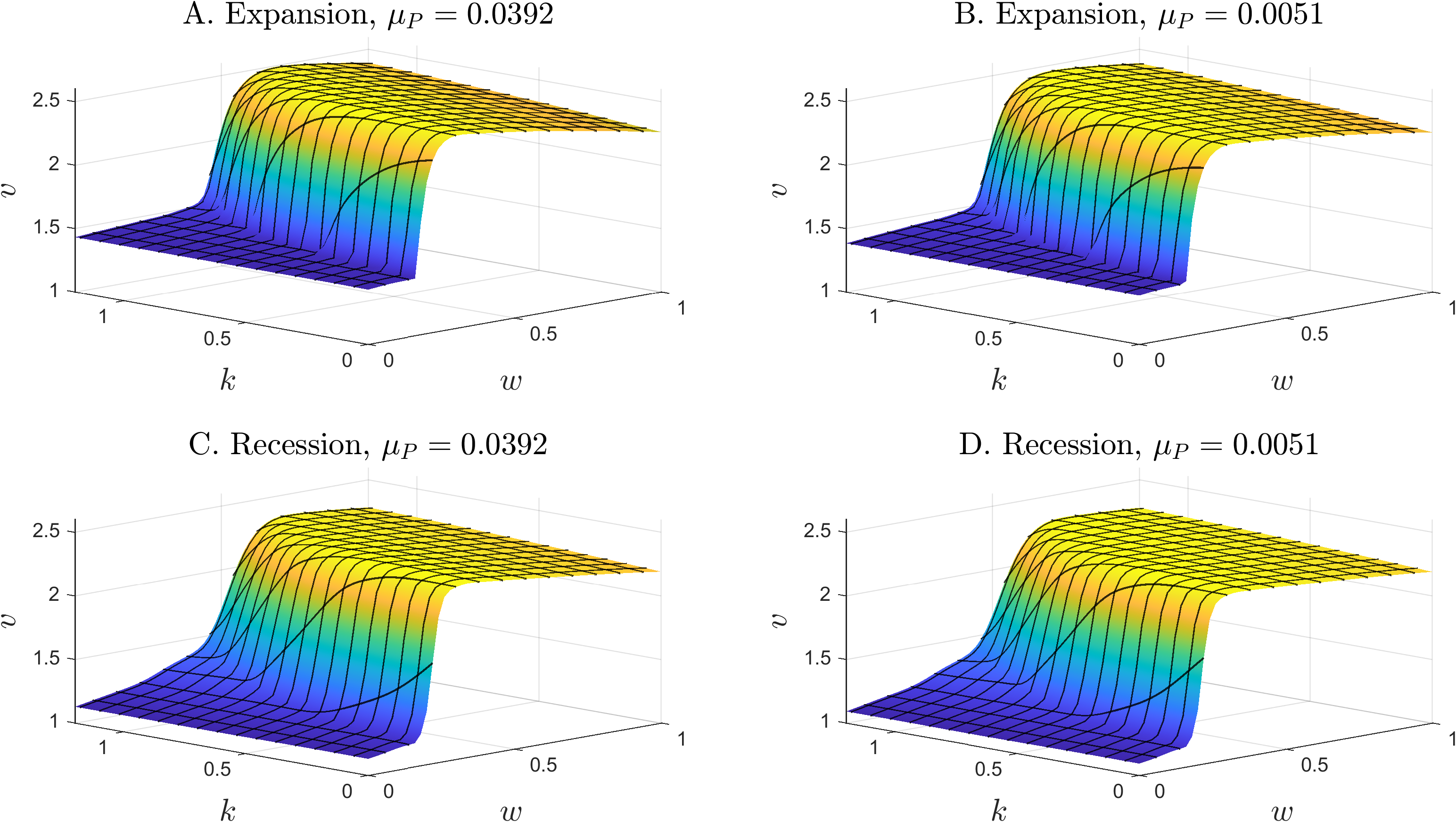}
    \caption{Surface plots of the value function}
    \label{fig: illus value fun 3D}
  \end{subfigure}

  \caption{\textbf{Scaled value function at $t=0$.} \small This figure illustrates the scaled value function before default $v(t,w,k,\mu_P,s)$ at $t=0$. In both subfigures, the first (second) row displays the value function in the expansionary (recessionary) state. The first (second) column sets $\mu_P$ to its mean conditional on the expansionary (recessionary) state.}
  \label{fig: illus value fun}
\end{figure}

\Cref{fig: illus value fun 2D} illustrates the scaled value function $v(t,w,k,\mu_P,s)$ at $t=0$. Panel A plots the value function in the expansionary state ($s=2$) when the expected return on PE is equal to its mean conditional on $s=2$. We see that the value function $v$ is a non-monotonic function of the liquid fraction of total wealth $w$. Specifically, $v$ is first increasing in $w$ when $w$ is sufficiently small (e.g., for $w<0.30$ when $k=0$) and then decreasing in $w$ when $w$ is sufficiently large (e.g., for $w>0.30$ when $k=0$).

\begin{figure}[t]
  \centering
  \includegraphics[width=1.0\textwidth]{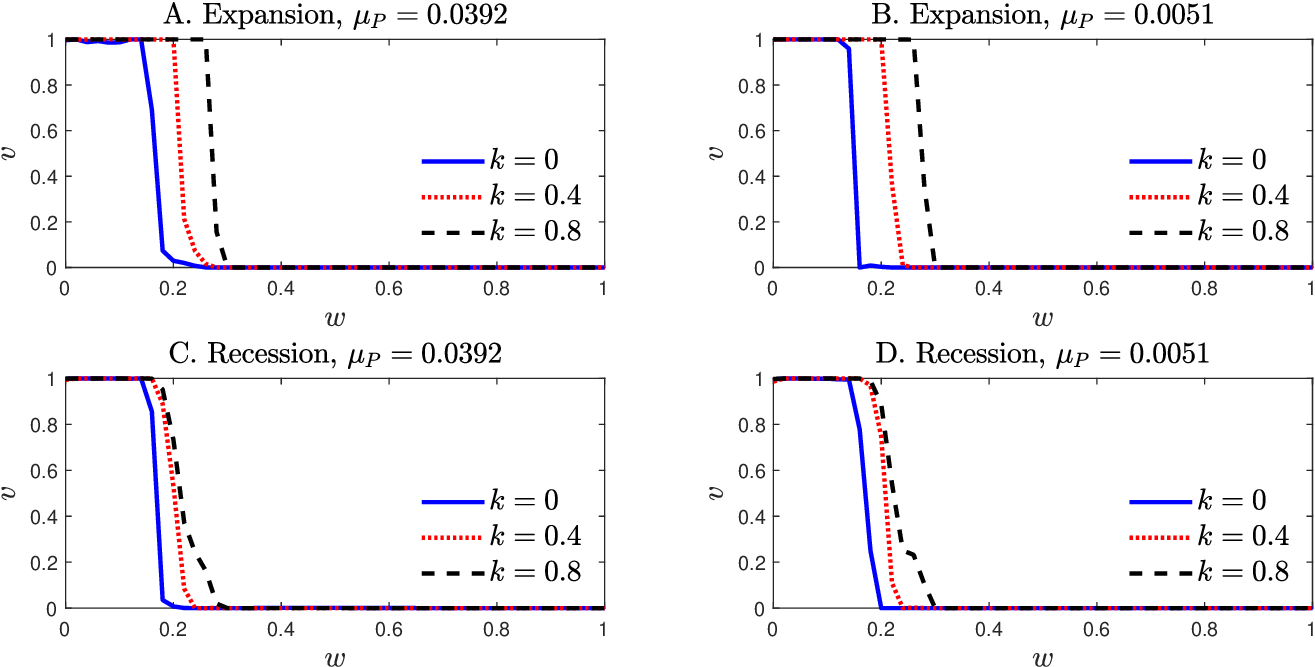}
  \caption{\textbf{One quarter ahead default probabilites at $t=0$.} \small This figure illustrates the one period ahead default probability at $t=0$. The first (second) row displays the default probabilities in the expansionary (recessionary) state. The first (second) column sets $\mu_P$ to its mean conditional on the expansionary (recessionary) state.}\label{fig: illus def prob1}
\end{figure}

The reason for this non-monotonicty is as follows. Defaulting becomes likely when $w$ is small; this is illustrated in \Cref{fig: illus def prob1} which plots the one quarter ahead default probabilities at $t=0$. When the current period's liquid wealth is low, the LP becomes less likely to have enough liquid wealth next period to meet its capital commitments and to pay its risk budget cost (see equation \eqref{eq: law of motion W}). Hence, the value function becomes increasing in $w$ when $w$ is small, as having more liquidity helps the LP avoid default. Instead, when $w$ is sufficiently large so that there is no risk of default, the LP's value function becomes decreasing in $w$ because the LP would be able to achieve better investment outcomes had it allocated more of its wealth to PE.

%\begin{figure}[htbp!]
%  \centering
%  \includegraphics[width=1.0\textwidth]{figures/illus_value_fun_2D.eps}
%  \caption{\textbf{Scaled value function at $t=0$.} \small This figure illustrates the scaled value function before default $v(t,w,k,\mu_P,s)$ at $t=0$. The first (second) row displays the value function in the expansionary (recessionary) state. The first (second) column sets $\mu_P$ to its mean conditional on the expansionary (recessionary) state.}\label{fig: illus value fun}
%\end{figure}
%
%\begin{figure}[htbp!]
%  \centering
%  \includegraphics[width=1.0\textwidth]{figures/illus_value_fun_3D.eps}
%  \caption{\textbf{Scaled value function at $t=0$.} \small This figure uses surface plots to illustrate the scaled value function before default $v(t,w,k,\mu_P,s)$ at $t=0$.}\label{fig: illus value fun 3D}
%\end{figure}

Similarly, the value function $v$ is non-monotonic in uncalled commitments relative to total wealth $k$. For example, in panel A of \Cref{fig: illus value fun 2D} we see that $v$ is decreasing in $k$ for small $w$ (e.g., when $w=0.2$) while $v$ is increasing in $k$ for large $w$ (e.g., when $w=0.8$). The reason is as follows. When the LP has little liquidity, default concerns become first order and having more uncalled commitments increases the likelihood of default, thereby decreasing $v$. In contrast, when the LP has sufficient liquidity and default is no longer a concern, higher uncalled commitments $k$ increases the LP's PE exposure which increases $v$. This is because the presence of adjustment costs means that it takes time for the LP to build up its PE exposure through capital commitments.

Panel B of \Cref{fig: illus value fun 2D} plots the value function in the expansionary state when the PE expected return is low while panels C and D illustrate the value function in the recessionary state. The value function is monotonically increasing in the PE expected return $\mu_P$, as higher expected PE returns unambiguously translate into higher expected utility. Moreover, the value function is lower during recessions (panels C and D) compared to expansions (panels A and B), reflecting the less favorable investment opportunities during economic downturns. Despite these level differences, we see that the shape of the value function remains similar across different macroeconomic states and PE expected returns. \Cref{fig: illus value fun 3D} illustrates further details of the value function using three-dimensional surface plots.

\paragraph{New commitments.}

%\Cref{fig: illus policy n 2D} and \Cref{fig: illus policy S 2D} illustrate the optimal policies for new PE commitments $N$ and allocation to stocks $S$, respectively, at $t=0$. Both policies are scaled with respect to total wealth $W+P$.

\begin{figure}[t]
  \centering
  \includegraphics[width=1.0\textwidth]{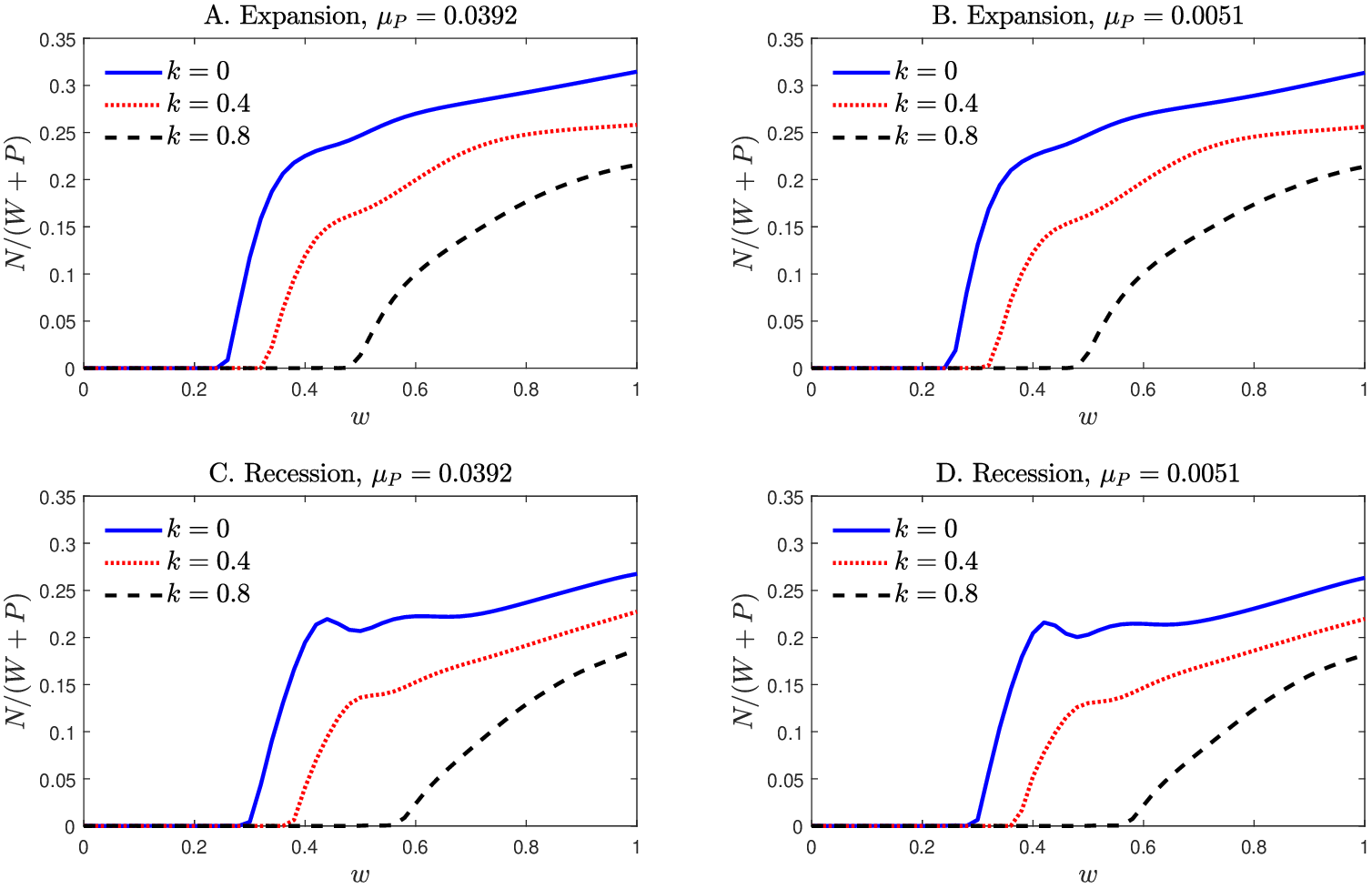}
  \caption{\textbf{New commitments at $t=0$.} \small This figure illustrates new commitments relative to total wealth, $N/(W+P)$, at $t=0$. The first (second) row displays the results for the expansionary (recessionary) state. The first (second) column sets $\mu_P$ to its mean conditional on the expansionary (recessionary) state.}\label{fig: illus policy n 2D}
\end{figure}

\Cref{fig: illus policy n 2D} illustrate the optimal policy for new PE commitments, scaled with respect to total wealth, at $t=0$. We see that new PE commitments are increasing in the liquid fraction of total wealth $w$ and decreasing in the uncalled commitments relative to total wealth $k$. The LP becomes less willing to commit new capital to PE when default concerns are high as defaulting leads to costly PE liquidations. This is why new PE commitments are low when $w$ is low or $k$ is high. Instead, when default is unlikely, the LP commits more capital to PE in order to increase its PE exposure and thereby lower its future liquid wealth share (as discussed above, the value function becomes decreasing in $w$ in the absence of default concerns).

Comparing panels A and B of \Cref{fig: illus policy n 2D}, and panels C and D of \Cref{fig: illus policy n 2D}, we see that current PE expected returns $\mu_P$ have surprisingly little impact on new commitments across both expansionary and recessionary states. This muted response stems from the inherent delay between making commitments and their eventual conversion into PE investments. Because PE expected returns exhibit low quarterly autocorrelation in the baseline calibration ($\varrho_P=0.20$), initial differences in $\mu_P$ largely dissipate before committed capital is called. Consequently, optimal commitment levels depend primarily on the expected conditional return $\mathbb{E}[\mu_{P,t}|s_t]$ rather than the current level of $\mu_P$, since only invested capital (NAV) ultimately generates PE returns.

Finally, comparing across macroeconomic states, we see from \Cref{fig: illus policy n 2D} that new commitments are higher during expansions than recessions. \Cref{fig: illus policy n 3D} in \autoref{sec: numerical implementation} uses three-dimensional surface plots to illustrate further details regarding new commitments.

\paragraph{Allocation to stocks.}

\begin{figure}[t]
  \centering
  \includegraphics[width=\textwidth]{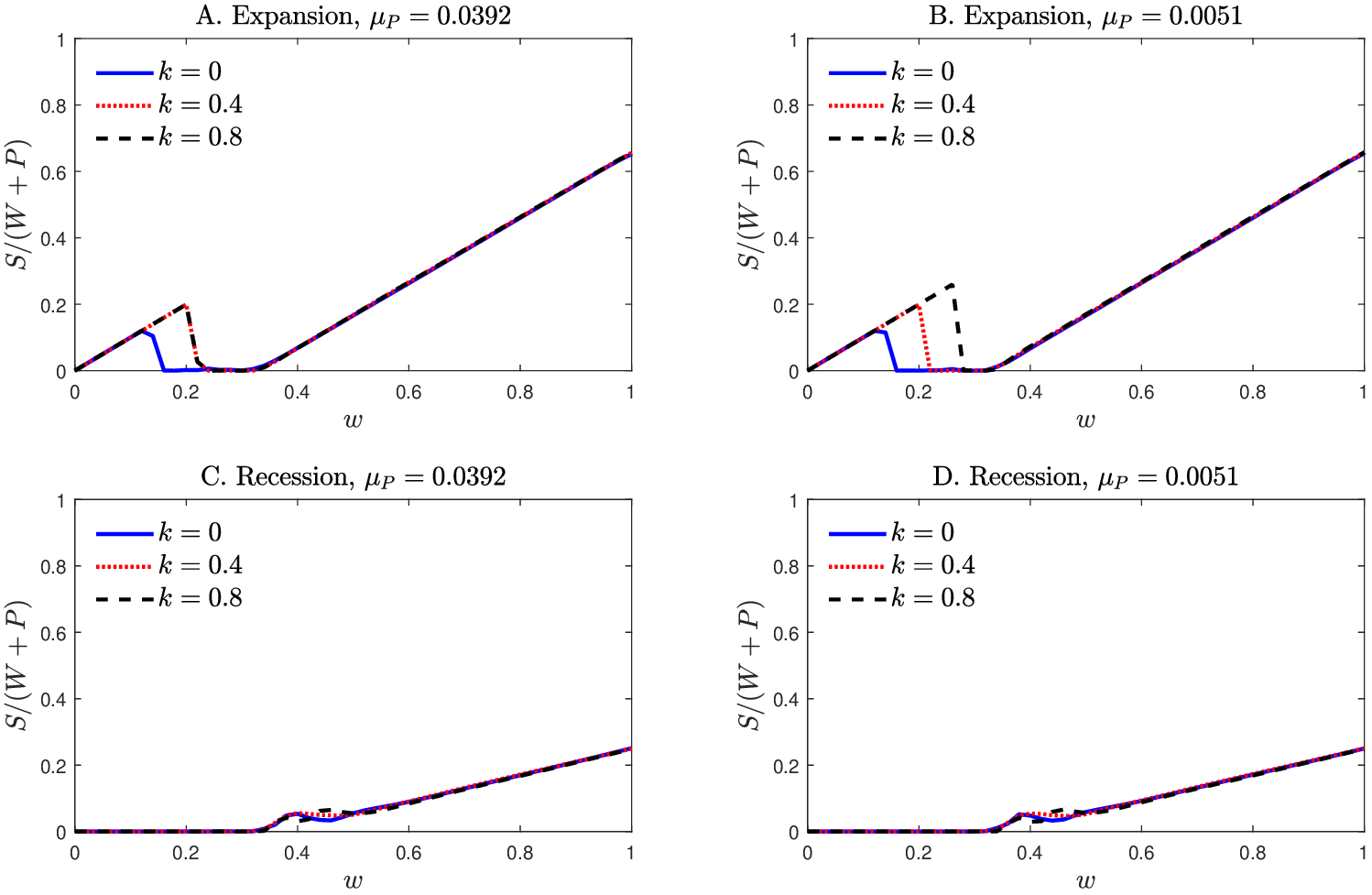}
  \caption{\textbf{Allocation to stocks at $t=0$.} \small This figure illustrates the optimal allocation to public stocks, expressed as a fraction of total wealth, at $t=0$. The first (second) row displays the optimal allocation in the expansionary (recessionary) state. The first (second) column sets $\mu_P$ to its mean conditional on the expansionary (recessionary) state.}\label{fig: illus policy S 2D}
\end{figure}

\Cref{fig: illus policy S 2D} illustrates the optimal allocation to public stocks as a fraction of total wealth at $t=0$ (see \Cref{fig: illus policy S 3D} in \autoref{sec: numerical implementation} for surface plots that display further details). Panel A depicts this allocation during the expansionary state, with PE expected returns at their mean conditional on $s=2$. When liquid wealth $w$ exceeds 0.3, making default risk negligible, two patterns emerge. First, stock allocation increases with $w$, reflecting a rebalancing mechanism: as liquid wealth rises and PE exposure ($1-w$) mechanically falls, the LP maintains its desired risk exposure by increasing its allocation to public stocks. For example, the portfolio shifts from 50\% PE, 17\% stocks, and 33\% bonds at $w=0.5$ to 20\% PE, 46\% stocks, and 34\% bonds at $w=0.8$. Second, stock allocation remains independent of uncalled commitments $k$. This is because, absent default risk, the value function depends primarily on next period's total wealth growth rate. Since uncalled commitments affect total wealth only after their conversion to PE investments (NAV), which occurs with a lag, they have no immediate impact on portfolio growth (see equation \eqref{eq: total wealth growth not default}) and therefore do not interact with the allocation decision for public stocks.

When default risk is non-negligible, a different mechanism is at play. For example, default occurs with certainty when $w=0.1$ for the levels of $k$ considered here (see \Cref{fig: illus def prob1}). In this case, the LP makes its stock allocation decision knowing that it will default on its PE commitments and incur a PE liquidation loss next period. The LP chooses to maximize its stock allocation to compensate for this loss when macroeconomic conditions are favorable (see panels A and B of \Cref{fig: illus policy S 2D}). Conversely, it takes a conservative approach and does not allocate anything to public stocks when macroeconomic conditions are unfavorable (see panels C and D of \Cref{fig: illus policy S 2D}).

However, this default mechanism is rarely at work along the optimal path when we simulate the model. The LP tries to avoid default since defaulting is costly. The cumulative default rate over the LP's 10 year investment horizon is only 0.1\%.

\subsubsection{Marginal value of liquidity}

The illiquid nature of PE assets makes liquidity management central. Indeed, the LP must finance its capital calls out of its liquid wealth or suffer the consequences of default. The marginal value of liquidity at time $t$, measured in units of certainty equivalent terminal wealth, is given by
\begin{equation}\label{def: marginal value of liquidity}
  \frac{\partial}{\partial W}(W+P)v(t,w,k,\mu_P,s)=v+(1-w)v_w-kv_k.
\end{equation}
On the left-hand side of equation \eqref{def: marginal value of liquidity}, $(W+P)v(t,w,k,\mu_P,s)$ expresses the value function $V(t,w,k,\mu_P,s)$ in certainty equivalent units of terminal wealth (see equation \eqref{eq: def scaled value fun not default}). Here, $W+P$ is current total wealth and $v(t,w,k,\mu_P,s)$ is the certainty equivalent value per unit of wealth. On the right-hand side, the $v$ term captures the direct effect of an additional unit of wealth. The $(1-w)v_w$ and $-kv_k$ account for compositional effects due to changes in the liquid fraction of wealth $w$ and uncalled commitments relative to total wealth $k$. The formulation of the marginal value of liquidity \eqref{def: marginal value of liquidity} is analogous to the marginal value of cash in the liquidity management model of \citet{bolton/chen/wang:11, bolton/chen/wang:13}.

%Our model quantifies the marginal value of liquidity in certainty equivalent terms as follows. From equation \eqref{eq: def scaled value fun not default}, the certainty equivalent value of wealth at time $t$ is given by $(W+P)v(t,w,k,\mu_P,s)$ where $v$ is the certainty equivalent value of terminal wealth per unit of current total wealth. 

\begin{figure}[htbp]
  \centering
  % Subfigure 1 (top)
  \begin{subfigure}[t]{\textwidth}
    \centering
    % Replace 'example-image-a' with your actual image file name
    \includegraphics[width=\textwidth]{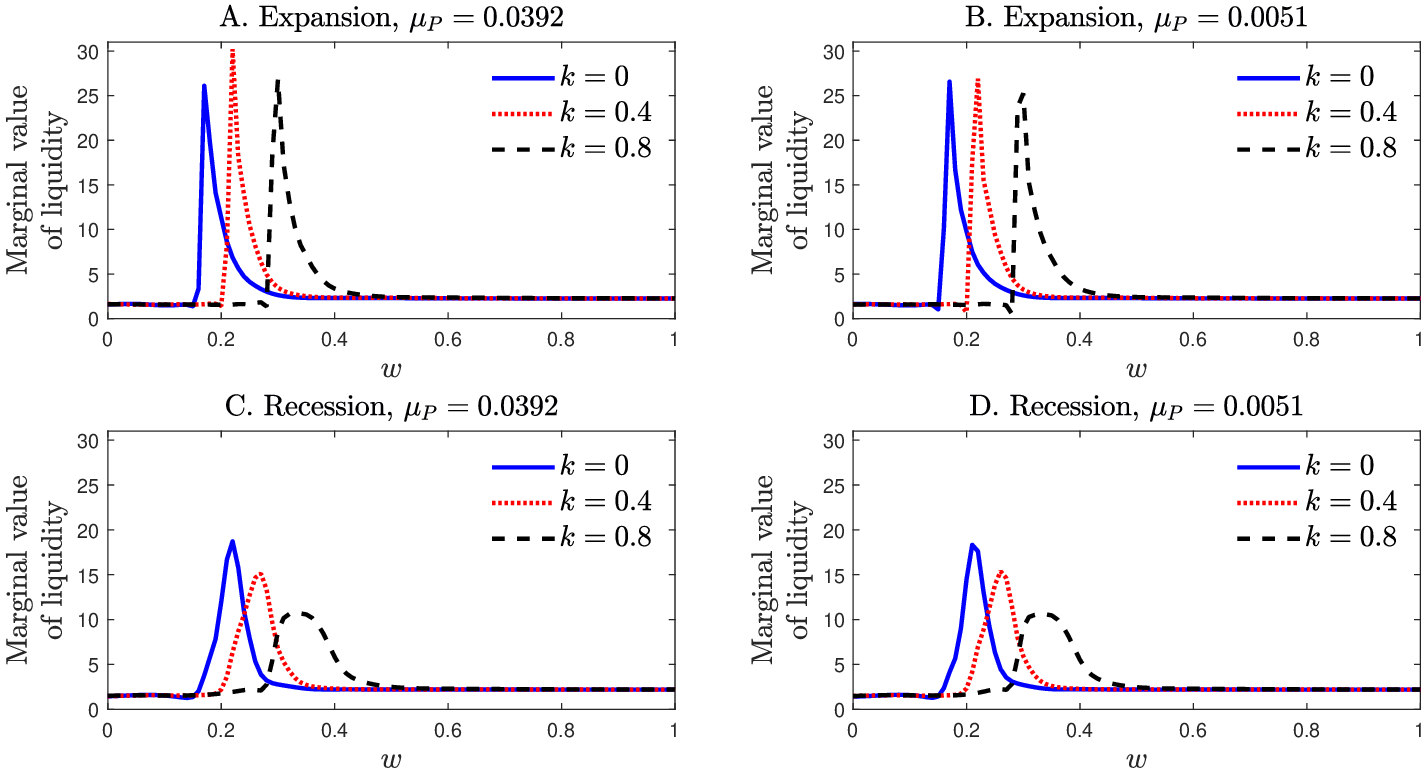}
    \caption{Slices of the marginal value of liquid wealth along the $w$ dimension.}
    \label{fig: illus mval cash 2D}
  \end{subfigure}

  \vspace{1em} % Optional extra vertical space between subfigures

  % Subfigure 2 (bottom)
  \begin{subfigure}[b]{\textwidth}
    \centering
    % Replace 'example-image-b' with your actual image file name
    % \includegraphics[width=\textwidth]{figures/illus_value_fun_3D.eps}
    \includegraphics[width=\textwidth]{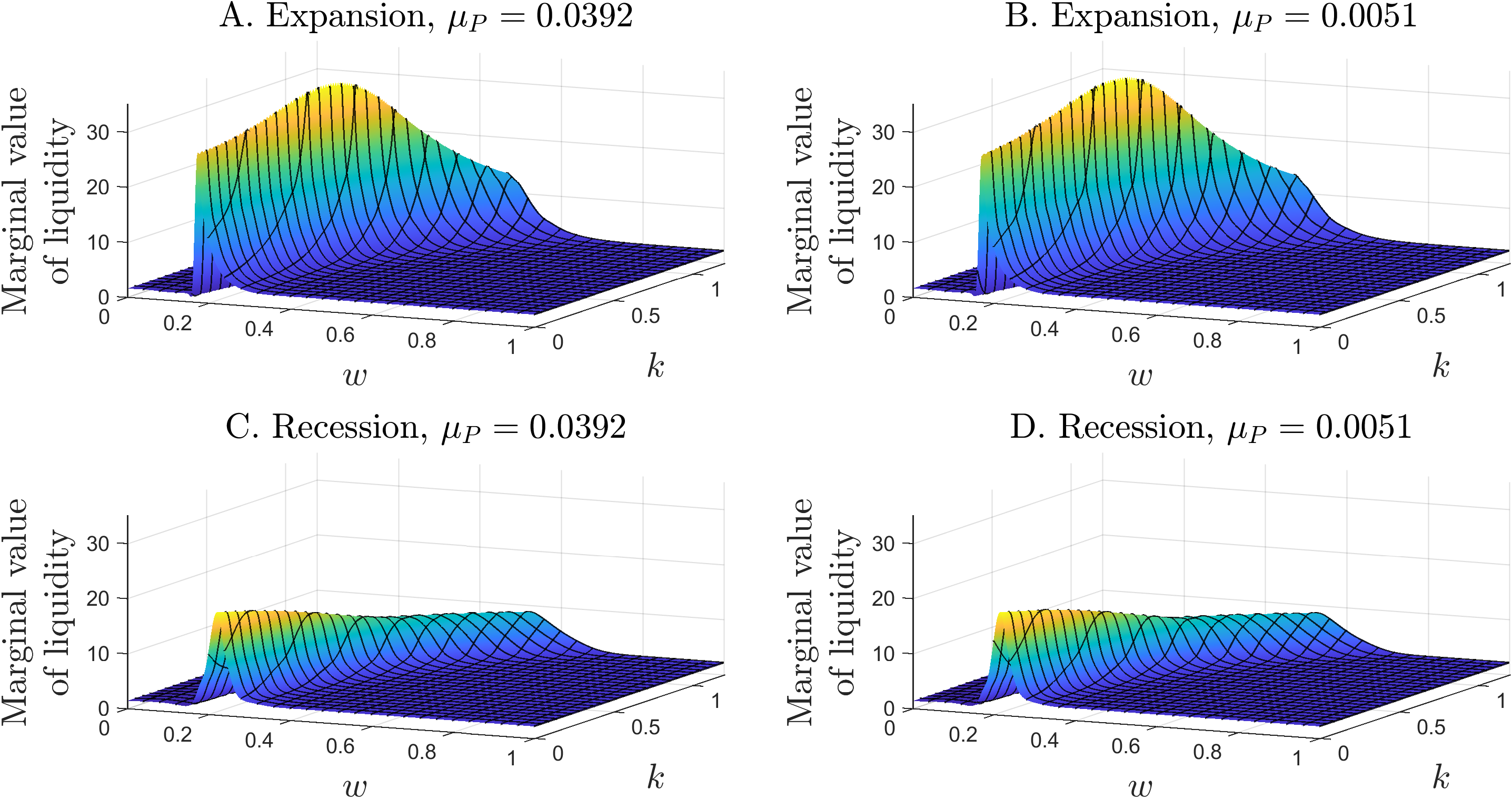}
    \caption{Surface plots of the marginal value of liquid wealth}
    \label{fig: illus mval cash 3D}
  \end{subfigure}

  \caption{\textbf{Marginal value of liquid wealth at $t=0$.} \small This figure illustrates the marginal value of liquid wealth \eqref{def: marginal value of liquidity} at $t=0$. In both \Cref{fig: illus mval cash 2D} and \Cref{fig: illus mval cash 3D}, the first (second) row displays the value function in the expansionary (recessionary) state; the first (second) column sets $\mu_P$ to its mean conditional on the expansionary (recessionary) state.}
  \label{fig: illus mval cash}
\end{figure}

\Cref{fig: illus mval cash 2D} illustrates the marginal value of liquid wealth \eqref{def: marginal value of liquidity} at $t=0$; \Cref{fig: illus mval cash 3D} uses surface plots to provide additional details. We see that the marginal value of liquid wealth is highly sensitive to default risk. As the liquid fraction of wealth $w$ decreases, default probabilities increase (see \Cref{fig: illus def prob1}), leading to a spike in the marginal value of liquidity. 

For instance, panel A of \Cref{fig: illus mval cash 2D} shows that when $k=0.4$, the macroeconomic state is expansionary, and $\mu_P$ equals its conditional mean in the expansionary state, the marginal value of liquidity equals 2.28 when $w=1$ and default is not a concern. However, as $w$ drops to $0.22$, the marginal value of liquidity rises to 30.21; at this point, the one-quarter-ahead default probability is 0.21 and rapidly approaches 1 as $w$ declines further. Eventually, as $w$ continues to decrease, the marginal value of liquidity levels off because default becomes inevitable, so that marginal increases in liquid wealth no longer helps to avert default. When $w=0$ and default is certain, the marginal value of liquidity falls to 1.62\textemdash this lower certainty equivalent value reflects the lower growth rate of wealth after the LP defaults and loses access to PE investing.

The plots further demonstrate how the marginal value of liquidity varies over the state space. For example, the location of the peak in the marginal value of liquidity depends strongly on the ratio of uncalled capital to wealth $k$; higher values of $k$ result in the peak occurring at larger values of $w$.

%strongly depends on default. As $w$ decreases and default probabilities increase, the marginal value of liquidity spikes. Benchmark: when $w=1$ so that default is not a concern, the marginal value of liquidity is 2.28. That is, a dollar at $t=0$ is equivalent to $2.28$ dollars due in 10 years time with certainty. As $w$ decreases and default risk increases, the marginal value of liquidity spikes. For example, in Panel A, the marginal value spikes to 30.21 (for $k=0.4)$ when $w=0.22$. The one-quarter ahead default probability is 0.21 at this point and rapidly increases towards 1 as $w$ falls further.
%
%When $w$ is low so that default is all but assured, the marginal value of liquidity falls. Example: at $w=0$, marginal value of liquidity is 1.62. This is just the value of default.

\subsection{Life cycle dynamics}
\label{sec: life cycle dynamics}

\begin{figure}[t]
  \centering
  \includegraphics[width=1.0\textwidth,trim=1cm 0cm 1.5cm 0cm, clip]{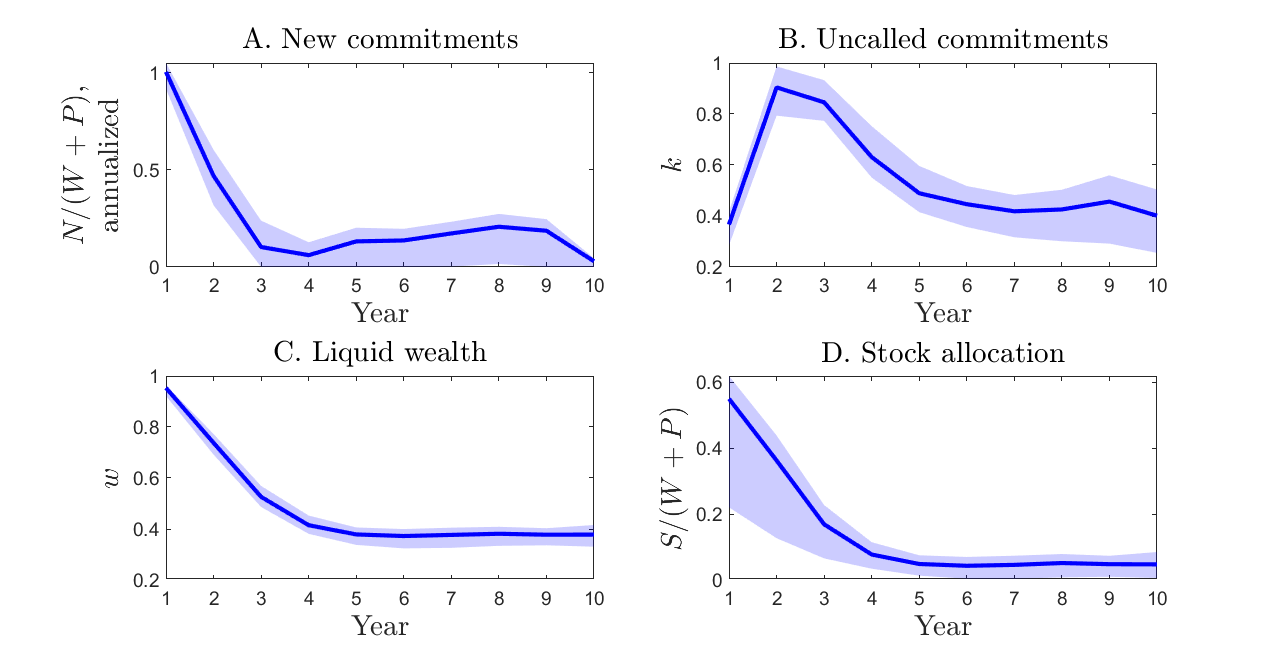}
  \caption{\textbf{Life cycle dynamics.} \small This figure illustrates the life cycle dynamics of the LP's portfolio. The solid line plots the average outcome in each year while the shaded region provides the 95\% confidence interval. Panels A through D plot the outcomes for new commitments, uncalled commitments, liquid wealth, and the stock allocation, respectively. All outcomes are shown relative to total wealth.}\label{fig: lifecycle uncond baseline}
\end{figure}

We now turn to the life cycle dynamics of the LP's portfolio. We assume that the LP starts out at $t=0$ without any PE investments (i.e., $P_0=K_0=0$) and a units of liquid wealth $W_0=1$ so that $k_0=0$ and $w_0=1$. The initial macroeconomic state $s_0$ and PE expected return $\mu_{P,0}$ are drawn from their stationary distributions. We then simulate the LP's portfolio allocation over its investment horizon of 10 years. \Cref{fig: lifecycle uncond baseline} shows the outcome\textemdash the solid line plots the average outcome at each point in time while the shaded region plots the 95\% confidence interval around the average.

% Starting out, the LP aggressively makes new capital commitments in the first two years to rapidly increase its PE exposure (see panel A of \Cref{fig: lifecycle uncond baseline}). For example, new commitments average 100\% of total wealth during the first year before dropping to 47\% and 10\% of total wealth on average in the second and third years, respectively. As a result, uncalled capital increases and peaks at an average of 90\% and 85\% of total wealth during the second and third years (see panel B). The liquid portion of total wealth drops to 74\% and 53\% on average before reaching a steady level of 38\% on average by year 5, remaining at this level until the fund matures at year 10 (see panel C). This drop in liquid wealth reflects the LP's increased PE exposure, which rises to 26\% and 47\% of total wealth on average in the second and third years, respectively, before reaching 62\% if total wealth by year 5. From year 5 onwards, the LP is in maintenance mode, making new commitments to maintain its PE exposure at an average of 62\% of total wealth until the investment horizon is reached at year 10. Stock allocation initially starts high at 55\% of total wealth on average in the first year. This is because the LP can only achieve its target aggregate risk exposure through public stocks before its PE exposure is built up. The stock allocation subsequently drops to 5\% of total wealth on average during the LP's maintenance phase.

Starting out, the LP aggressively makes new capital commitments in the first two years to rapidly increase its PE exposure (see panel A of \Cref{fig: lifecycle uncond baseline}). For example, new commitments average 100\% of total wealth during the first year before dropping to 47\% and 10\% of total wealth on average in the second and third years, respectively. As a result, uncalled capital increases and peaks at an average of 90\% and 85\% of total wealth during the second and third years (see panel B). This is accompanied by a drop in the liquid portion of wealth (see panel C) or, equivalently, an increase in the LP's PE exposure. Specifically, the LP's NAV in PE investments rises to 26\% and 47\% of total wealth on average in the second and third years, respectively, before reaching 62\% of total wealth by year 5. From year 5 onwards, the LP is in maintenance mode, making new commitments to maintain its PE exposure at an average of 62\% of total wealth until the investment horizon is reached at year 10.

Stock allocation initially starts high (see panel D); for example, stock allocation averages 55\% of total wealth in the first year. This is because the LP can only achieve its target aggregate risk exposure through public stocks before its PE exposure is built up. The wide confidence interval for stock allocation during this initial phase is due to differences in stock allocation across the business cycle which we discuss in \Cref{sec: business cycle dynamics}. The stock allocation subsequently drops to 5\% of total wealth on average during the LP's maintenance phase.

\begin{figure}[t]
  \centering
  \includegraphics[width=1.0\textwidth,trim=3cm 0cm 3cm 0cm, clip]{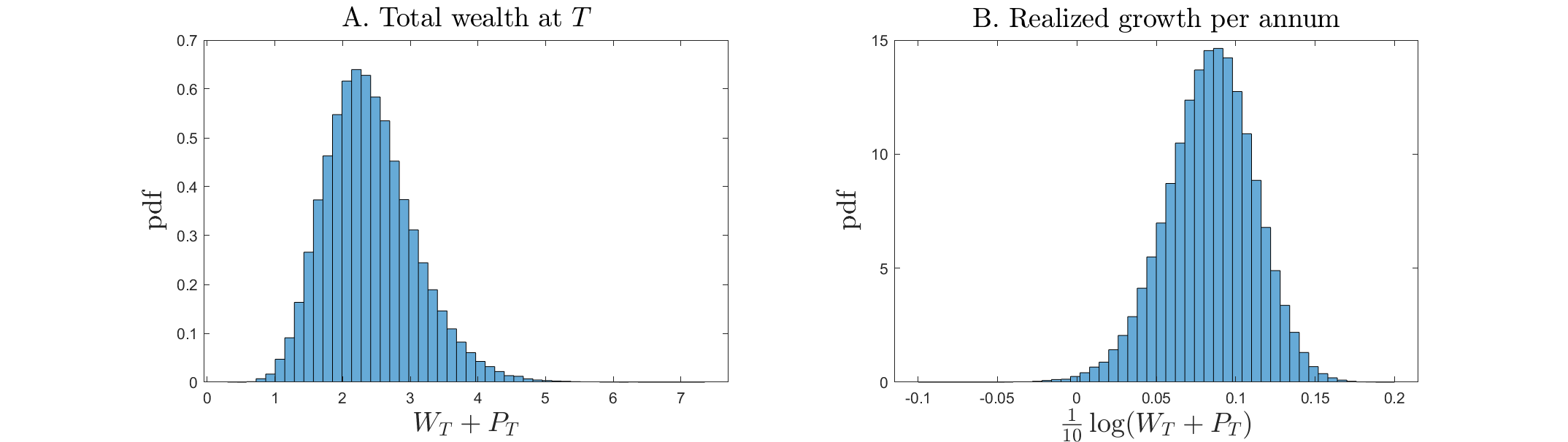}
  \caption{\textbf{Total wealth outcome distribution.} \small This figure illustrates the distribution of total wealth outcomes at the end of the investment horizon $t=T$. The initial total wealth is $W_0+P_0=1$. Panel A plots the distribution for total wealth $W_T+P_T$. Panel B plots the distribution for the realized growth in total wealth per annum, $\log(W_T+P_T)/10$.}\label{fig: illus total wealth baseline}
\end{figure}

\Cref{fig: illus total wealth baseline} shows the distribution of outcomes at the terminal date $T$. Panel A plots the distribution of for total wealth $W_T+P_T$. Starting from a unit initial total wealth, the LP's terminal wealth equals 2.41 on average with a standard deviation of 0.66. This wealth distribution translates into a certainty equivalent wealth of $\mathbb{E}[v(t=0,w_0,k_0,\mu_P,s)\left|w_0=1,k_0=0\right.]=2.25$ on the initial date. Panel B illustrates the distribution of final outcomes in terms of the annualized realized total returns $\frac{1}{10}\log((W_T+P_T)/(W_0+P_0))$. The annualized realized total return over the LP's investment horizon has a mean of 8.41\% and a standard deviation of 2.78\%. Note that this seemingly low standard deviation is because we are reporting the standard deviation of annualized long-horizon returns (see \Cref{footnote: remark annualized vol}).

\subsubsection{Heuristic versus dynamic private asset allocation}

In practice, investors often use heuristics to simplify private asset allocation decisions. One common heuristic is to split the allocation problem into two steps. In the first step, investors determine a long-run target portfolio by solving a static portfolio choice problem that abstracts from liquidity risk and the timing lags inherent in PE investing (e.g., a traditional Markowitz approach). In the second step, they adjust their allocations to reach the targets from the first step by accounting for the delays between capital commitments, capital calls, and eventual distributions (see, for example, \citealt{TA:02}). We demonstrate that portfolio allocations derived from our dynamic model can differ dramatically from these heuristic approaches. Specifically, we show that portfolio outcomes from our dynamic model, starting from year 5 when the LP has reached the maintenance phase, substantially differ from the long-run target portfolio obtained under the heuristic approach. We provide details below.

Within our framework, the first step of the heuristic approach corresponds to solving
\begin{equation}\label{eq: heuristic problem}
    V_{\mbox{heuristic}}(t,W_{total},s) = \max_{S,B,P\geq0} \mathbb{E}\left[V_{\mbox{heuristic}}(t+1,W^\prime_{total},s^\prime)\left|W_{total},s\right.\right]
\end{equation}
subject to
\begin{align}
W^\prime_{total} & = R_P^\prime P R_S^\prime S+ R_f(s)B - W_{total}\Gamma(\theta), \nonumber\\
    \theta &=\frac{\theta_B B + \theta_S S + \theta_P P}{B+S+P},\nonumber\\
    W_{total} & = S + B + P,\nonumber
\end{align}
and the terminal condition $V_{\mbox{heuristic}}(T,W,s)=W^{1-\gamma}/(1-\gamma).$ The heuristic problem \eqref{eq: heuristic problem} resembles the full problem \eqref{eq: nondefault problem} but reduces the total number of state variables by abstracting away from the illiquidity of PE investments. In implementing the heuristic problem \eqref{eq: heuristic problem}, we further set $\mu_P=\mathbb{E}[\mu_{P,t}|s_t=s]$ to its mean conditional on the macroeconomic state. In \autoref{app: scaled value functions}, we show that the solution to the heuristic problem \eqref{eq: heuristic problem} amounts to optimizing the certainty-equivalent growth rate of total wealth at each point in time. The solution is static in the sense that the optimal portfolio weights depend solely on the macroeconomic state $s$ and do not otherwise change over time.

\begin{figure}[t]
  \centering
  \includegraphics[width=1.0\textwidth]{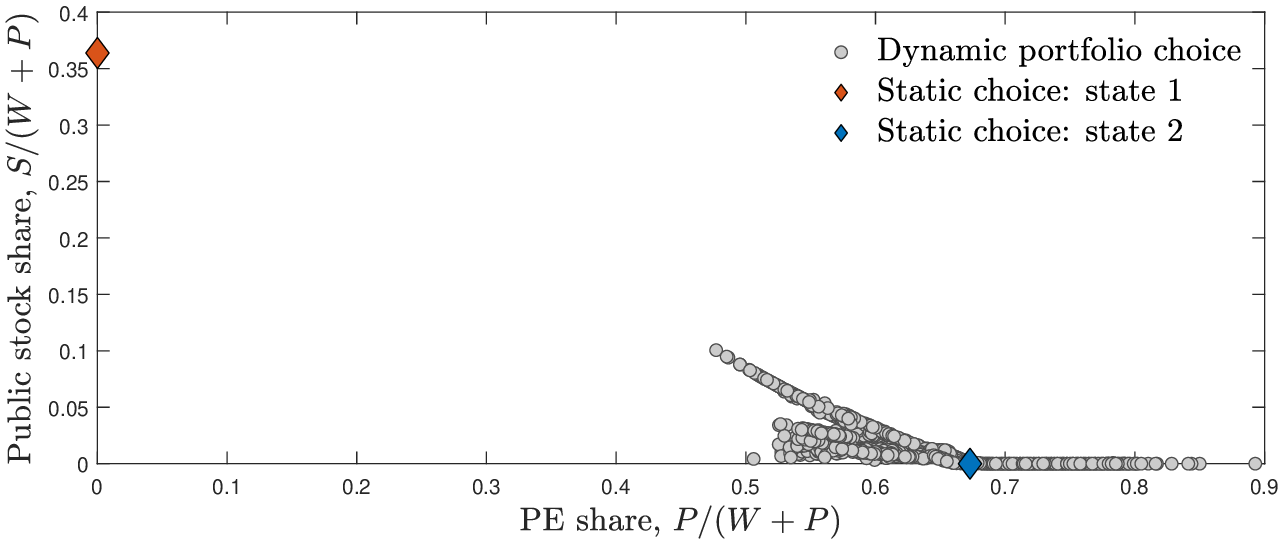}
  \caption{\textbf{Portfolio choice: heuristic vs. dynamic private asset allocation.} \small This figure compares the portfolio outcomes under the heuristic problem \eqref{eq: heuristic problem} (shown by the diamonds) against the outcomes under the full dynamic allocation problem \eqref{eq: nondefault problem} (shown in the circles). Outcomes for the latter correspond are from year 5 onwards during which the maintenance phase has been reached.}\label{fig: static vs dynamic allocations}
\end{figure}

\Cref{fig: static vs dynamic allocations} compares the portfolio outcomes under the heuristic approach \eqref{eq: heuristic problem} with those under our full dynamic model \eqref{eq: nondefault problem}. The circles represent the portfolio allocations derived from our dynamic model for year 5 onwards—when the limited partner (LP) has entered its maintenance phase—while the diamonds show the static portfolio choices obtained from solving \eqref{eq: heuristic problem} for each macroeconomic state.

Under our baseline calibration (\Cref{tbl: parameters}), the heuristic approach yields an optimal portfolio allocation of 67.3\% in private equity (PE) and 0\% in stocks (with the remainder in bonds) in the expansionary state, and 0\% in PE and 36.4\% in stocks in the recessionary state. In contrast, the optimal dynamic allocation maintains a PE allocation of at least 45\% during the maintenance phase, while the stock allocation does not exceed 10\%. This difference in outcomes is especially stark during the recessionary state, where the heuristic outcome falls entirely outside the range of values generated by the optimal dynamic allocation.

\subsection{Business cycle dynamics}
\label{sec: business cycle dynamics}

\begin{figure}[t]
  \centering
  \includegraphics[width=1.0\textwidth,trim=1cm 0cm 1.8cm 0cm, clip]{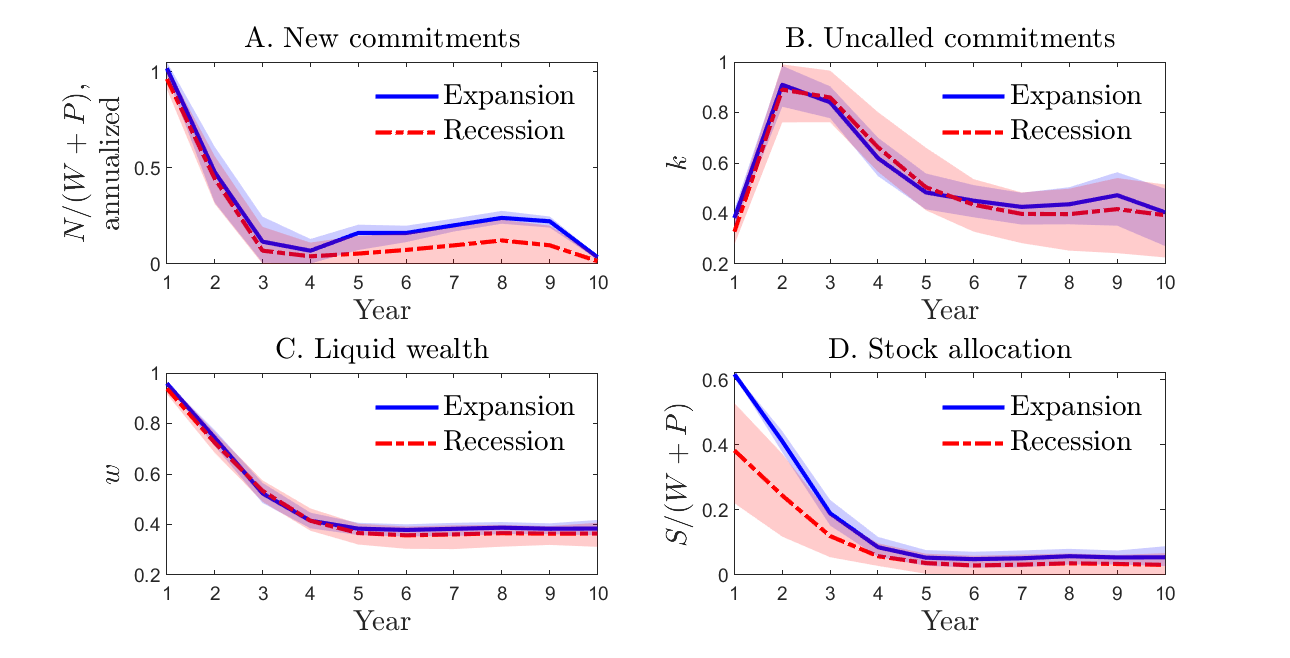}
  \caption{\textbf{Life cycle dynamics over the business cycle.} \small This figure illustrates the life cycle dynamics of the LP's portfolio after conditioning on business cycle conditions. Panels A through D plot the outcomes for newcommitments, uncalled commitments, liquid wealth, and the stock allocation, respectively, where all outcomes are shown relative to total wealth. The solid (dashed) line shows the average outcome conditional on a expansionary (recessionary) state. Shaded regions indicated 95\% confidence intervals, conditional on the macroeconomic state.}\label{fig: lifecycle cond baseline}
\end{figure}

\Cref{fig: lifecycle cond baseline} illustrates how business cycle conditions affect the life cycle dynamics of the LP's portfolio. The solid and dashed lines plot the average outcomes conditional on the macroeconomic state being in a expansion and a recession, respectively; the shaded 95\% confidence intervals are also conditional on the macroeconomic state. Overall, we see that the allocation patterns across the business cycle are similar compared to the unconditional case discussed in \Cref{sec: life cycle dynamics}.

% Points to talk about:
% 1. Comparing \Cref{fig: lifecycle uncond baseline} and \Cref{fig: lifecycle cond baseline}, we see that the overall allocation patterns are similar.
% 2. The main difference is with respect to the LP's stock allocation. Most pronounced during the ramping up phase. Panel D: For example, year 1: stock allocations of 62\% vs 38\% of total wealth in expansions and recessions, respectively. By the maintenance phase from year 5 beyond: the difference is about 2\% of total wealth (higher during expansions).
% 3. There is a smaller difference in new commitments (panel A). Ramping up phase: difference in new commitments averages around 5\% of total wealth in the first 3 years. Panel B: difference in uncalled commitment over business cycle is similarly smaller on average (compared to stock allocation differences). Overall, the difference in PE allocation (NAV to total wealth) is about 2\% higher over the lifetime of the fund (panel C).
% 4. Explain why the main difference is due to stock allocation and not PE: PE allocation involves adjustment costs whereas adjustment costs for stocks are negligible. Therefore: the LP keep the PE allocation stable across the business cycle and mainly adjust its exposure to aggregate risks through stocks.

The most notable difference across the business cycle is the difference in the LP's stock allocation, particularly during the ramping-up phase (see panel D of \Cref{fig: lifecycle cond baseline}). For instance, in year 1, stock allocations average 62\% of total wealth during expansions compared to 38\% during recessions. This difference narrows to about 2\% of total wealth by the maintenance phase from year 5 onwards.

There is a smaller difference in new commitments across the business cycle (see panel A). For example, During the ramping-up phase in the first three years, new commitments are lower during recessions compared to expansions by about 5\% of total wealth on average. Similarly, panels B and C show that the difference in uncalled commitments and liquid wealth over the business cycle is smaller on average compared to the differences in stock allocation. For example, the difference in PE allocation across the business cycle averages 2\% of total wealth over the lifetime of the fund.

The main reason for the pronounced difference in stock allocation across the business cycle, as opposed to PE allocation, is due to adjustment costs. PE allocation involves significant adjustment costs, whereas adjustment costs for stocks are negligible. Therefore, the LP keeps the PE allocation relatively stable across the business cycle and primarily adjusts its exposure to aggregate risks through stock allocation.

\begin{figure}[t]
  \centering
  \includegraphics[width=1.0\textwidth,trim=0.5cm 0cm 1.2cm 0cm, clip]{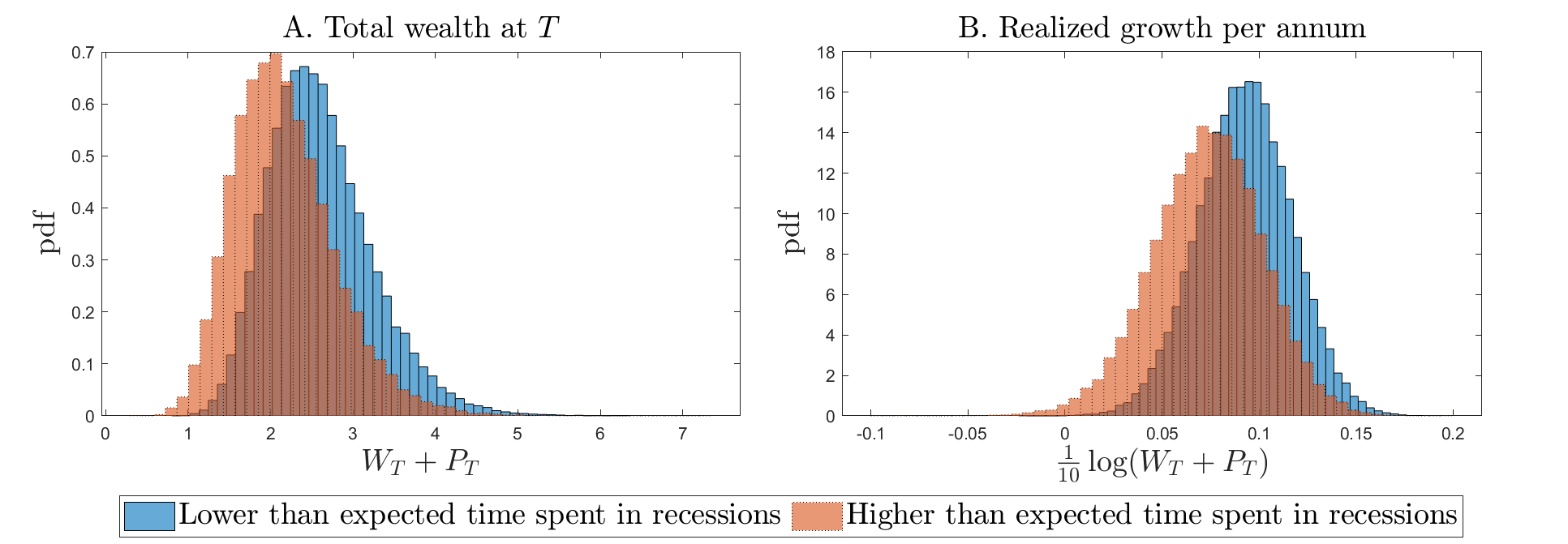}
  \caption{\textbf{Business cycles and total wealth outcomes.} \small This figure illustrates the distribution of total wealth outcomes at the end of the investment horizon $t=T$. The initial total wealth is $W_0+P_0=1$. Panel A plots the distribution for total wealth $W_T+P_T$. Panel B plots the distribution for the realized growth in total wealth per annum, $\log(W_T+P_T)/10$. The solid (dashed) bars plot the outcomes conditional on the LP experiencing a lower (higher) than expected number of recessions over its investment horizon.}\label{fig: illus total wealth baseline bus cycle}
\end{figure}

% points to address:
% 1. We conduct a separate but related exercise to illustrate the role of business cycle conditions. We investigate a special case where the LP does not account for business cycle conditions and always assumes that the economy is in an expansionary state. This is equivalent to setting the transition probabilities $p_{12}=p_{21}=0$ and the initial macroeconomic state $s_0=2$. We then compare the life cycle dynamics under this assumption to the baseline case where the LP accounts for business cycle conditions. The results are shown in \Cref{fig: lifecycle uncond comp statics always expansion}.
% 2. Point out key differences. Without accounting for the possibility of recessions: PE allocation becomes significantly more aggessive. Example: in year 1 new commitments average 117\% of total wealth compared to 100\% in the baseline case. In the maintenance phase: PE allocation average 67% as opposed to 62\% in the baseline that accounts for the possibility of business cycles.

\Cref{fig: illus total wealth baseline bus cycle} plots the distribution of outcomes at the terminal date $T$ conditional on business cycle conditions encountered over the LP's investment horizon. The transition probabilities for the macroeconomic state from \Cref{tbl: parameters} imply that recessions occur 1/6 of the time. The solid blue (dashed red) bars plot outcome distributions conditional on the LP encountering recessions for less (more) than 1/6 of the time over its investment horizon. Panel A plots the outcomes for wealth. On average, the LP ends up with a terminal wealth of 2.60 and 2.17 conditional on experiencing a lower- and higher-than expected number of recessions, respectively. The corresponding standard deviations of terminal wealth are similar: 0.63 and 0.61 conditional on experiencing a lower- and higher-than expected number of recessions, respectively. The outcomes in terms of annualized realized total returns are shown in panel B. Conditional on experiencing a lower than expected number of recessions, the realized annualized returns average 9.29\% with a standard deviation of 2.40\%. The annualized return averages 7.35\% with a standard deviation of 2.83\% if a higher than expected number of recessions is encountered instead.

\subsubsection{Cost of ignoring business cycle when making PE allocations}

\begin{figure}[t]
  \centering
  \includegraphics[width=1.0\textwidth]{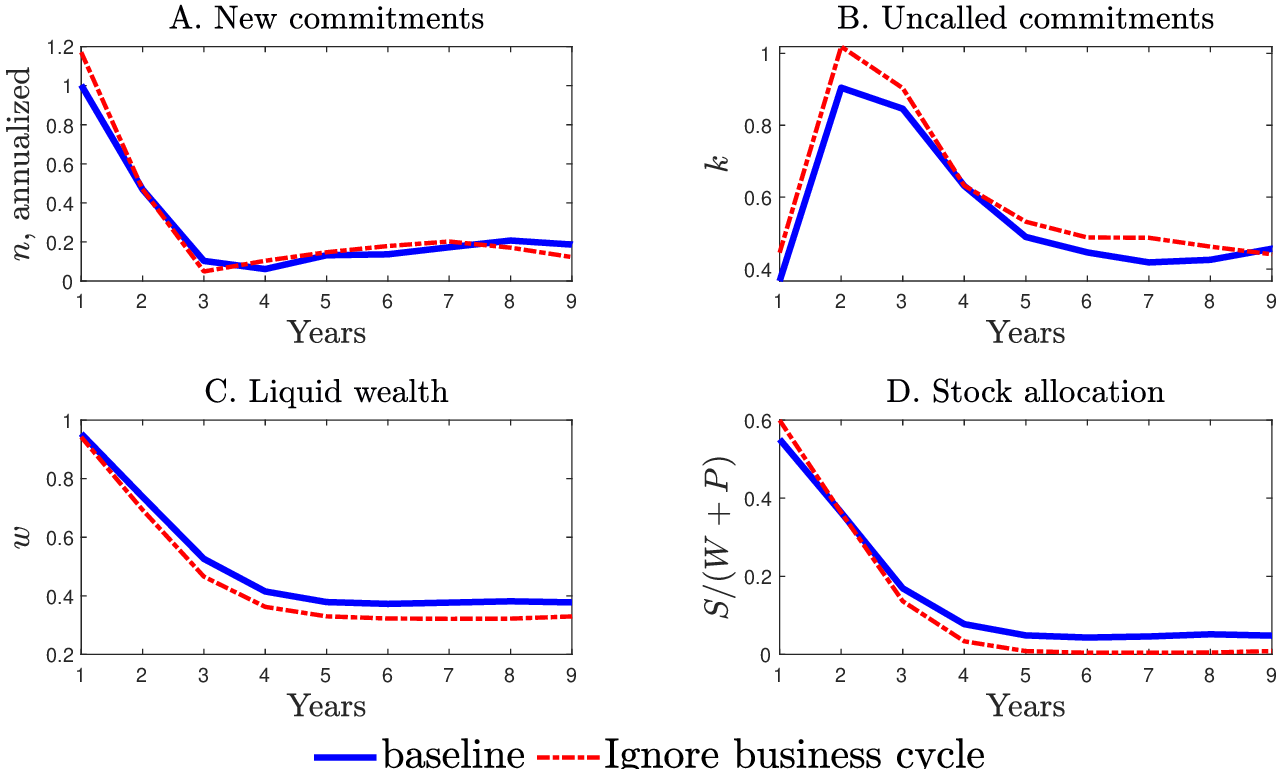}
  \caption{\textbf{Life cycle outcomes without accounting for business cycles.} \small The solid and dashed lines plot the average outcomes when business cycles are accounted for and not accounted for, respectively.}\label{fig: lifecycle uncond comp statics always expansion}
\end{figure}

Next, we illustrate the importance of accounting for business cycle conditions through the following exercise. We consider a naive LP whose decisions are based on a model that ignores business cycles. Specifically, this naive LP solves the model under the assumption that the economy is always in an expansionary state (i.e., this LP sets $s_0=2$, $p_{21}=0$, and uses the remaining parameters from \Cref{tbl: parameters}). The naive LP then applies the resulting policy functions, which ignore business cycle fluctuations, in a setting where such fluctuations are present. We compare the life cycle dynamics of this naive LP to the baseline case, where the LP optimally accounts for business cycle conditions. The results are shown in \Cref{fig: lifecycle uncond comp statics always expansion}.

We observe that without considering the possibility of recessions, the LP's PE allocation is significantly more aggressive compared to the optimal policy that accounts for business cycle fluctuations. For instance, in year 1, new commitments average 117\% of total wealth compared to 100\% in the baseline case. During the maintenance phase, the PE allocation averages 67\% of total wealth, as opposed to 62\% in the baseline case. Consequently, the naive LP defaults substantially more frequently\textemdash the naive LP's 10-year cumulative default probability is 13.6\% compared to only 0.1\% under the optimal policy.

\begin{figure}[t]
  \centering
  \includegraphics[width=1.0\textwidth,trim=0.5cm 0cm 1.5cm 0cm, clip]{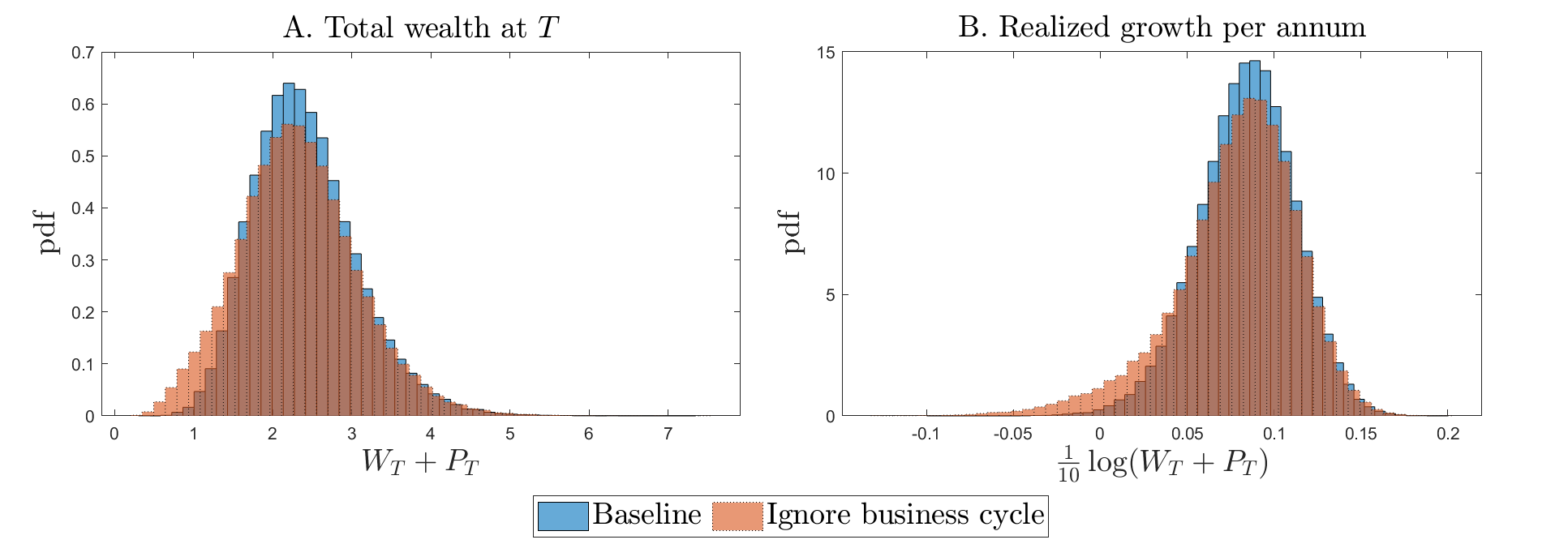}
  \caption{\textbf{Total wealth outcome without accounting for business cycles.} \small This figure illustrates the distribution of total wealth outcomes at the end of the investment horizon $t=T$. The initial total wealth is $W_0+P_0=1$. Panel A plots the distribution for total wealth $W_T+P_T$. Panel B plots the distribution for the realized growth in total wealth per annum, $\log(W_T+P_T)/10$. The solid and dashed bars plot the outcomes for the baseline and a naive investor that ignores business cycles, respectively.}\label{fig: compare outcomes ignore bus cycle}
\end{figure}

This large increase in default frequency has a significant negative impact on the naive LP's final portfolio outcomes. \Cref{fig: compare outcomes ignore bus cycle} compares the outcome distributions for the naive investor against the baseline; panels A and B compare outcomes for terminal total wealth and annualized realized returns, respectively. Compared to the baseline investor that takes business cycles into account, the wealth total wealth distribution for the naive LP has a lower mean (2.32 vs. 2.41) and a higher volatility (0.75 vs 0.66). Similarly, the naive LP's annualized realized returns have a lower mean (7.83\% vs 8.41\%) and a higher volatility (3.58\% vs 2.78\%). The biggest difference in outcomes, however, is heavier left-tail encountered by the naive investor. For example, the first and fifth percentiles for the naive LP's realized annualized returns equal -3.06\% and 1.14\%, as opposed to 1.34\% and 3.68\% for the baseline case. The naive LP's suboptimal policies translate into a 9.3\% loss in initial total wealth\textemdash the naive LP's certainty equivalent value is 2.04 compared to 2.25 for the baseline case.

These results underscore the importance of considering business cycle conditions when making optimal PE allocation decisions.

\subsection{Does serial correlation in PE returns matter for \emph{long-term} investors?}
\label{sec: serial correlation in returns}

% points to discuss:
%
% 1. Point out that PE returns are autocorrelated. The unconditional autocorrelation is $$
% 2. what we do: we set the conditional autocorrelation of PE returns to zero, $\varrho_P=0$, and recalibrate the PE expected return process so that the expected PE return and volatility remain the same for both states. That is, we recalibrate $\nu_P(1)=0.0051$ and $\nu_P(2)=0.0392$ and $\sigma_P(1)=0.0772$ and $\sigma_P(2)=0.0427$. This preserves long run moments and volatilities for PE returns conditional on the state $\mathbb{E}[\log R_{P,t+1}|s_t]$ and $\sigma(\log R_{P,t+1}|s_t)$. Therefore, long run unconditional moments remain the same too. But gets rid of the autocorrelation in PE returns.
% 3. under this setting,

Positive serial correlation is a hallmark feature of the observed returns of illiquid alternative asset, PE investments included. A large literature investigates the sources of this serial correlation and its implications for investors (see, e.g., \citealt{geltner1991smoothing} and \citealt{getmansky/lo/makarov:2004}). A likely explanation is that serial correlation can arise from having to mark-to-market illiquid assets, which can lead to observed returns that appear smooth even when the underlying true returns do not have serial correlation \citep{getmansky/lo/makarov:2004}. As a result, the literature has argued for ``unsmoothing'' such returns to obtain a more accurate picture of the underlying asset's risk and return characteristics; for example, naively using smoothed returns can lead to overestimation of Sharpe ratios and faulty portfolio allocations.

In this section, we revisit the question of whether unsmoothing is necessary for optimal portfolio choice. We differ from the existing literature in that we take the perspective of a \emph{long-term} investor. Our headline result is that for long-term investors, whether or not returns must be first unsmoothed may be a moot point. The intuition is as follows. First, long-term investors care about properties of long-horizon returns, which are much less affected by short-term mark-to-market fluctuations. Second, even if serial correlation is a true feature of short run PE returns, it may not be exploitable by investors after taking realistic implementation lags and adjustment costs into account. We provide details below.

Consider the following thought experiment. Suppose, for argument's sake, that the quarterly autocorrelation of PE expected returns of $\varrho_{P,1}+\varrho_{P,2}=0.20$ in our baseline calibration is entirely due to smoothing. The LP may then want to base its portfolio choice on unsmoothed returns by considering a model without serial correlation: $\varrho_{P,1}=\varrho_{P,2}=0$. To keep the overall properties of PE returns unchanged, this LP simultaneously recalibrates the other parameters of the PE returns process so that the expected PE return $\mathbb{E}[\log R_{P,t+1}|s_t]$ and volatility $\sigma(\log R_{P,t+1}|s_t)$ remain identical to that of the baseline calibration in both of the macroeconomic states. This involves setting $\nu_P(1)=0.0051$, $\nu_P(2)=0.0392$, $\sigma_P(1)=0.0772$, and $\sigma_P(2)=0.0427$ while keeping the remaining parameters unchanged from their baseline values in \Cref{tbl: parameters}.

\begin{figure}[tb]
  \centering
  \includegraphics[width=1.0\textwidth]{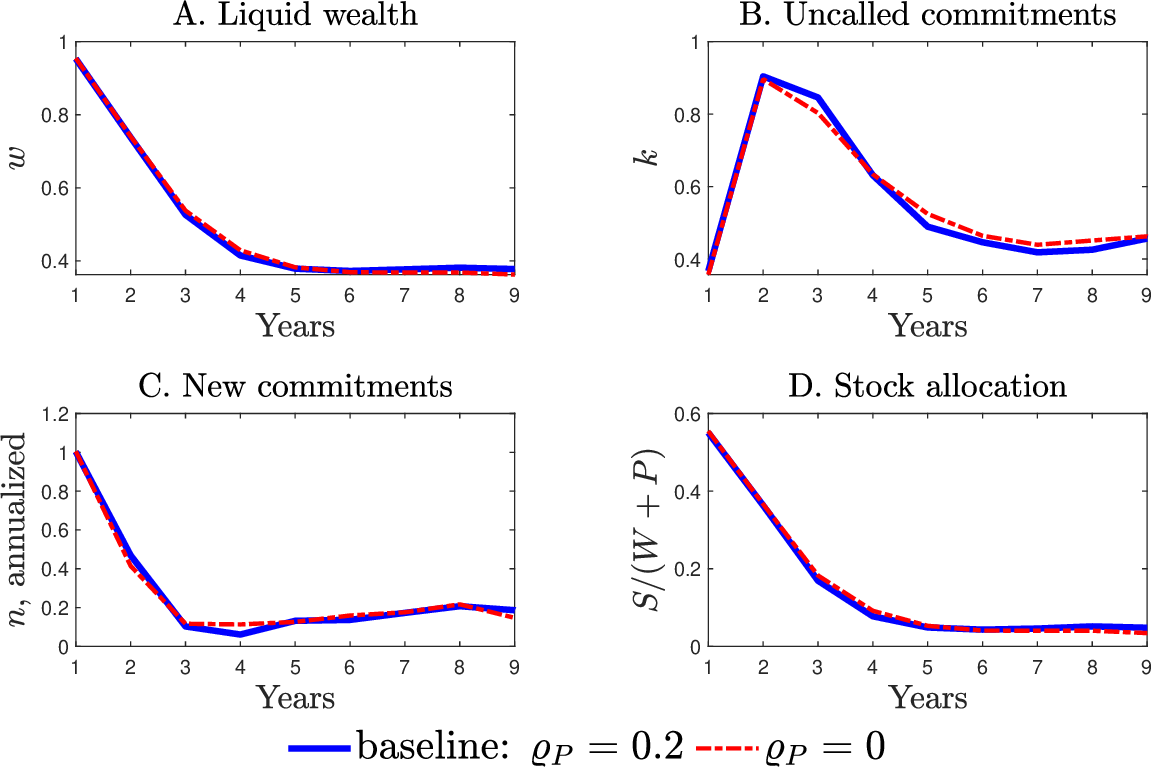}
  \caption{\textbf{PE return autocorrelation and life cycle dynamics.} \small The solid and dashed lines plot the average lifecycle outcomes under (1) our baseline model, and (2) a model where the PE returns process has zero autocorrelation but otherwise have the same return moments.}\label{fig: lifecycle uncond comp statics PE autocorr}
\end{figure}

\Cref{fig: lifecycle uncond comp statics PE autocorr} compares the resulting difference in life cycle dynamics between the baseline case and the case where the LP ignores serial correlation in PE returns. We see that the LP's portfolio allocations appear near-identical across the two cases. Furthermore, the two outcomes remain near-identical when we further condition outcomes on the macroeconomic state (see \Cref{fig: lifecycle cond comp statics PE autocorr} in \autoref{app: additional figures}).

The results in this section show that for long-term investors, the presence of serial correlation in PE returns may not be a critical factor in determining optimal portfolio allocations.

\subsection{Risk budget}
\label{sec: risk budget}

% \begin{enumerate}
%   \item Higher risk aversion: $\gamma=4$
%   \item Higher PE adjustment costs: $\gamma_N=0.2$
%   \item Differences in risk budget
%     \begin{enumerate}
%       \item Version 1: $\theta_S=\theta_P=2$
%       \item Version 2: $\theta_S=1.5$ and $\theta_P=2$
%     \end{enumerate}
%   \item Higher cost of liquidation: $\alpha(1)=\alpha(2)=0.66$
%   \item Different correlation between PE and stocks: $\rho(1)=\rho(2)=0.4575$
%   \item No business cycle: always in the expansionary state ($p_{12}=p_{21}=0$ and $s_0=2$)
%   \item PE returns are uncorrelated: set the conditional autocorrelation to zero, $\varrho_P=0$ and recalibrate $\nu_P(s)$ and $\sigma_P(s)$ so that $\mathbb{E}[\log R_{P,t+1}|s_t]$ and $\sigma(\log R_{P,t+1}|s_t)$ remain the same
% \end{enumerate}

The risk weights in our baseline calibration $\theta_S=\theta_P=1.5$ correspond to 50\% risk charges for PE and stocks (the risk budget threshold $\overline{\theta}$ is normalized to 1). In this section, we conduct comparative static exercises to investigate how the LP's outcomes depend on these risk charges. This is a useful exercise in at least two contexts. First, while risk charges of 50\% are representative numbers on average, risk charges can vary widely depending on asset quality. For example, S\&P Global assesses anywhere between 35\% to 99\% market risk charges for equities depending on asset quality \citep[Table 14]{spglobal:23}. Second, individual LPs may assess different risk charges based on their own individual circumstances. For example, an insurer may assess higher risk charges for its PE investments depending on properties of its insurance portfolio; the risk weights $\theta_P$ and $\theta_S$ can then be interpreted as effective risk charges whose values depend on the riskiness of other components of the insurer's balance sheet.

We demonstrate the effects of risk charges through two scenarios. In the first scenario, we set $\theta_S=\theta_P=2$, so that risk charges are uniformly higher. In the second scenario, we increase the risk charge for PE to $\theta_P=2$ while keeping $\theta_S=1.5$ unchanged. For both scenarios, we take the remaining parameters from the baseline calibration in \Cref{tbl: parameters}.

%\begin{figure}[p]
%  \centering
%  \begin{subfigure}{\textwidth}
%      \centering
%      \caption{\textbf{Scenario 1: $\theta_S=\theta_P=2$.}}
%      \includegraphics[width=0.9\textwidth]{figures/comparison_lifecycle_uncond_risk_budget1.eps}
%      \label{fig: lifecycle uncond comp statics risk budget 1}
%  \end{subfigure}
%
%  \vspace{2em} % Adjust vertical spacing if needed
%
%  \begin{subfigure}{\textwidth}
%      \centering
%      \caption{\textbf{Scenario 2: $\theta_S=1.5,\;\;\theta_P=2$.}}
%      \includegraphics[width=0.9\textwidth]{figures/comparison_lifecycle_uncond_risk_budget2.eps}
%      \label{fig: lifecycle uncond comp statics risk budget 2}
%  \end{subfigure}
%
%  \caption{\textbf{Life cycle dynamics under different risk charges.} \small We consider outcomes under alternative risk charges. Scenario 1 sets $\theta_S=\theta_P=2$ while scenario 2 sets $\theta_S=1.5$ and $\theta_P=2$.}
%  \label{fig:lifecycle uncond comp statics risk budget}
%\end{figure}

\begin{figure}[t]
  \centering
    \includegraphics[width=1.0\textwidth]{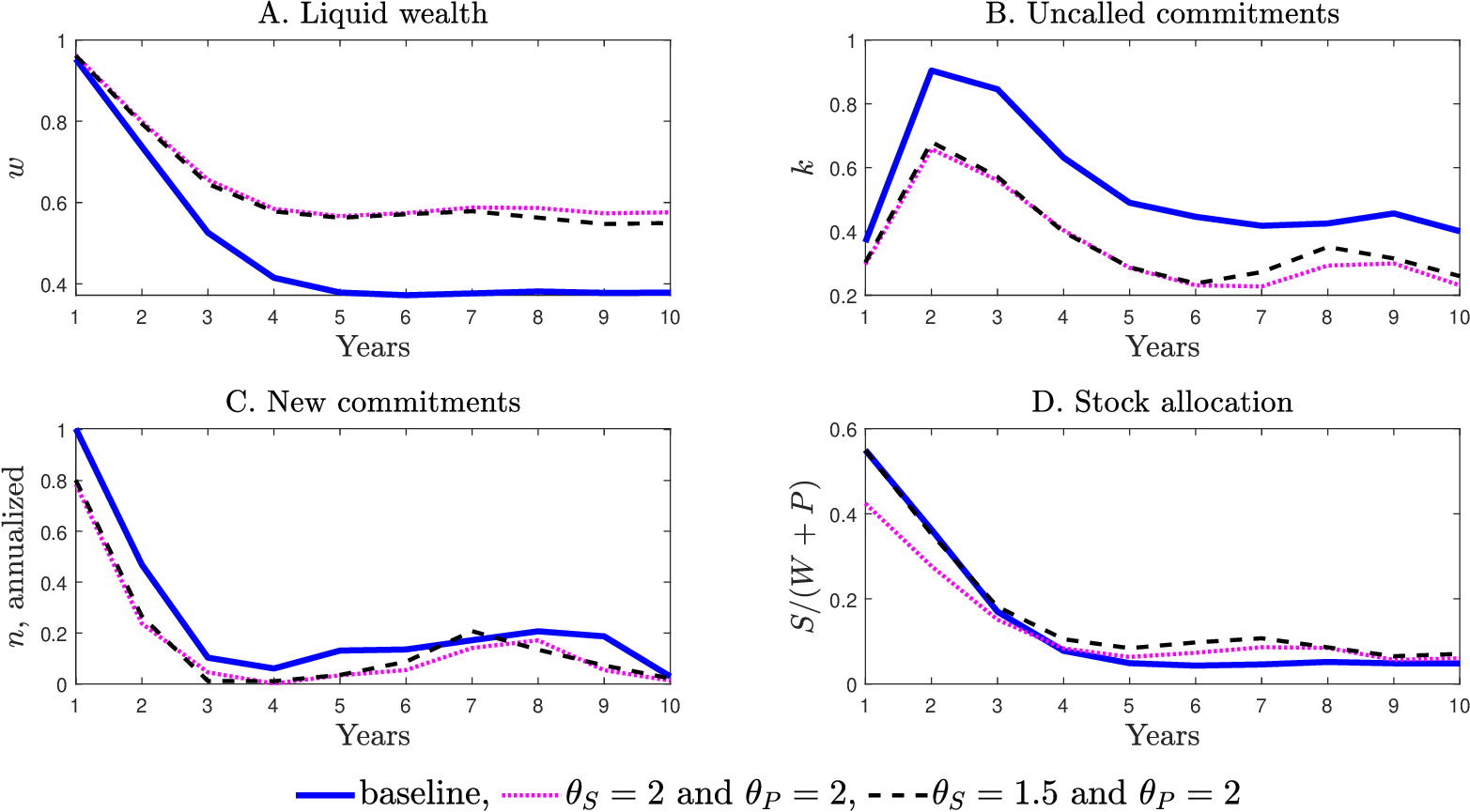}
  \caption{\textbf{Life cycle dynamics under different risk charges.} \small This figure displays average life cycle outcomes under alternative risk charges. For reference, the solid line plots outcomes under the baseline calibration in which $\theta_P=\theta_S=1.5$. The dotted line plots outcomes when both $\theta_S$ and $\theta_P$ equal 2, and the dashed line plots outcomes when only $\theta_P$ is increased to 2.}
  \label{fig: lifecycle uncond comp statics risk budget}
\end{figure}

% compare_tot_wealth_dist_risk_budget

\Cref{fig: lifecycle uncond comp statics risk budget} illustrates the impact of risk charges on average life cycle outcomes. Compared to the baseline outcome (solid line), we see that overall allocations to PE and stocks are lower under the first scenario (dotted line) in which the risk charges for both PE and stocks are increased to $\theta_P=\theta_S=2$. For instance, during the initial ramp-up phase, uncalled commitments peak at 66\% of total wealth in year 2, as opposed to 90\% in the baseline calibration. Similarly, stock allocation in year 1 is reduced to 42\% of total wealth on average compared to 55\% in the baseline. In the maintenance phase from year 5 onwards, the PE allocation is 42\% of total wealth, down from 62\% in the baseline. The LP does, however, compensates with a slightly higher stock allocation during the maintenance phase: 7\% of total wealth on average compared to 5\% on average for the baseline. \Cref{fig: lifecycle cond comp statics risk budget 1} in \Cref{app: additional figures} additionally displays outcomes for the first scenario after further conditioning on the business cycle.

The second scenario isolates the effect of risk charges by only increasing $\theta_P$. In this case, we see from \Cref{fig: lifecycle uncond comp statics risk budget} that only PE allocations are affected to first order. Specifically, the average outcome for stock allocation remains roughly unchanged from the baseline (compare the solid and dashed lines in panel D) while the outcomes for PE allocation are similar to that under the first scenario (compare the dotted and dashed lines in panels A, B, and C). \Cref{fig: lifecycle cond comp statics risk budget 2} in \Cref{app: additional figures} additionally displays outcomes for the second scenario after further conditioning on the business cycle.

\begin{figure}[t]
  \centering
    \includegraphics[width=1.0\textwidth]{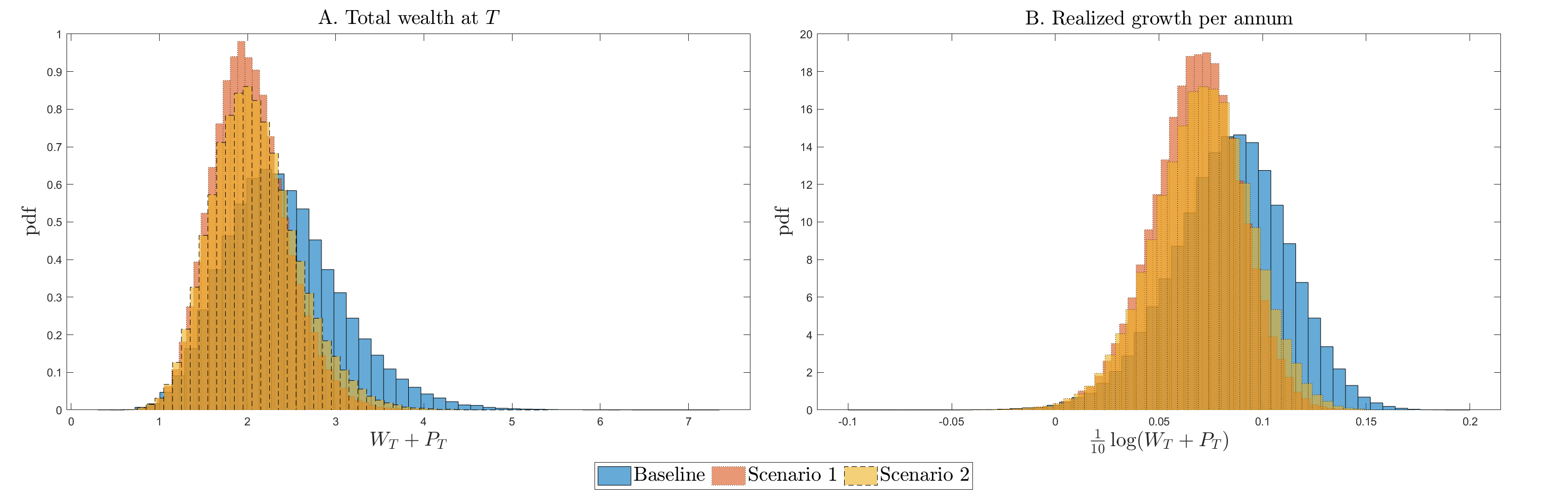}
  \caption{\textbf{Total wealth outcomes under different risk charges.} \small This figure illustrates the distribution of total wealth outcomes at the end of the investment horizon $t=T$. The initial total wealth is $W_0+P_0=1$. Panel A plots the distribution for total wealth $W_T+P_T$. Panel B plots the distribution for the realized growth in total wealth per annum, $\log(W_T+P_T)/10$. Both panels compare outcomes under (1) the baseline in which $\theta_P=\theta_S=1.5$, (2) scenario 1 in which both $\theta_P$ and $\theta_S$ are increased to 2, and (3) scenario 2 in which only $\theta_P$ is increased to 2.}
  \label{fig: wealth distribution comp statics risk budget}
\end{figure}

\Cref{fig: wealth distribution comp statics risk budget} compares the distribution of final outcomes under the different risk charge scenarios. We see that higher risk charges do indeed decrease the riskiness of the LP's overall portfolio. For example, the standard deviation of annualized realized returns decreases to 2.35\% (from the baseline value of 2.78\%) under the second scenario when $\theta_P$ increases to 2; it further decreases to 2.14\% under the first scenario when both $\theta_P$ and $\theta_S$ increase to 2. This decrease in risk does come at the expense of lower returns. The average annualized return decreases to 7.08\% under the second scenario (down from 8.41\% for the baseline), and further decreases to 6.79\% under the first scenario.

The costs of imposing higher risk charges in certainty equivalent terms are as follows. Increasing $\theta_P$ alone to 2 (i.e., scenario 2) reduces the certainty equivalent wealth from the baseline value of 2.25 to 2.00 or, equivalent, a 11\% reduction in  certainty equivalent wealth. When we additionally increase $\theta_S$ to 2 (i.e., scenario 1), the certainty equivalent wealth further decreases to 1.92 for a 15\% reduction in certainty equivalent wealth compared to the baseline.

\section{Conclusion}

%Exciting recent developments in solution techniques involving various machine learning methods have enabled us to tackle portfolio choice problems with higher dimensions, more realistic trading frictions, and more complex portfolio constraints. We use Deep Kernel Gaussian Processes to accurately characterize the optimal policies in a private asset allocation model featuring many realistic factors that challenge private equity investors.

Exciting recent developments in machine learning have enabled the solution of portfolio choice problems featuring higher dimensionality, more realistic trading frictions, and more complex constraints. In this paper, we employ Deep Kernel Gaussian Processes to accurately characterize the optimal policies in a private asset allocation model that incorporates a host of real-world complexities, including illiquidity, commitment lags, serial correlation in returns, business cycle fluctuations, and regulation-induced constraints. Our approach not only provides novel insights into private equity investing but also offers a flexible blueprint for tackling similarly challenging problems in economics and finance that involve substantial nonlinearities and a large number of state variables.

\clearpage

\bibliographystyle{rfs}
\bibliography{references_PE}

\clearpage

\appendix

\begin{center}
  {\noindent{\Large \textbf{Appendix}}}
  \end{center}

  \setstretch{1.0}

\begin{small}

  \renewcommand{\thetable} {A.\arabic{table}}
\setcounter{table}{0}
\renewcommand{\thefigure} {A.\arabic{figure}}
\setcounter{figure}{0}
\renewcommand{\theequation}{A.\arabic{equation}}
\setcounter{equation}{0}
\section{Specification for expected PE returns}\label{app: expected PE returns}

In this section, we use the \citet{getmansky/lo/makarov:2004} model of smoothed returns for illiquid alternative investments to motivate our specification \eqref{eq: muP AR1 process} for PE expected returns.

Applied to our context, the \citet[equation 21]{getmansky/lo/makarov:2004} model posits that the observed return on PE investments $\log(R_{P,t+1})$ is a weighted average of contemporaneous and lagged values of the true but unobserved return on PE investments:
\begin{equation}
    \log(R_{P,t+1}) = \omega_0 \log(\widetilde{R}_{P,t+1})+\omega_1\log(\widetilde{R}_{P,t})+...+\omega_{k}\log(\widetilde{R}_{P,t+1-k}),\label{eq: GLM model}
\end{equation}
In terms of notation for this section, we use a tilde to denote quantities associated with the true but unobserved returns. For example, $\log(R_{P,t+1})$ is the observed PE return at $t+1$ while $\log(\widetilde{R}_{P,t+1})$ is the unobserved contemporaneous true return. The coefficients $\omega_j$ in equation \eqref{eq: GLM model} are positive weights that sum to 1. We use the geometric weighting scheme considered in \citet[equation 43]{getmansky/lo/makarov:2004}:
\begin{equation}
    \omega_0=(1-\phi)/(1-\phi^{k+1}),\quad \omega_j = \omega_0 \phi^j \mbox{ for } j=1,...,k,\label{eq: GLM model weights}
\end{equation}
where $\phi\in(0,1)$. This weighting scheme leads to the convenient AR(1) process for the expected returns on observed PE returns \eqref{eq: muP AR1 process}.

Specifications \eqref{eq: GLM model} and \eqref{eq: GLM model weights} imply that the expected return for observed PE returns $\mu_{P,t}\equiv\mathbb{E}_t[\log(R_{P,t+1})]$ evolves as follows:
\begin{align}
    \mu_{P,t} & = \omega_0 \mathbb{E}_{t}[\log(\widetilde{R}_{P,t+1})]+\phi\mathbb{E}_{t}[\omega_0\log(\widetilde{R}_{P,t})+...+\omega_0\phi^{k-1}\log(\widetilde{R}_{P,t+1-k})]\nonumber\\
    &=\omega_0 \mathbb{E}_{t}[\log(\widetilde{R}_{P,t+1})]+\phi\mathbb{E}_{t}[\log(R_{P,t})-\omega_0\phi^{k}\log(\widetilde{R}_{P,t-k})]\nonumber\\
    &= \omega_0 \widetilde{\mu}_{P,t} + \phi \log(R_{P,t}) - \omega_0\phi^{k+1}\mathbb{E}_{t}[\log(\widetilde{R}_{P,t-k})]\label{eq: muP derivation 1}\\
    &\approx \omega_0 \widetilde{\mu}_{P,t} + \phi \log(R_{P,t})\label{eq: muP derivation 2}\\
    &=\omega_0 \widetilde{\mu}_{P,t} + \phi \mu_{P,t-1} + \phi\sigma_{P,t-1}\varepsilon_{P,t}.\label{eq: muP derivation 3}
\end{align}
Equation \eqref{eq: muP derivation 1} is a direct consequence of specifications \eqref{eq: GLM model} and \eqref{eq: GLM model weights}; $\widetilde{\mu}_{P,t}\equiv \mathbb{E}_{t}[\log(\widetilde{R}_{P,t+1})]$ is the expected return for the true return process. The approximation \eqref{eq: muP derivation 2} holds if $k$ is large enough or if $\phi$ is small enough. Equation \eqref{eq: muP derivation 3} writes the observed return as $\log(R_{P,t})=\mu_{P,t-1}+\sigma_{P,t-1}\varepsilon_{P,t}$ where $\mu_{P,t-1}\equiv \mathbb{E}_{t-1}\left[\log(R_{P,t})\right]$, $\sigma_{P,t-1}\equiv \sqrt{Var_{t-1}\left[\log(R_{P,t})\right]}$, and $\varepsilon_{P,t}\equiv \left(\log(R_{P,t})-\mu_{P,t-1}\right)/\sigma_{P,t-1}$. Comparing equation \eqref{eq: muP derivation 3} to equation \eqref{eq: muP AR1 process}, we see that the AR(1) process for expected PE returns \eqref{eq: muP AR1 process} is a more flexible version of equation \eqref{eq: muP derivation 3} in which the loading on the $\sigma_{P,t-1}\varepsilon_{P,t}$ term is allowed to differ from the autoregressive coefficient.

We make further functional form assumptions in our implementation. Specifically, the specification in equations \eqref{eq: asset returns specification} and \eqref{eq: muP AR1 process} assumes (1) the expected true return $\widetilde{\mu}_{P,t}=\widetilde{\mu}_{P}(s_t)$ depends only on the macroeconomic state (hence $\nu_P(s_t)$ in equation \eqref{eq: muP AR1 process} corresponds to the $\omega_0\widetilde{\mu}_{P,t}$ term from equation \eqref{eq: muP derivation 3}), (2) the volatility of observed PE returns $\sigma_{P,t-1}=\sigma_P(s_{t-1})$ depends only on the macroeconomic state, and (3) the shock to observed PE returns $\varepsilon_{P,t}\sim\mathcal{N}(0,1)$ is normally distributed. 

\renewcommand{\thetable} {B.\arabic{table}}
\setcounter{table}{0}
\renewcommand{\thefigure} {B.\arabic{figure}}
\setcounter{figure}{0}
\renewcommand{\theequation}{B.\arabic{equation}}
\setcounter{equation}{0}
  %\section{Model details}\label{app: model details}

\section{Scaled value functions}\label{app: scaled value functions}

\paragraph{Scaled choice variables.}

We work with the following scaled versions of the choice variables:
\begin{align}
    n&\equiv\frac{N}{W+P},\\
    \widetilde{\phi}_S&\equiv \frac{S+\gamma_S(W+P)\left(\frac{S}{W+P}\right)^2}{W-\gamma_N (W+P)\left(n-\overline{n}\right)^2}.\label{eq: def phiS 1}
\end{align}
Here, $n$ is the new PE commitments scaled by total wealth. The interpretation for $\widetilde{\phi}_S$ is as follows. The denominator in equation \eqref{eq: def phiS 1} is the amount of liquid wealth that is available to be invested in stocks and bonds after deducting adjustment costs for new PE commitments from current liquid wealth. Of this amount, $\widetilde{\phi}_S$ is the fraction invested in stocks inclusive of stock adjustment costs. We also denote by
\begin{equation}
    \phi_S\equiv \frac{S}{W-\gamma_N (W+P)\left(n-\overline{n}\right)^2}.\label{eq: def phiS 2}
\end{equation}
the fraction invested in stocks excluding stock adjustment costs. The relation between $\widetilde{\phi}_S$ and $\phi_S$ is
\begin{equation}
    \phi_S = \frac{-1+\sqrt{1 + 4\gamma_S[w-\gamma_N(n-\overline{n})^2]\widetilde{\phi}_S}}{2\gamma_S[w-\gamma_N(n-\overline{n})^2]}.\label{eq: phiS relation}
\end{equation}

\paragraph{Scaled value function in default.}

The recursive formulation for the scaled value function in default is
\begin{equation}\label{eq: scaled value function default}
    v_D(t,s) = \max_{\widetilde{\phi}_S\in[0,\widetilde{\phi}_{S,max}(s)]}\mathbb{E}\left[\left(g_D^\prime \right)^{1-\gamma}|s\right]^{\frac{1}{1-\gamma}}\mathbb{E}\left[v_D(t+1,s^\prime)^{1-\gamma}|s\right]^{\frac{1}{1-\gamma}}
\end{equation}
where
\begin{equation}\label{eq: growth rate default gD'}
    g_D^\prime\equiv\frac{W^\prime}{W}=R_S^\prime \phi_S + R_f(s)\left(1 - \widetilde{\phi}_S\right)-\Gamma(\theta_S \widetilde{\phi}_S)
\end{equation}
is the growth rate of total wealth in default and
\begin{equation*}
    v_D(T,s)=1
\end{equation*}
is the terminal condition. Note that when computing the scaled value function \eqref{eq: scaled value function default}, equations \eqref{eq: def phiS 1}, \eqref{eq: def phiS 2}, and \eqref{eq: phiS relation} for $\widetilde{\phi}_S$, $\phi_S$, and their relation, respectively, are computed with $P=0$ and $n=0$. In addition, note that while the $\mathbb{E}\left[u_D(t+1,s^\prime)^{1-\gamma}|s\right]^{\frac{1}{1-\gamma}}$ on the right-hand side of equation \eqref{eq: scaled value function default} affects the scaled value function, it does not affect the optimal portfolio choice. This is because the growth rate of total wealth \eqref{eq: growth rate default gD'} does not depend on $s^\prime$. As a result, the optimal policies in default only depend on the macroeconomic state $s$.

The upper bound $\widetilde{\phi}_{S,max}(s)\equiv\sup\left\{\widetilde{\phi}_S\in[0,1]:\;\inf_{R_S^\prime} g_D^\prime = R_f(s)\left(1 - \widetilde{\phi}_S\right)-\Gamma(\theta_S \widetilde{\phi}_S)\right\}$ appearing in problem \eqref{eq: scaled value function default} ensures that the growth in wealth $g_D^\prime$ is strictly positive so that utilities are well-defined. The upper bound equals
\begin{equation}
    \widetilde{\phi}_{S,max}(s)=\left\{
    \begin{array}{ll}
        1 & \mbox{if } \theta_S\leq \overline{\theta}, \\
        \min\left\{\frac{\overline{\theta}}{\theta_S}+\frac{-R_f(s)+\sqrt{R_f(s)^2+4\kappa\theta_SR_f(s)(\theta_S-\overline{\theta})}}{2\kappa\theta_S^2},1\right\} &  \mbox{if } \theta_S> \overline{\theta}.
    \end{array}
    \right.
\end{equation}
Similarly, the lower bound $\widetilde{\phi}_S\geq 0$ rules out shorting which is also necessary to ensure that utilities are well-defined.

\paragraph{Scaled value function before default.}

The recursive formulation for the scaled value function before default is
\begin{equation}\label{eq: scaled value function not default}
    v(t,w,k,\mu_P,s)=\max_{n,\widetilde{\phi}_S}\mathbb{E}\left[\max\left\{ g_{D}^{\prime}v_D\left(t+1,s^{\prime}\right),g^{\prime}v\left(t+1,w^{\prime},k^{\prime},\mu_{P}^{\prime},s^{\prime}\right)\right\} ^{1-\gamma} \left|w,k,\mu_P,s\right. \right]^{\frac{1}{1-\gamma}}
\end{equation}
subject to
\begingroup
\allowdisplaybreaks
\begin{align}
g_{D}^{\prime} =&\,\left(1-w\right)\left[\lambda_D(s^\prime)+\alpha(s^\prime)(1-\lambda_D(s^\prime))\right]R_{P}^{\prime}-\Gamma(\theta_D)\nonumber\\
&+\left[w-\gamma_N(n-\overline{n})^2\right]\left[\phi_S R_S^\prime + (1-\widetilde{\phi}_S)R_f(s)\right],\nonumber\\
\theta_D=&\,\frac{\theta_S\left[w-\gamma_N(n-\overline{n})^2\right]\widetilde{\phi}_S}{1-\gamma_N(n-\overline{n})^2},\nonumber\\
g^\prime=&\,(1-w)R_P^\prime + \left[w-\gamma_N(n-\overline{n})^2\right]\left[\phi_S R_S^\prime + (1-\widetilde{\phi}_S)R_f(s)\right]-\Gamma(\theta),\label{eq: total wealth growth not default}\\
\theta=&\,\frac{\theta_P(1-w)+\theta_S\left[w-\gamma_N(n-\overline{n})^2\right]\widetilde{\phi}_S}{1-\gamma_N(n-\overline{n})^2},\nonumber\\
k^\prime=&\,\frac{[1-\lambda_K(s^\prime)]k+[1-\lambda_N(s^\prime)]n}{g^\prime},\nonumber\\
w^\prime=&\,\frac{
\left\{
\begin{array}{c}
     \lambda_D(s^\prime)R_P^\prime(1-w)-\lambda_K(s^\prime)k-\lambda_N(s^\prime)n-\Gamma(\theta)  \\
     +\left[w-\gamma_N(n-\overline{n})^2\right]\left[\phi_S R_S^\prime + (1-\widetilde{\phi}_S)R_f(s)\right]
\end{array}
\right\}
}{g^\prime},\nonumber\\
\mu_P^\prime=&\,\varrho_{P,1}\mu_P+\varrho_{P,2}\log R_P^\prime + \nu_P(s^\prime),\nonumber\\
n\in&\,[n_{min}(w), n_{max}(w)], \quad \widetilde{\phi}_S\in[0,\widetilde{\phi}_{S,max}(w,n,s)],\label{eq: choice bounds not default}
\end{align}
\endgroup
with terminal condition
\begin{equation*}
v\left(T,w,k,\mu_{P},s\right)=1.
\end{equation*}
Here, $g_D^\prime=W_D^\prime/(W+P)$ and $g^\prime=(W^\prime+P^\prime)/(W+P)$ denote the growth rates of total wealth conditional on defaulting and not defaulting, respectively.

The limits on the choice variables \eqref{eq: choice bounds not default} ensure that utility is well-defined. To see why, note that the log returns on PE and the stock are normally distributed and are therefore unbounded. Hence, this would require $\max\{g_D^\prime,g^\prime\}>0$ for all realization of return shocks. This condition implies
$\inf g_D^\prime=[w-\gamma_N(n-\overline{n})^2](1-\widetilde{\phi}_S)R_f(s)-\Gamma(\theta_D)\geq0$ where the infimum is taken over both $R_P^\prime$ and $R_S^\prime$. From this, we obtain
\begin{equation*}
    n_{min}(w)=\max\left\{\overline{n}-\sqrt{\frac{w}{\gamma_N}},0\right\},\quad n_{max}(w)=\overline{n}+\sqrt{\frac{w}{\gamma_N}},
\end{equation*}
and
\begin{align*}
    &\widetilde{\phi}_{S,max}(w,n,s)\\
    =&\left\{
    \begin{array}{ll}
      1   & \mbox{if } \widetilde{\theta}_S(w,n)\leq \overline{\theta}, \\
      \min\left\{\frac{\overline{\theta}}{\widetilde{\theta}_S(w,n)}+\frac{-\widetilde{R}_f(w,n,s)+\sqrt{\widetilde{R}_f(w,n,s)^2+4\kappa\widetilde{\theta}_S(w,n)\widetilde{R}_f(w,n,s)\left[\widetilde{\theta}_S(w,n)-\overline{\theta}\right]}}{2\kappa\widetilde{\theta}_S(w,n)^2},1\right\}   & \mbox{if } \widetilde{\theta}_S(w,n)> \overline{\theta},
    \end{array}
    \right.
\end{align*}
where
\begin{equation*}
    \widetilde{\theta}_S(w,n)\equiv \frac{\theta_S[w-\gamma_N(n-\overline{n})^2]}{1-\gamma_N(n-\overline{n})^2},\quad\mbox{and }\widetilde{R}_f(w,n,s)\equiv[w-\gamma_N(n-\overline{n})^2]R_f(s).
\end{equation*} 

\paragraph{Scaled value function for the heuristic problem \eqref{eq: heuristic problem}.} 

The value function for the heuristic problem \eqref{eq: heuristic problem} can be scaled as $V_{\mbox{heuristic}}(t,W_{total},s)=\left[W_{total} v_{\mbox{heuristic}(t,s)}\right]^{1-\gamma}/(1-\gamma)$ where $v_{\mbox{heuristic}}$ solves
\begin{equation}\label{eq: scaled value function heuristic}
    v_{\mbox{heuristic}}(t,s) = \max_{\phi_S,\phi_P}\mathbb{E}\left[\left(g_{total}^\prime \right)^{1-\gamma}|s\right]^{\frac{1}{1-\gamma}}\mathbb{E}\left[v_{\mbox{heuristic}}(t+1,s^\prime)^{1-\gamma}|s\right]^{\frac{1}{1-\gamma}}
\end{equation}
subject to
\begin{align*}
  g_{total}^\prime\equiv\frac{W_{total}^\prime}{W_{total}}&=R_S^\prime \phi_S +R_P^\prime \phi_P + R_f(s)\left(1 - \phi_S-\phi_P\right)-\Gamma(\theta_S\phi_S+\theta_P\phi_P),\\
  \phi_S,\phi_P&\in[0,1],\\
  \phi_S+\phi_P&=1
\end{align*}
and the terminal condition $v_{\mbox{heuristic}}(T,s)=1$.

  \renewcommand{\thetable} {C.\arabic{table}}
\setcounter{table}{0}
\renewcommand{\thefigure} {C.\arabic{figure}}
\setcounter{figure}{0}
\renewcommand{\theequation}{C.\arabic{equation}}
\setcounter{equation}{0}
\section{Numerical Implementation}
\label{sec: numerical implementation}

We first solve the scaled problem in default \eqref{eq: scaled value function default}. In doing so, we note that the the value function can be split as
\begin{equation}
    v_D(t,s) = \mathbb{E}\left[\left(u_D(t+1,s^\prime)\right)^{1-\gamma}|s\right]^{\frac{1}{1-\gamma}} \max_{\widetilde{\phi}_S\in[0,\widetilde{\phi}_{S,max}(s)]} \mathbb{E}\left[\left(g_D^\prime\right)^{1-\gamma}|s\right]^{\frac{1}{1-\gamma}}.
\end{equation}
This is because $g_D^\prime$ does not depend on $s^\prime$ (see equation \eqref{eq: growth rate default gD'}). Hence, the optimal portfolio choice problem depends $s$ alone and is obtained by solving
\begin{equation}
    \max_{\widetilde{\phi}_S\in[0,\widetilde{\phi}_{S,max}(s)]} \mathbb{E}\left[\left(g_D^\prime\right)^{1-\gamma}|s\right]^{\frac{1}{1-\gamma}}.
\end{equation}
To ensure numerical accuracy, we use adaptive quadrature in to compute expectations.

After obtaining $v_D(t,s)$, we solve the scaled problem before default (see equation \eqref{eq: scaled value function not default}) using backward induction. The procedure is as follows:
\begin{enumerate}
    \item Initiate the value function at $t=T$ using the terminal condition $v\left(T,w,k,\mu_{P},s\right)=1$.
    \item Obtain sample points for the state variables $(w,k,\mu_{P})$. In doing so, we sample 800 points via a Halton sequence over the hypercube $\{(w,k,\mu_{P}):\;w\in[0, 1], k\in[0 1.5], \mu_P\in[\mu_{P,lb},\mu_{P,ub}]\}$ where the bounds $\mu_{P,lb}$ and $\mu_{P,ub}$ are set to be the 0.1st and 99.9th percentiles of the stationary distribution for $\mu_P$, respectively.
    \item Given a value function $v_{GP}\left(t+1,w,k,\mu_{P},s\right)$ that has been previously fitted via Deep Kernel Gaussian Process Regression, do the following:
        \begin{enumerate}
            \item For each $s$ at time $t$, solve problem \eqref{eq: scaled value function not default} for the sample points for $(w,k,\mu_{P})$ obtained in step 2. When doing so, we use the fitted value function $v_{GP}\left(t+1,\cdot\right)$ as the continuation value when not defaulting next period. As in the problem after default, we use adaptive quadrature to compute expectations to ensure numerical accuracy. To ensure that we find a global maximum, we first consider a coarse grid of policy points. We then run an interior-point optimization algorithm starting from the best policy based on the coarse grid. The end result is a set of observations for the value and policy functions at the sample points.
            \item We use the observations from step (a) to fit deep kernel Gaussian processes for the value and policy functions at time $t$ for each state $s$. Our choice for the deep kernel \eqref{eq: def deep kernel} is as follows. The neural network component $NN$ takes the three-dimensional input $(w,k,\mu_P)$ and passes it through 3 hidden layers with 64, 32, and 16 units, respectively, before outputting 2 features (the shape of the hidden layers follow a so-called ``decoder structure''); we use Gaussian error linear unit (GELU) activation. These 2 output features are then passed through a Matern 5/2 kernel \eqref{eq: ardmatern52}. We train the parameters of the deep kernel Gaussian processes by maximizing the marginal likelihood \eqref{eq: GP marginal likelihood} using the Adam optimizer \citep{kingma2015adam} with a learning rate of 0.001. To ensure that the training converges to a global optimum, we run the Adam optimizer 10 times from different initial points. We always use the trained parameters from step $t+1$ as one of the initial points; the remaining initial points are randomly drawn with Xavier initialization for the neural network weights. We standardize both input and output variables when training.
        \end{enumerate}
    \item Repeat step 3 until $t=0$.
\end{enumerate}

\Cref{fig: illus value fun 3D} and \Cref{fig: illus policies 3D} illustrate the value function and policy functions at $t=0$, respectively.

\begin{figure}[tbh!]
  \centering
  \begin{subfigure}{0.8\textwidth}
  \centering
  \caption{\textbf{New commitments.}\label{fig: illus policy n 3D}}
  \includegraphics[width=\textwidth]{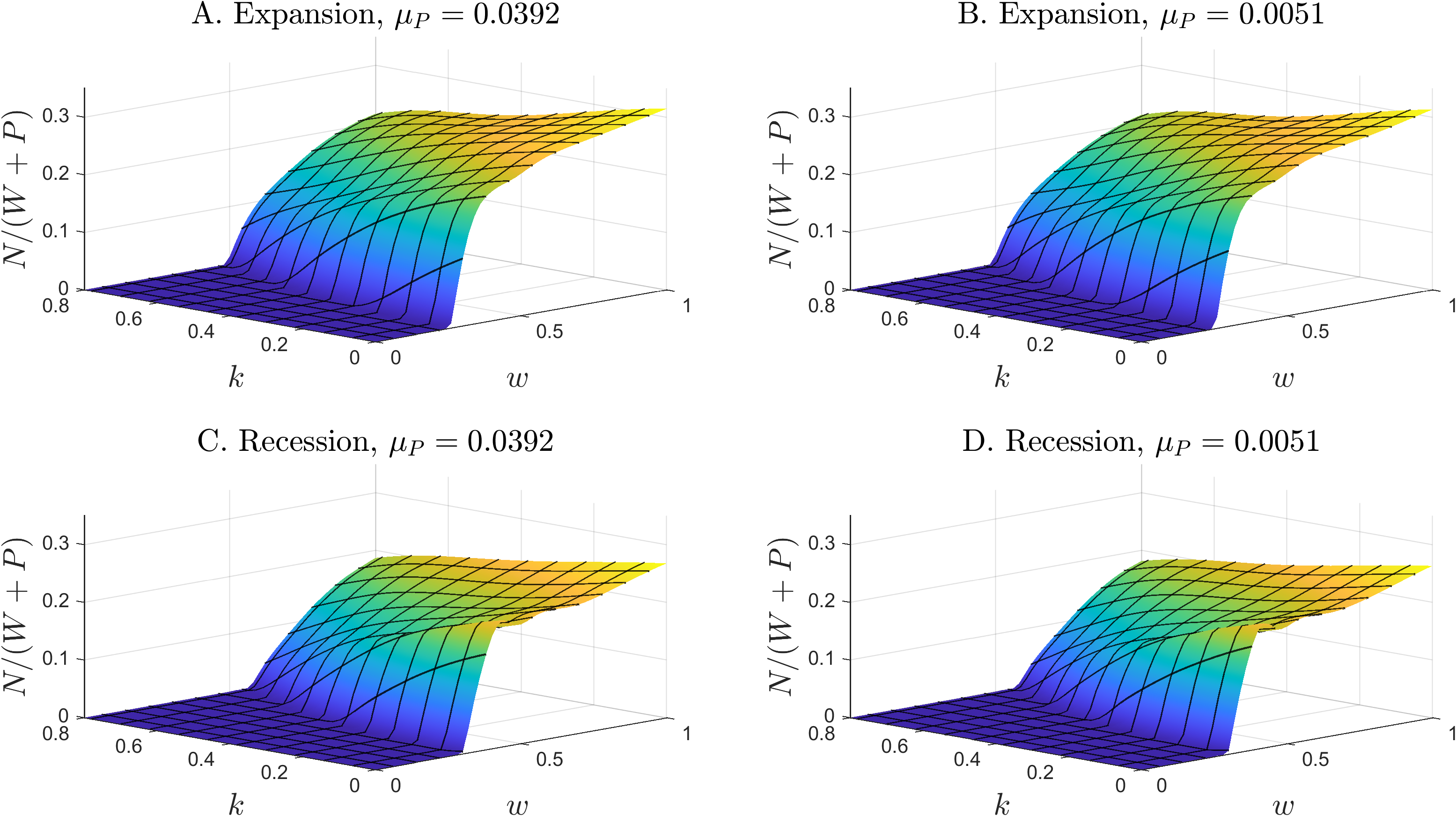}
  \end{subfigure}

  \begin{subfigure}{0.8\textwidth}
  \centering
  \caption{\textbf{Stock allocation.}\label{fig: illus policy S 3D}}
  \includegraphics[width=\textwidth]{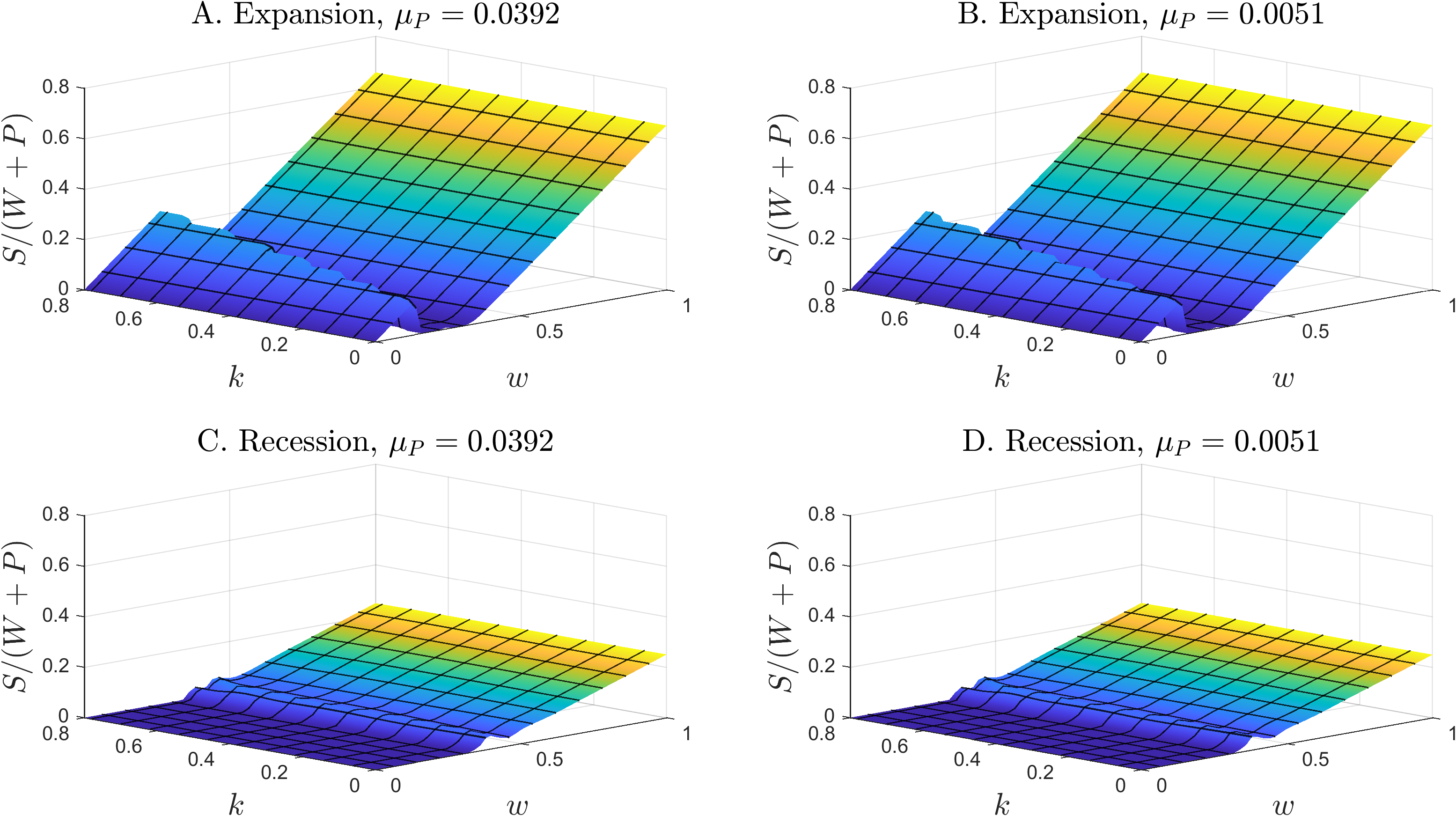}
  \end{subfigure}
  \caption{\textbf{Policies at $t=0$.}}\label{fig: illus policies 3D}
\end{figure}

%\begin{figure}[tbhp!]
%  \centering
%  \includegraphics[width=0.9\textwidth]{figures/illus_policy_n_3D.eps}
%  \caption{\textbf{Policy for new commitments at $t=0$.}}\label{fig: illus policy n 3D}
%\end{figure}
%
%\begin{figure}[tbhp!]
%  \centering
%  \includegraphics[width=0.9\textwidth]{figures/illus_policy_S_3D.eps}
%  \caption{\textbf{Policy for stock allocation at $t=0$.}}\label{fig: illus policy S 3D}
%\end{figure}
%\clearpage 

\paragraph{Heuristic portfolio choice problem.}

\renewcommand{\thetable} {D.\arabic{table}}
\setcounter{table}{0}
\renewcommand{\thefigure} {D.\arabic{figure}}
\setcounter{figure}{0}
\renewcommand{\theequation}{D.\arabic{equation}}
\setcounter{equation}{0}
\section{Additional Figures}
\label{app: additional figures}

\begin{figure}[htb!]
  \centering
  \includegraphics[width=0.9\textwidth]{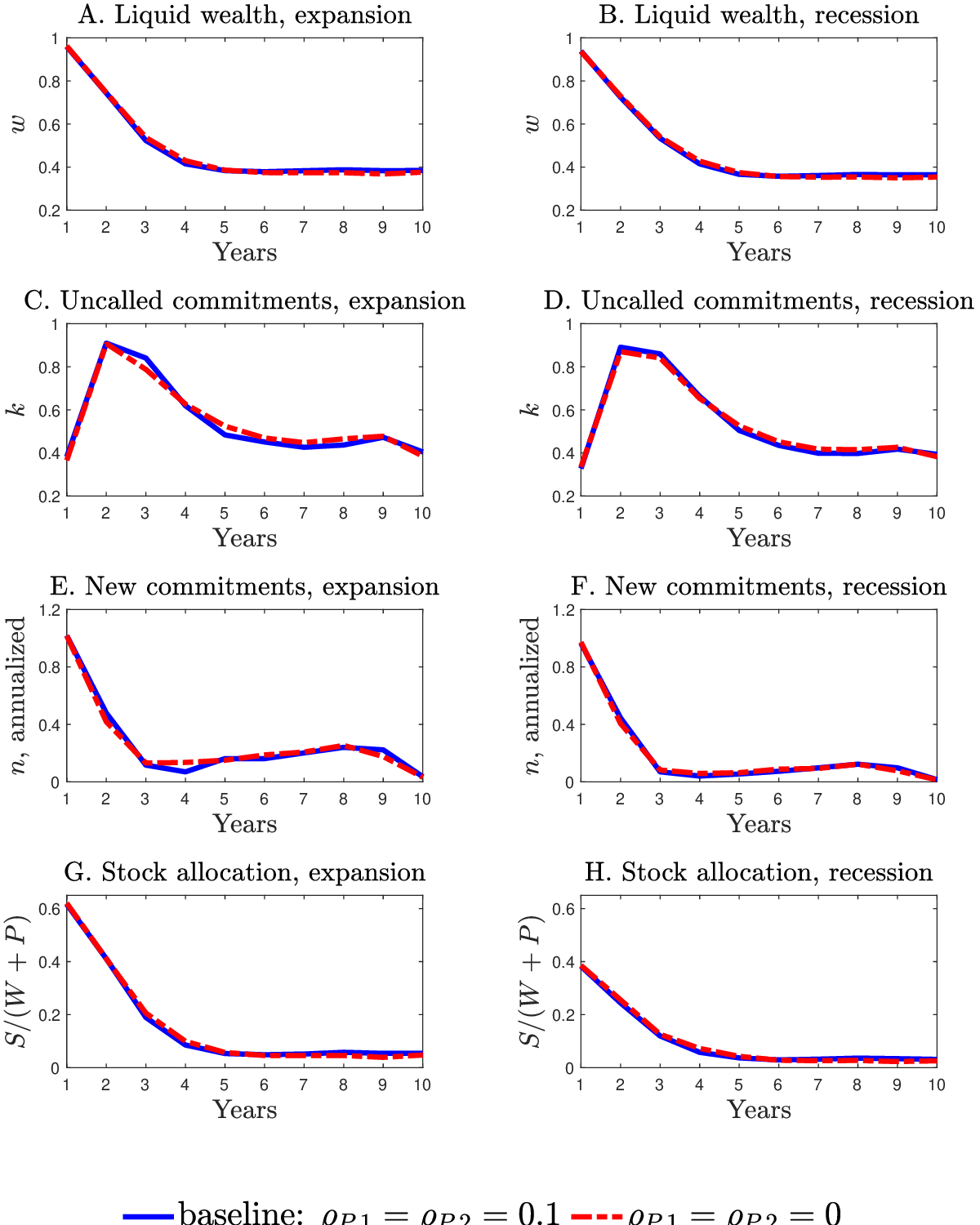}
  \caption{\textbf{Life cycle comparison (conditional), PE return autocorrelation.} \small The solid and dashed lines plot the average lifecycle outcomes under (1) our baseline model, and (2) a model where the PE returns process has zero autocorrelation but otherwise have the same return moments.}\label{fig: lifecycle cond comp statics PE autocorr}
\end{figure}
\clearpage

\begin{figure}[htb!]
  \centering
  \includegraphics[width=1.0\textwidth]{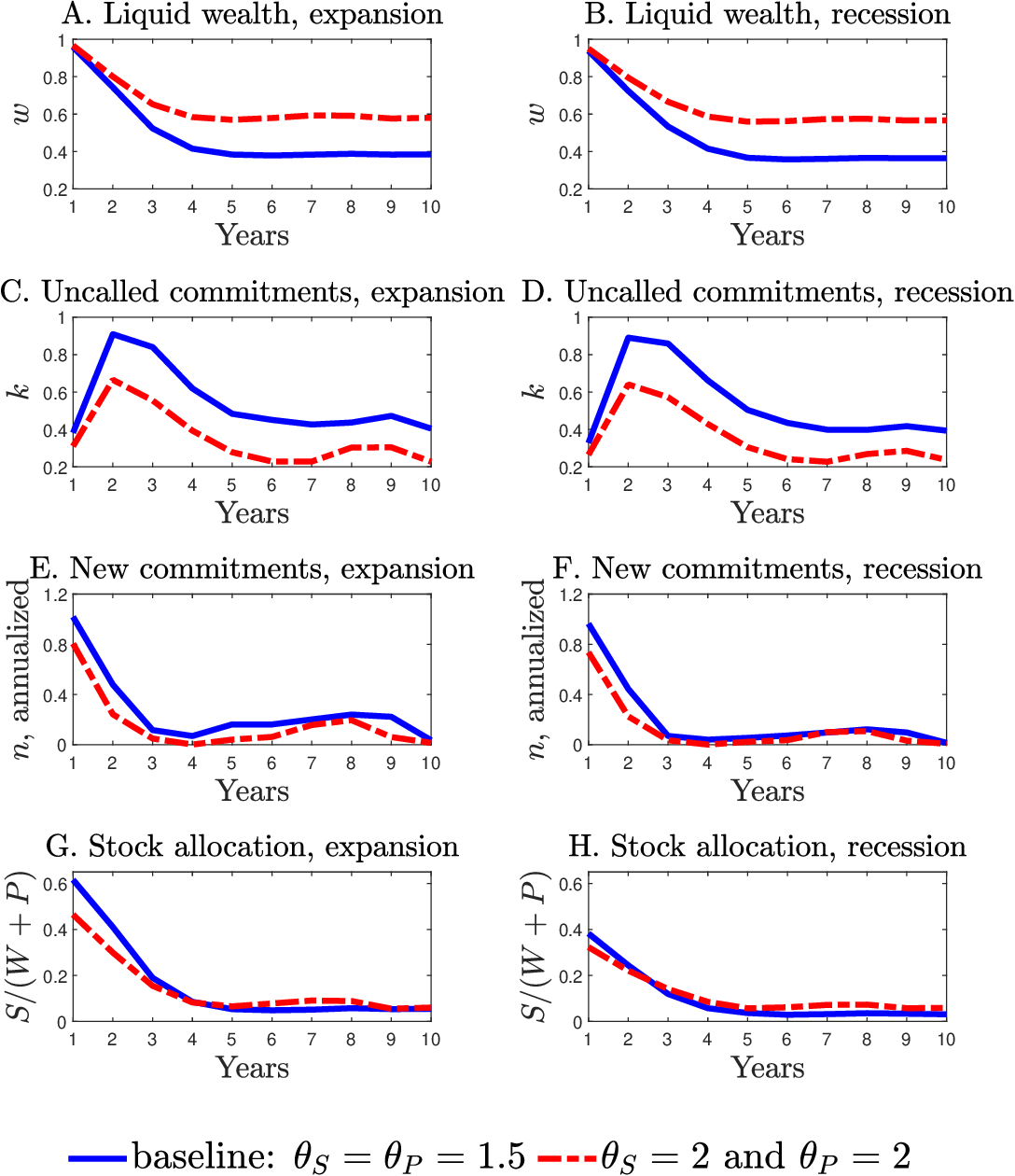}
  \caption{\textbf{Life cycle comparison (conditional), risk budget 1.}}\label{fig: lifecycle cond comp statics risk budget 1}
\end{figure}
\clearpage

\begin{figure}[htb!]
  \centering
  \includegraphics[width=1.0\textwidth]{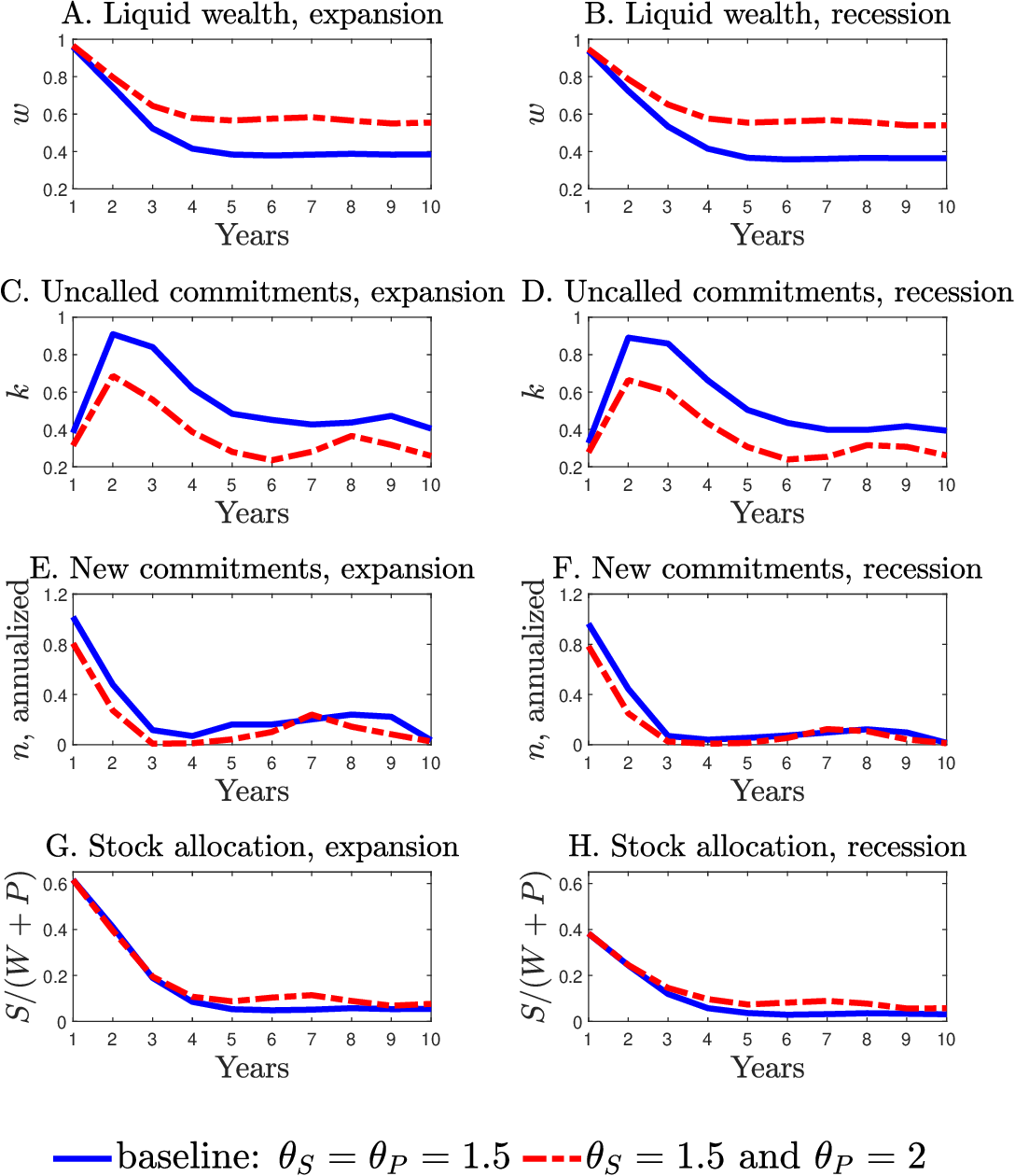}
  \caption{\textbf{Life cycle comparison (conditional), risk budget 2.}}\label{fig: lifecycle cond comp statics risk budget 2}
\end{figure}

\clearpage

\end{small}

%% remember: comment out for public version
%\input{to_do_list.tex}

\end{document}